\documentclass[twocolumn]{aa}
\usepackage{ulem}
\usepackage{graphicx}
\usepackage{longtable}
\usepackage{array}%
\usepackage{txfonts}
\begin{document}
   \title{The ACS Survey of Galactic Globular Clusters. XII.
	Photometric Binaries along the Main-Sequence}

\author{
A. \,P. \,Milone\inst{1,2,3},
G. \,Piotto\inst{3,4},
L. \,R. \,Bedin\inst{4,5},
A. \,Aparicio\inst{1,2},
J. \,Anderson\inst{5},
A. \,Sarajedini\inst{6},
A. \,F. \, Marino\inst{7},
A. \,Moretti\inst{4},
M. \,B. \,Davies\inst{8},
B. \,Chaboyer\inst{9},
A. \,Dotter\inst{5,10},
M. \,Hempel\inst{11},
A. \,Mar\'\i n-Franch\inst{12,1,2},
S. \,Majewski\inst{13},
N. \,E. \,Q. \,Paust\inst{5},
I. \,N. \,Reid\inst{5},
A. \,Rosenberg\inst{1,2},
M. \,Siegel\inst{14} 
          }
   \offprints{A.\ P.\ Milone}

\institute{
Instituto de Astrof\`\i sica de Canarias, E-38200 La
	      Laguna, Canary Islands, Spain
\and
Department of Astrophysics, University of La Laguna,
           E-38200 La Laguna, Tenerife, Canary Islands, Spain
\and
Dipartimento  di   Astronomia,  Universit\`a  di
  Padova, Vicolo dell'Osservatorio 3, Padova, I-35122, Italy
\and
INAF-Osservatorio Astronomico di Padova, Vicolo dell'Osservatorio 5,
I-35122 Padova, Italy 
\and
Space Telescope Science Institute, 3700 San Martin Drive,
Baltimore, MD 21218, USA
\and
Department of Astronomy, University of Florida,
	      211 Bryant Space Science Center, Gainesville, FL 32611, USA
\and
Max Plank Institute for Astrophysics, Postfach 1317, D-85741 Garching, Germany
\and
Lund Observatory, Box 43, SE 221-00, Lund, Sweden
\and
Department of Physics and Astronomy, Dartmouth College,
	      6127 Wilder Laboratory, Hanover, NH 03755, USA
\and
Department of Physics and Astronomy, University of Victoria, PO Box
3055, STN CSC, Victoria, BC, V8W 3P6 Canada
\and
P. Universidad Cat\'olica de Chile, Departamento de
Astronom\'ia y Astrof\'isica, Casilla 306, Santiago 22, Chile
\and
Centro de Estudios de F\'isica del Cosmos de Arag\'on (CEFCA), 44001, Teruel, Spain
\and
Dept. of Astronomy, University of Virginia,
P.O. Box 400325, Charlottesville, VA 22904-4325
\and
Department of Astronomy and Astrophysics, Pennsylvania State University, 525 Davey
Laboratory, State College, PA 16801
}

   \date{Received Xxxxx xx, xxxx; accepted Xxxx xx, xxxx}

%
   \abstract{

 {\it Context.} The fraction of binary stars is an important ingredient to interpret
 globular cluster dynamical evolution and their stellar population.\\
{\it Aims.}
   We investigate the properties of main-sequence binaries
   measured in a uniform 
 photometric
sample of 59 Galactic globular
   clusters that were observed by {\it HST} WFC/ACS as a part
   of the Globular Cluster Treasury project.\\
{\it Methods.}
We measured the fraction of binaries and the distribution of mass-ratio
 as a function of radial location within the cluster, from the central
 core to  beyond 
the half-mass radius.
We studied the radial distribution of binary stars, and the distribution
of stellar mass ratios.
We investigated monovariate relations between the fraction  of binaries and
the main parameters of their host clusters.\\
{\it Results.}
We found that  in nearly all the clusters, the total
fraction of binaries
is significantly smaller than the fraction of binaries in the
field, with a few exceptions only.
Binary stars are significantly more centrally concentrated than single MS stars
in most of the clusters studied in this paper.
The distribution of the mass ratio is generally flat (for mass-ratio
parameter {\it q}$>$0.5).
We found a significant anti-correlation between the binary fraction in
a cluster and its absolute luminosity (mass). Some, less significant
correlation with the collisional parameter, the central stellar
density, and the central velocity dispersion are present. There is no
statistically significant relation between the binary fraction and
other cluster parameters. We confirm the correlation between the binary
fraction and the fraction of blue stragglers in the cluster.
}

   \titlerunning{The Main-Sequence Binary Population in 59 Globular Clusters}
   \authorrunning{Milone et al.}
\maketitle
%
\section{Introduction}
The knowledge of the binary frequency in Globular Clusters (GCs)
is of fundamental importance in many astrophysical studies.
Binaries play an important role in the cluster dynamical evolution,
as they represent an important source of heating. They are also
important for the interpretation of the stellar populations in GCs.
A correct determination  of the stellar mass and luminosity functions in GCs
requires accurate measure of the fraction of binaries. Stellar evolution
in a binary system can be different from isolated stars in the field.
Exotic stellar  objects, like Blue Stragglers (BSSs),  cataclysmic variables,
millisecond pulsars  and  low mass  X-ray binaries
represent late evolutionary  stages    of close binary    systems. The
determination of the fraction   of binaries plays a fundamental  role
towards the understanding of the
origin and evolution of these peculiar objects.

There are three main techniques used in literature to measure the fraction
of binaries in GCs (Hut et al.\ 1992).
The first  one identifies binaries  by measuring their radial velocity
variability (e.\ g.\ Latham 1996). This method  relies on the detection
of  each  individual  binary  system   but,  due  to actual limits  in
sensitivity of  spectroscopy, these studies  are possible only for the
brightest GC stars. Moreover, this technique is sensitive  to binaries
with short orbital periods,  and the estimated  fraction  of binaries
depends  on the eccentricity distribution.
The second approach is based  on the search for photometric  variables
(e.\ g.\ Mateo 1996). As in  the  previous case, it  is  possible to infer
specific properties of each binary system (like the measure of orbital
period, mass ratio, orbital  inclination). Unfortunately,  this method
is  biased towards binaries  with   short  periods and  large  orbital
inclination. The estimated   fraction of binaries  depends on  the
assumed distribution of orbital periods, eccentricity and mass ratio.
Both of these  techniques have  a   low discovery efficiency  and are  very
expensive in  terms  of telescope time  because   of the necessity  to
repeat measures in different epochs.

A third method, based on  the analysis  of the  number of
stars located on the red side of the  MS fiducial line, may represent a
more efficient approach to measure the fraction of binaries in a cluster
for several reasons:
\begin{itemize}
\item{The availability of a large number (thousands) of stars makes it a
  statistically robust method;}
\item{It is efficient in terms of observational time: two filters are
  enough for detecting binaries, and repeated measurements are not needed.}
\item{It is sensitive to binaries with any orbital period and
  inclination.}
\end{itemize}
This latter approach has been used by many authors (e.\ g.\ Aparicio
et al.\ 1990, 1991, Romani \& Weinberg 1991, Bolte 1992, 
Rubenstein \& Baylin 1997, Bellazzini et al.\ 2002, Clark,
Sandquist  \& Bolte 2004,  Richer et al.\ 2004,  Zhao  \& Baylin 2005,
Sollima et al.\ 2007, 2009, 
Bedin et al.\ 2008, Milone et al.\ 2009, 2010, 2011) to study the
populations of binaries in  individual stellar clusters. The
relatively small   number   of clusters that have been  analyzed is a
consequence of the intrinsic difficulties of the method: 
\begin{itemize}
\item{High photometric quality is required and high resolution is
  necessary to minimize the fraction of blends in the central regions of GCs;}
\item{Differential reddening (often present) spreads the MS and
  makes it more difficult to distinguish the binary sequence from the single-star MS population;}
\item{An accurate analysis of photometric errors as well as a correct
  estimate of field contamination are necessary to
  distinguish real binaries from bad photometry stars and field objects.}
\end{itemize}

 The first study of binaries in a large sample of GCs comes from
  Sollima et al.\ (2007), who investigated the global properties of
  binaries in 13 low-density GCs. These authors found that the total
  fraction of    binaries ranges from 0.1 to 0.5 in the core depending on the
  cluster, thus confirming   
  the deficiency of binaries in GCs compared to the field where more
  than half of stars are in binary systems (Mayor et al.\ 1992,
  Dunquennoy \& Mayor 1991, Fischer \& Marcy 1992, Halbwachs et
  al.\ 2003,  Rastegaev et al.\ 2010, Raghavan et al.\ 2010). 
 At variance with the high fraction of binaries in field sdB stars (Masted et al.\ 2001, Napiwotzki et al.\ 2004), a  lack of close binaries among GC hot
horizontal branch stars (the cluster counterpart of field sdBs)  has
been confirmed by Moni Bidin et al.\ (2006, 2009).
 
 Sollima et al.\ (2010) extended the study of binaries to five
 high-latitude open clusters with ages between $\sim$0.3-4.3 Gyr and
 found that the fraction of binaries are generally larger than in GCs
 and range between $\sim$0.3 and 0.7 in the core.
 Very high binary fractions have been observed also in some young star
 clusters and for pre-main sequence T-Tauri stars, where the total
 binary fraction might be as high as 0.9 (e.\ g.\ Prosser et
 al.\ 1994, Petr et al.\ 1998, McCaughrean 2001, Duch{\^e}ne 1999).
 
 These findings suggest that the star formation condition, as well as the
 environment, could play a fundamental role on the evolution of binary
 systems. 
 The binary populations in star clusters has been investigated in
 detail, mainly through Monte-Carlo and Fokker-Plank simulations
 (e.\ g.\ Giersz \& Spurzem 2000, Fregeau et al.\ 2003, Ivanova et
 al.\ 2005), N-body (e.\ g.\ Shara \& Hurley 2002, Trenti, Heggie, \&
 Hut 2007, Hurley,  Aarseth, \& Shara (2007), Fregeau et al.\ 2009,
 Marks, Kroupa, \& Oh 2011) and fully analytical computations (Sollima 2008). 

 While the evolution of binaries stimulated by interactions with
 cluster stars could play the major role, there are many processes that
 also influence the binary population in stellar systems. For instance binary
 systems can form by tidal-capture (e.\ g.\ Hut et al.\ 1992, Kroupa
 1995a). Destruction of binaries may occur via coalescence of
 components through encounters or tidal dissipation between
 the components (Hills 1984, Kroupa 1995b, Hurley \& Shara
 2003). Stellar evolutionary processes can significantly effect the
 property of binaries and binary-binary interaction can led to
 collisions and mergers (e.\ g.\ Fregeau et al.\ 2004).  
The comparison of simulation results with
observed binary fraction is hence a powerful tool to shed light on
both the cluster and the binaries evolution. 
   
In this  paper, we report the  observational results of our
search for photometric binaries 
among GCs present in the  {\it HST} Globular Cluster Treasury
catalog (Sarajedini  et al.\ 2007, Anderson et al.\ 2008),
 which is based on {\it HST} ACS/WFC data We
exploited both  the    homogeneity of  this  dataset,  and    the high
photometric accuracy   of  the measures  to   derive  the  fraction of
binaries in the densest regions of 59 GCs. We deserve to future
   works any attempt to interpret the empirical findings presented in
   this paper. 

\section{Observations and data reduction}
\label{sec:data}
Most of  the data used in this  paper come from the
{\it HST}  ACS/WFC images taken for  GO 10775  (PI Sarajedini), an
{\it HST} Treasury project, where a total of 66 GCs were observed
through the F606W and F814W filters. For 65 of them, the database 
 consists in four or five F606W and F814W deep exposures plus a short
 exposure in each band. The pipeline used for the data reduction
allowed us to obtain precise photometry from nearly the tip of the
red giant branch (RGB) to several magnitudes below the main sequence turn-off (MSTO), typically reaching $\sim$0.2 ${\it m}_{\odot}$.

The GO\ 10775 data set as well  as the methods used for its photometric
reduction  have been  presented and described  in
papers II and IV of this series
(Sarajedini et al.\ 2007 and Anderson et al.\ 2008).
\footnote {
Due to a partial guiding failure, we only obtained part of the NGC 5987 data.
In this case the dataset  consists in three long exposures
in F814W and five in F606W, while only the F606W short exposure was
successfully obtained. For this cluster we obtained useful magnitudes
and colors for stars fainter than the sub giant branch and
with masses larger then $\sim$0.2 ${\it m}_{\odot}$. }

The uniform and deep photometry offers a database with unprecedented
quality that made possible a large number of studies (see
e.\ g.\ Sarajedini et al.\ 2010 and references therein).

In this paper we study the main sequence binary population in a subset
 of 59 GCs. We excluded three clusters (Lynga 7, NGC 6304, and NGC
 6717) that are strongly contaminated by field stars
 and for which there exist no archive {\it HST} data which could allow
 us to obtain reliable proper motions and separate them from cluster
 members. We also excluded Palomar 2 because of its high differential
 reddening, and NGC 5139 ($\omega$ Centauri), and NGC 6715
because of the multiple main sequences (Siegel et al.\ 2007, Bellini
et al.\ 2010 and references therein).
   The triple MS of NGC 2808 made the binary-population extremely
   complicated and we presented it in a separate
   paper (Milone et al.\ 2011a). 

In addition, we also  used archive {\it HST} WFPC2, WFC3 and  ACS/WFC
images from other programs to obtain proper motions, when images
overlapping the GO10775 images were available.
Table~1 summarizes the archive data used in the present paper.

The recipes of  Anderson et al.\  (2008) have been
used to reduce the archive  ACS/WFC data.  The WFPC2 data are analyzed 
by using  the programs  and the techniques  described in  Anderson \&
King (1999, 2000, 2003).  We measured star positions and fluxes on the WFC3
  images with a software mostly based on img2xym\_WFI (Anderson et
  al.\ 2006). Details on this program will be given in a stand-alone
  paper. Star positions and fluxes have been corrected for geometric
  distortion and pixel-area using the solutions provided by Bellini \&
Bedin (2009).
%
%
\begin{table*}[ht!]
\scriptsize{
\begin{tabular}{llccccr}
\hline\hline  ID & DATE & N$\times$EXPTIME & FILT & INSTRUMENT & PROGRAM & PI  \\
\hline
\hline
 ARP 2    & May 11 1997 		     & 1$\times$260s$+$5$\times$300s   & F814W  & WFPC2   &   6701 & Ibata, R.    \\
          & May 11 1997         	     &  5$\times$300s$+$1$\times$350s  & F606W  & WFPC2   &   6701 & Ibata, R. \\
\hline
 NGC 104  & Sep 30 2002 - Oct 11 2002        & 1$\times$10s$+$6$\times$100s$+$3$\times$115s & F435W  & ACS/WFC &   9281 & Grindlay, G.    \\
          & Jul 07 2002                      & 6$\times$60s$+$1$\times$150s    & F475W  & ACS/WFC &   9443 & King, I.\ R.    \\
          & Jul 07 2002                      & 20$\times$60s                   & F475W  & ACS/WFC &   9028 & Meurer, J.    \\
\hline
 NGC 362  & Dec 04 2003                      & 4$\times$340s                   & F435W  & ACS/WFC &  10005 & Lewin, W.    \\
	  & Dec 04 2003                      & 2$\times$110s$+$2$\times$120s   & F625W  & ACS/WFC &  10005 & Lewin, W.    \\
          & Sep 30 2005                      & 3$\times$70s$+$20$\times$340s   & F435W  & ACS/WFC &  10615 & Anderson, S.    \\
\hline
 NGC 5286 & Jul 07 1997                      & 3$\times$140s$+$1$\times$100s   & F555W  & WFPC2   &   6779 & Gebhardt, K.    \\
          & Jul 07 1997                      & 3$\times$140s$+$1               & F814W  & WFPC2   &   6779 & Gebhardt, K.    \\
\hline
 NGC 5927 & May 08 1994			     & 6$\times$50s$+$8$\times$600s    & F555W  & WFPC2   &   5366 & Zinn, R.  \\
          & May 08 1994			     & 6$\times$70s$+$8$\times$800s    & F814W  & WFPC2   &   5366 & Zinn, R.  \\
          & Aug 06 2002			     & 30s+500s	                       & F606W  & ACS/WFC &   9453 & Brown, T. \\
          & Aug 06 2002			     & 15s+340s	                       & F814W  & ACS/WFC &   9453 & Brown, T. \\
          & Aug 28 2010                      & 50s$+$2$\times$455s             & F814W  &UVIS/WFC3&  11664 & Brown, T. \\
          & Aug 28 2010                      & 50s$+$2$\times$665s             & F555W  &UVIS/WFC3&  11664 & Brown, T. \\
\hline
 NGC 6121 & Jun 19 2003                      & 15$\times$360s                  & F775W  & ACS/WFC &   9578 & Rhodes, J. \\
\hline
 NGC 6218 & Jun 14 2004 		     & 4$\times$340s		       & F435W  & ACS/WFC &  10005 & Lewin, W. \\
 	  & Jun 14 2004 		     & 2$\times$40s$+$2$\times$60s     & F625W  & ACS/WFC &  10005 & Lewin, W. \\
\hline
 NGC 6352 & Mar 29 1995                      &  7$\times$160s                  & F555W  & WFPC2   &   5366 & Zinn, R. \\
          & Mar 29 1995                      &  6$\times$260s                  & F814W  & WFPC2   &   5366 & Zinn, R. \\
\hline
 NGC 6388 & Jun 30 - Jul 03 2010             & 6$\times$880s                   & F390W  &UVIS/WFC3&  11739 & Piotto, G. \\
\hline
 NGC 6397 & Aug 01 2004 - Jun 28 2005	     & 5$\times$13s$+$5$\times$340s    & F435W  & ACS/WFC &  10257 & Anderson, J. \\
\hline
 NGC 6441 & Aug 04-08 2010                   & 6$\times$880s                   & F390W  &UVIS/WFC3&  11739 & Piotto, G. \\
\hline
 NGC 6496 & Apr 01 1999                      &  2$\times$1100s+4$\times$1300s  & F606W  & WFPC2   &   6572 & Paresce, F. \\
          & Apr 01 1999                      &  2$\times$1100s+4$\times$1300s  & F814W  & WFPC2   &   6572 & Paresce, F. \\
\hline
 NGC 6535 & Aug 04 1997                      &  8$\times$140s                  & F555W  & WFPC2   &   6625 & Buonanno, R. \\
          & Aug 04 1997                      &  9$\times$160s                  & F814W  & WFPC2   &   6625 & Buonanno, R. \\
\hline
 NGC 6624 & Oct 15 1994 		     & 6$\times$50s$+$8$\times$600s    & F814W  & WFPC2   &   5366 & Zinn, R. \\
\hline
 NGC 6637 & Mar 31 1995 		     & 6$\times$60s$+$8$\times$700s    & F814W  & WFPC2   &   5366 & Zinn, R. \\
\hline
 NGC 6652 & Set 05 1997 		     & 12$\times$160s                  & F814W  & WFPC2   &   6517 & Chaboyer, B. \\
\hline
 NGC 6656 & Feb 22 1999 - Jun 15 1999	     & 192$\times$260s                 & F814W  & WFPC2   &   7615 & Sahu, K. \\
          & Feb 22 1999 - Jun 15 1999	     &  72$\times$260s                 & F606W  & WFPC2   &   7615 & Sahu, K. \\
\hline
 NGC 6681 & May 09 2009 		     & 32$\times$300s                  & F450W  & WFPC2   &  11988 & Chaboyer, B \\
\hline
 NGC 6838  & May 21 2000                      &   2$\times$100s                & F439W  & WFPC2   &   8118 & Piotto, G. \\
          & May 21 2000                      &   2$\times$30s                  & F555W  & WFPC2   &   8118 & Piotto, G. \\
\hline
 TERZAN 7 & Mar 18 1997         	     &  1$\times$260s$+$5$\times$300s  & F814W  & WFPC2   &   6701 & Ibata, R. \\
          & Mar 18 1997         	     &  5$\times$300s$+$1$\times$350s  & F606W  & WFPC2   &   6701 & Ibata, R. \\
\hline
\end{tabular}
}
\label{tabelladati}
\caption{Description of the {\it HST} additional archive data sets used in this
  paper, other than those from GO-10775. }
\end{table*}
\subsection{Selection of the star sample}
\label{sec:selections}
Binaries that   are able  to   survive in the   dense environment  of a
GC  are so  close  that even the Hubble  Space Telescope
{\it HST} is  not able to resolve the single components.   For this
reason, light  coming from each  star   will combine, and the binary
system will appear as a single point-like source.
In this paper we will take advantage from this
fact to search for binaries by carefully studying the
region in the CMD where their combined light puts them.

If we consider the two  components of a binary system
and indicate with $m_{1}$,  $m_{2}$,   $F_{1}$, and $F_{2}$ their
magnitudes and fluxes, the binary  will appear as a single
object with a magnitude:
\begin{center}
$m_{\rm bin}=m_{1}-2.5~log(1+\frac{F_{2}}{F_{1}})$.
\end{center}
In  the case of  a binary formed  by two  MS  stars (MS-MS binary) the
fluxes are related to  the two stellar  masses ($\mathcal{M}_{1}$,
$\mathcal{M}_{2}$), and its  luminosity depends  on  the  mass  ratio
$q=\mathcal{M}_{2}/\mathcal{M}_{1}$ (in  the following we will assume
$\mathcal{M}_{2} < \mathcal{M}_{1}$, $q<1$). The equal-mass binaries
form a sequence that is almost parallel to the MS, $\sim$0.75
magnitudes brighter. When the  masses of  the   two components  are
different, the binary will appear redder and brighter than the primary
and populate a CMD region on the red side of the MS ridge line (MSRL)
but below the equal-mass binary line.
%
   \begin{figure}[ht!]
   \centering
   \includegraphics[width=8.5 cm]{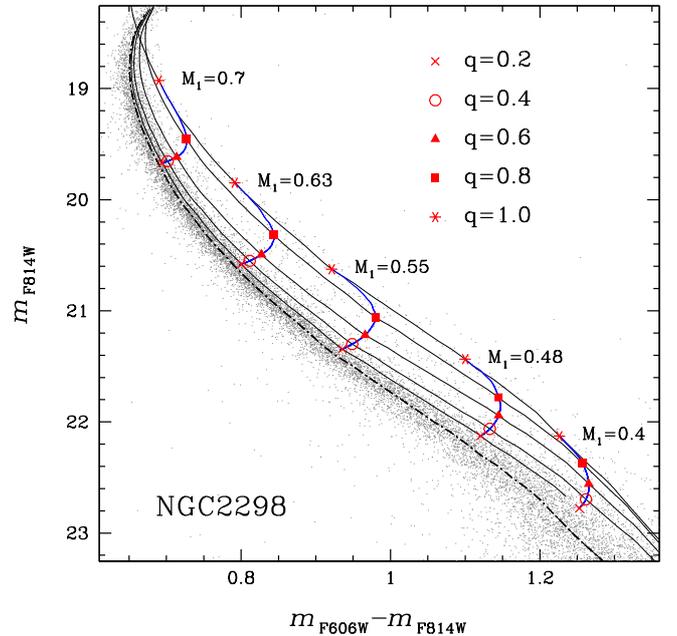}
      \caption{Model MS-MS binary sequences with different mass ratios for
      NGC 2298. The dashed-dotted line is the MSRL while, continuous
      black lines indicate the sequences of constant q and blue lines
      mark sequences of constant ${\mathcal M}_{1}$.
       }
         \label{f1}
   \end{figure}
%

In Fig.~\ref{f1} we  used  our empirical MSRL and   the mass-luminosity
relations  of Dotter et al.\ (2007)  to generate  sequences of  MS-MS
binary  systems   with different mass    ratios.  

An obvious  consequence of  this analysis is  that our   capability in
detecting  binaries mainly depends  on the photometric  quality of the
data.   Distinguishing the binary populations in clusters requires
   high-resolution images and high-precision
   photometry.  Not all stars in clusters can be measured equally well.
   Crowding, saturation, and image artifacts such as diffraction spikes,
   bleeding columns, hot pixels, and cosmic rays can
prevent certain stars from being measured well.  The first challenge to this project
   will be to identify which stars can be measured well and which are
   hopeless.

\begin{figure}[ht!]
   \centering
   \includegraphics[width=9.00 cm]{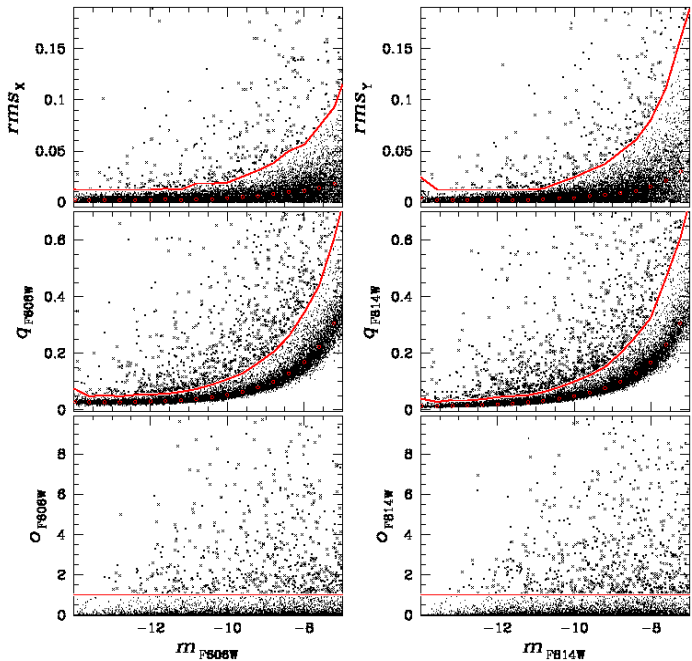}
      \caption{
	Diagnostic parameters that we have used  to select the sample
        of NGC 2298 stars  with  high-quality  photometry. The
        parameters are plotted  as  a  function of  the instrumental
        ${\it m}_{\rm  F606W}$ and ${\it m}_{\rm  F814W}$ magnitudes.
        Red circles indicate the median
${\it rms}_{\rm X, Y}$, and ${\it q}_{\rm F606W, F814W}$ per intervals
        of 0.4 magnitude.
Red  lines separate the well measured stars (thin points) from those
that are more likely to  have poorer  photometry (thick points). See text
for details.
	}
         \label{selezioni1}
   \end{figure}

    In addition to the basic stellar positions and photometry, the
    software described in Anderson et al.\ (2008) calculates 
several useful parameters that will help us reach this goal.  The
following  parameters are provided for every star:
\begin{itemize}
\item{The  rms of the  positions measured  in different  exposures and
  transformed into a common reference frame ( ${\it rms}_{\rm X}$ and ${\it rms_{\rm Y}}$) }
\item{The average residuals of the PSF fit for each star (${\it q}_{\rm F606W}$
   and ${\it q}_{\rm F814W}$ ) }
\item{
The total amount of flux in the 0.5 arcsec aperture from neighboring
stars relative to the star's own flux
(${\it o}_{\rm F606W}$   and ${\it o}_{\rm F814W}$).
	}
\end{itemize}
True binary stars will be so close to each other as to be
indistinguishable from single stars in our images, so the
    binarity has no impact on the above diagnostics.
\footnote{ As an example, in the closest GC, NGC 6121, 1 AU corresponds to
  $\sim$0.5 mas i.\ e. $\sim$0.01 ACS/WFC pixel.}
Therefore, it is safe to use the above diagnostics to indicate which stars
    are likely measured well and which ones are likely contaminated.
As an example, in  the six panels  of Fig.\ref{selezioni1}, we
show these parameters as a function of the instrumental
\footnote{The instrumental magnitude is calculated as $-$2.5 log(DN),
  where DN is the total number of digital counts above the local sky for the considered stars}
 ${\it m}_{\rm  F606W}$ and ${\it m}_{\rm  F814W}$
magnitudes, and  illustrate the criteria that we have used to select the sample of
stars with the best photometry for NGC 2298.

We note a clear trend in the quality fit  and the {\it rms} parameters
as a function  of the magnitude. At all magnitudes, there are  outliers
that are likely sources   with   poorer   photometry  and that need
to be removed before  any analysis.  Because  of  this,   we  adopted
the  following procedure to select the best measured stars.

   \begin{figure}[ht!]
   \centering
   \includegraphics[width=9.00 cm]{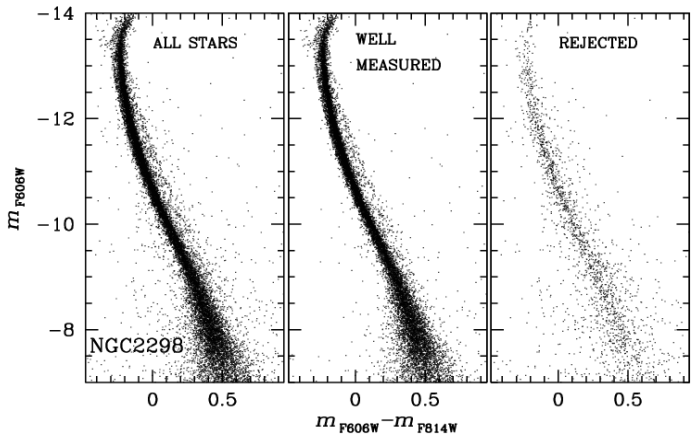}
      \caption{CMD of all the  measured stars (\textit{left}),
         of stars that passed our criteria of selection
         (\textit{middle}), and CMD of rejected stars (\textit{right}) }
         \label{selezioni2}
   \end{figure}
We began by  dividing all the stars of each cluster into bins  of 0.4 magnitude;
for each of them, we computed the median values of the parameters
${\it rms_{\rm X, Y}}$ and ${\it q_{\rm F606W, F814W}}$ defined above
and the $68.27^{\rm th}$ element of the percentile distribution
(hereafter $\sigma$).
We added to the median  of each bin four  times $\sigma$,
and fitted  these points with a  spline to obtain  the red lines of
 Fig.~\ref{selezioni1}. All stars below  the red line
have been flagged as `well-measured' according to that diagnostic.

The parameters ${\it o}_{\rm F606W}$ and ${\it o}_{\rm
 F814W}$ do not show a clear trend with  magnitude. We flagged as
`well-measured' all the stars with ${\it o}_{\rm F606W}<1$ and ${\it o}_{\rm F814W}<1$.

In  Fig.~\ref{selezioni2} we compare the color-magnitude diagram (CMD) of all the
measured  stars of  NGC 2298  ($left$), the CMD  of  stars that  pass all  the
selection criteria ($middle$), and the CMD of rejected stars ($right$).
The sample  of stars that have been  used in the analysis that follows
includes  stars flagged  as  `well-measured' with respect to
all the  parameters we used as diagnostics of  the photometric  quality.

The photometric catalog by Anderson et al.\ (2008) also provides
the  rms  of  the  ${\it m}_{\rm F606W}$ and ${\it m}_{\rm F814W}$ magnitude
 measures  made  in   different   exposures.
However, a star can have a large magnitude rms either because of poor
photometry or because it is a binary system with short period
photometric variability. 
 In order to avoid the exclusion of this class of binaries,
we  have not  used  the rms of magnitude measures  as
diagnostics  of the photometric  quality in  the selection of our stellar
sample.
\subsection{Artificial-star tests}
\label{sec:artsample}
Artificial-star (AS) tests played a  fundamental role  in this  analysis; they
allowed us  to determine the completeness  level of our  sample, and to
measure  the  fraction  of chance-superposition  ''binaries''.
   The GC Treasury reduction products (see Anderson et al.\ 2008)
   also contain a set of AS tests.  The artificial
   stars were inserted with a flat luminosity function in F606W and
   with colors that lie along the MSRL for each cluster.  Typically,
   $10^{5}$ stars were added for each cluster, with a
spatial density that was flat within the core, and declined as
$r^{-1}$ outside of the core.
The stars were added one at a time, and as such they
   will never interfere with each other.

Each  star in the  input AS catalog is added to each
image with  the appropriate position and flux. The AS routine measures the images
with  the same procedure  used for  real stars  and produces  the same
output  parameters as  in Sect.~\ref{sec:data}.  We  considered an
artificial star as recovered when the  input and the output fluxes differ
by less than 0.75 magnitudes and the positions by less than 0.5 pixel.
 We applied to the recovered ASs the same criteria of selection described in
Sect.~\ref{sec:data} for real stars and based on the rms in position
and on the ${\it q}_{\rm F606W, F814W}$ and ${\it o}_{\rm F606W, F814W}$ parameters.
 In what follows, including the completeness measure, we used only the
 sample of ASs that passed all the criteria of selection. 

\begin{figure*}[ht!]
	\centering
   \includegraphics[width=13cm]{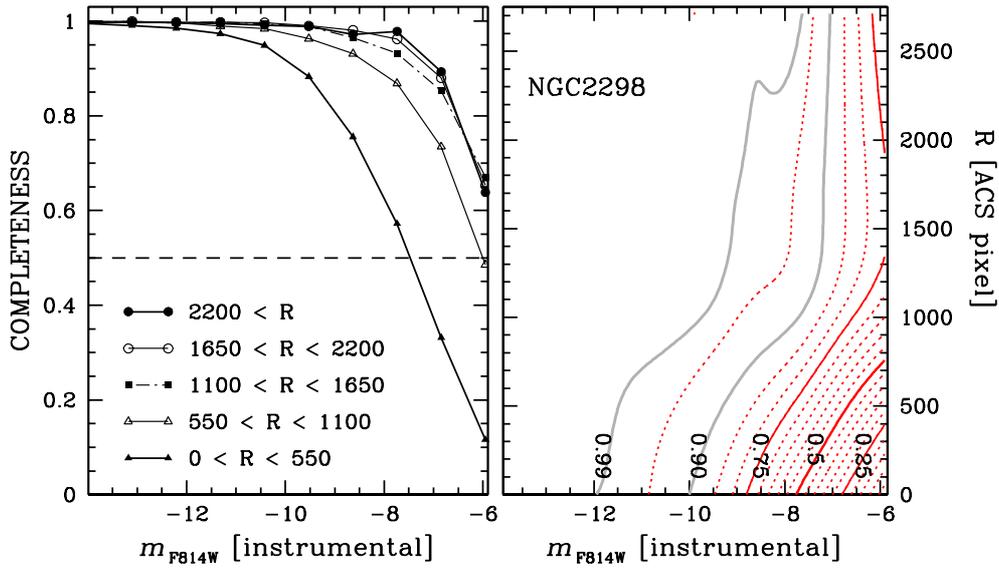}
      \caption{
        \textit{Left}:  Completeness  as  a function  of  the  $m_{\rm
          F814W}$  magnitude  in  five  annuli (the inner and outer
        radius of each annulus, in ACS  pixels, are quoted in the inset)
        for  NGC 2298.   \textit{Right}:
        Completeness  contours in the  radial distance  versus $m_{\rm
          F814W}$ magnitude plane.  The completeness levels
          corresponding to the red and gray continuous lines are
          quoted in the figure. Dotted lines indicate differences of
          completeness of 0.05 ranging from 0.05 to 0.95.     }
         \label{compl}
   \end{figure*}

Since  completeness  depends  on   crowding  as  well  as  on  stellar
luminosity, we measured it applying a procedure that accounts for both
the stellar  magnitude and the  distance from the cluster  center.  We
divided the  ACS field into 5  concentric annuli and,  within each of
them, we examined AS results in 9 magnitude bins, in the interval $-14
< {\it m}_{\rm F814W}< -5$.  For each of these 9 $\times$ 5 grid points we
calculated the completeness  as the ratio of recovered  to added stars
within that  range of radius and  magnitude. Finally, we
  interpolated  the  grid  points  and derived  the  completeness
  value associated with  each star. This grid  allowed us to 
estimate  the completeness  associated  to any  star  at any  position
within the cluster. 
 Results are shown in Fig.~\ref{compl} for NGC 2298.
 The stars used to measure the binary fraction have all
  completeness larger than 0.50.

%
\section{Photometric zero point variations}
\label{ZPvariations}
     In some clusters, the distribution of foreground dust can
     be patchy, which causes a variation of the reddening with
     position in the field, resulting in a non-intrinsic broadening
     of the stellar sequences on the CMDs.
In addition to these spreads, small unmodelable PSF
variations, mainly  due to focus  changes, can introduce  slight shifts
in the photometric zero point as a function of the star location
in the chip (see Anderson et  al.\ 2008 for details).  The 
 color variation due to inaccuracies in the PSF model
is  usually $\sim$0.005 (Anderson et al.\ 2008, 2009,
Milone et al.\ 2010).
In some  clusters,
differential  reddening effects may be  much  larger.   An appropriate
correction for these
effects is a fundamental step, as it can greatly sharpen
the MS, with a consequent
improved analysis of the MS binary fraction.

\subsection{Differential reddening}
\label{reddening}

%
   \begin{figure*}[ht!]
   \centering
   \includegraphics[width=14cm]{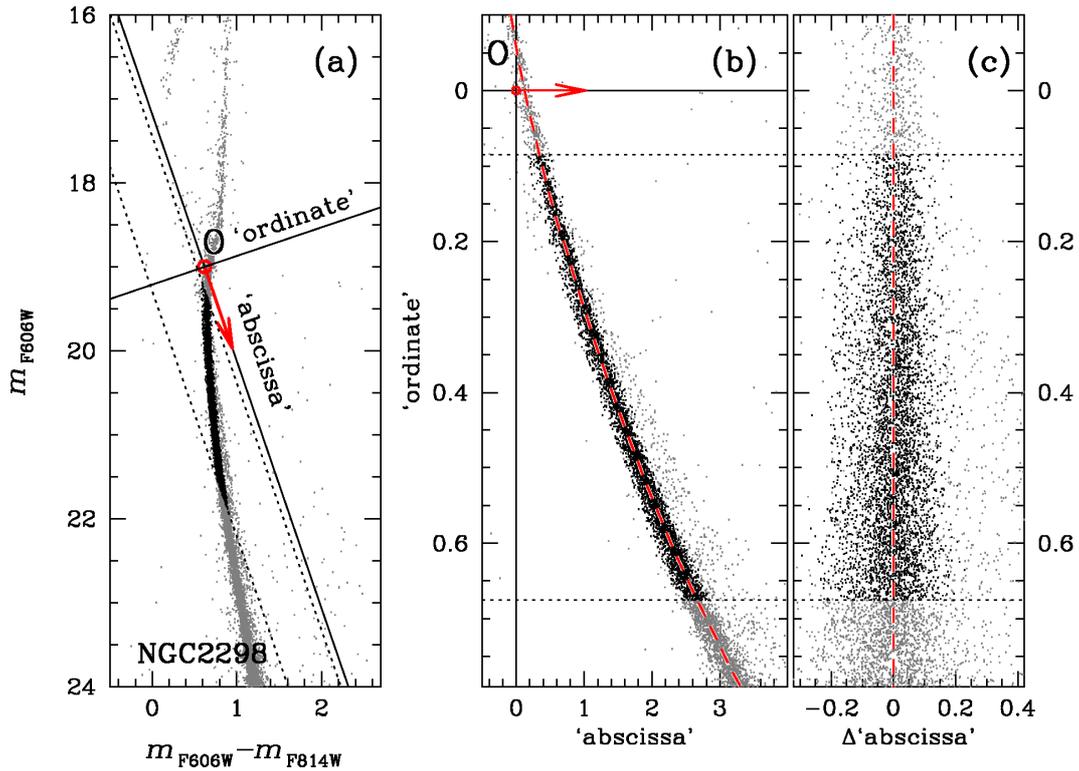}
      \caption{\textit{Panel a}: observed CMD of NGC 2298; the arrow
	indicates the direction of reddening.
	The continuous lines are the axes (`abscissa' and `ordinate')
	of the reference frame introduced in the
	procedure for the measurements of reddening variations.
	The position of NGC 2298 stars in this reference frame
	is shown in \textit{panel b} where we
	draw the fiducial line of the MS as a dashed red line.
	Stars between  the  dotted lines  (black  points)
	have   been  used  as  reference stars.
	\textit{Panel C} shows the rectified  `ordinate' vs. $\Delta$`abscissa' diagram.
       }
         \label{procred}
   \end{figure*}
%
   \begin{figure*}[ht!]
   \centering
   \includegraphics[width=13cm]{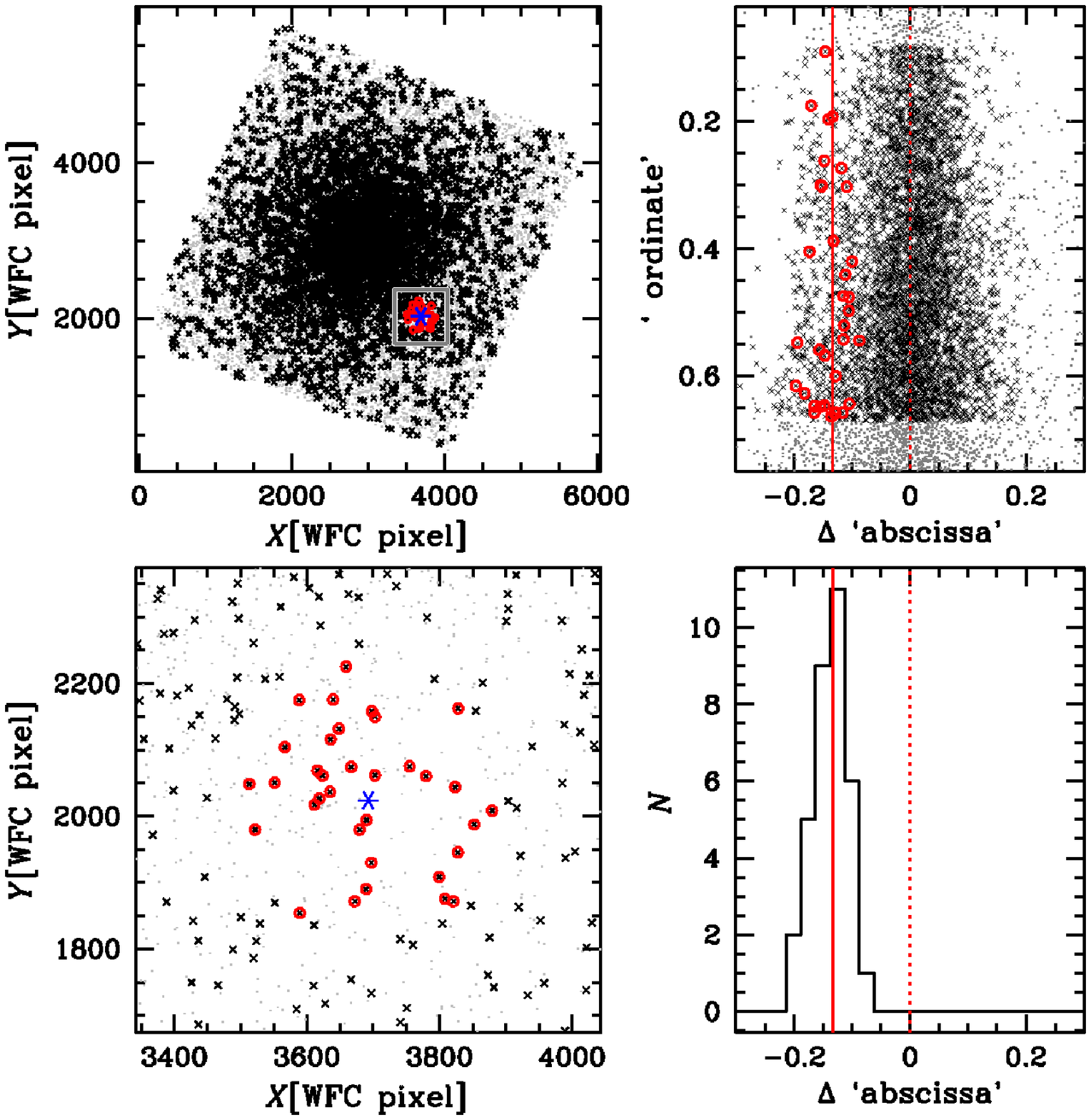}
      \caption{ Visualization of the local approach for the estimate
	of the differential reddening suffered by the  target
	star indicated with the blue
	asterisk. \textit{Upper left}: Position of NGC 2298
	stars in the ACS/WFC field of view (gray points).
  Red circles highlight the 35 reference stars in the {\it lower left} panel.
	Reference stars are indicated with black crosses, among them,
	 the 35 closest	neighbours (of the target star)
	 are marked with red circles.
	The \textit{lower left} panel is a zoom of a 700$\times$700 pixel
	centered on the target star.
	\textit{Upper right}: `ordinate' vs.\ $\Delta$ `abscissa' for all
	the stars in the NGC 2298 field of view. The median $\Delta$ `abscissa'
	of the 35 closest neighbor is indicated by the continuous red
	vertical line and corresponds to the differential reddening value
	suffered by the target star. The histogram of the $\Delta$ `abscissa'
	distribution of the 35 closest	neighbors is shown in the
	\textit{bottom right} panel.
	}
         \label{neigh}
   \end{figure*}
  In  order  to correct  for  differential  reddening,
	we started  by
  defining a  photometric reference  frame where the abscissa  is parallel  to the
  reddening line,  as shown in Fig.~\ref{procred} for  NGC 2298.  To do
  this, we have  first arbitrarily defined a point  ({\rm O}),
	near the MSTO in the
  CMD of $Panel~a$. Then  we have translated  the CMD such  that the origin of  the new
  reference  frame  corresponds to  {\rm  O}. Finally, we  have
  rotated the CMD counterclockwise by an angle:
\begin{center}
 $ \theta={\rm arctan} \frac{A_{\rm F606W}}{A_{\rm F606W}-A_{\rm F814W}} $
\end{center}
as shown in Fig.~\ref{procred}b.
The two quantities $A_{\rm F606W}$ and
$A_{\rm F814W}$ are the absorption coefficients in the F606W and F814W
ACS bands corresponding to the average reddening for each GC.
They are derived by assuming, for each GC, the average $E(B-V)$
listed in the Harris\ (1996, 2003) catalog and linearly interpolating
among the reddening and the absorption values given
in Table~3 of Bedin et al.\ (2005) for a cool  star.
The reason  for rotating the CMD is that it is much more intuitive to
determine a reddening difference  on  the horizontal  axis  rather
than  along the  oblique reddening line.

The  value of $\theta$ depends weakly on the stellar  spectral type,
but this variation can be ignored for our present purposes.
For simplicity, in this section,  we will  indicate as  `abscissa',
the  abscissa of  the rotated reference  frame, and  as  `ordinate',
its  ordinate. 

At this  point, we adopt an  iterative procedure that  involves the
following four steps: 
\begin{enumerate}
\item
We generate the red fiducial line shown in Fig.~\ref{procred}b.   In
order  to determine  this  line, we  used only MS stars. 
   We  divided  the  sample  of these MS reference stars into
  `ordinate' bins of 0.4 mag.  For each bin, we
  calculated the median `abscissa' that has been associated with the
  median `ordinate' of  the stars in the bin.   The fiducial has been
  derived  by  fitting these  median  points  with  a cubic spline.
  Here, it  is important to emphasize that the  use of the median
  allows us to minimize  the influence of  the outliers  as
  contamination by binary stars left in the sample,  field stars or
  stars with poor photometry. 
\item
For each star, we calculated the distance from the
  fiducial line along the reddening direction ($\Delta$~`abscissa').
In the right panel of Fig.~\ref{procred}, we plot `ordinate' vs. $\Delta$~`abscissa' for NGC 2298.
\item
We selected the sample of stars located  in the regions of the  CMD where the reddening line define a wide angle
with the fiducial line so that the shift in color and magnitude due to  differential reddening
 can be more easily separated from the random shift due to photometric errors.
These stars are used as reference stars to estimate reddening
variations associated to each star in the CMD and  are marked in
Fig.~\ref{procred} as heavy black points.

\item
The basic idea of our procedure, which is applied to each star (target) individually,
is to  measure  the differential  reddening
  suffered by the target star by  using the position in the `ordinate' vs.
 `abscissa' diagram
 of a local sample of reference stars located in a small spatial region
around the target with respect to the  fiducial  sequence.

   We must adopt an appropriate size for the comparison
    region in order to obtain the best possible reddening
    correction.  The optimal size is a compromise between
    two competing needs.  On one hand, we want to use the
    smallest possible spatial cells, so that the systematic
    offset between the  'abscissa' and the fiducial ridgeline
    will be
measured as accurately as possible for each
    star's particular location.  On the other hand, we want
    to use as many stars as possible, in order to reduce the
    error in the determination of the correction factor.

  As a compromise, for each star,  we typically selected the
  nearest 30-100 reference stars \footnote{The exact number 
    adopted for each cluster depends on the total number of reference stars with
    a larger number of stars used for the most populous clusters.}
and  calculate  the median  $\Delta$~`abscissa' that  is
  assumed as the reddening correction  for that star.
In this way, our
  differential  reddening correction  will be  done at  higher spatial
  frequencies in the more populated parts of the observed field.
In calculating the differential reddening suffered by a reference star, we excluded
this star in the computation of the median $\Delta$~`abscissa'.
  As an example, in Fig.~\ref{neigh} we illustrate this procedure for a star
	in the NGC 2298 catalog.
 The position of all the stars within the ACS/WFC field of view
 is shown in the upper-left panel where
 reference stars are represented by black crosses, and the remaining stars
 are indicated with gray points. Our target is plotted as a blue asterisk.
The 35 closest neighboring reference stars are marked
 with red circles.
The lower-left panel is a zoom showing the location of
  the selected stars in a 700$\times$700 pixel box centered on the target.
The positions of the 35 closest neighboring reference stars in the
  `ordinate' vs.\ $\Delta$ `abscissa' plane are shown in the upper right
  panel, and their histogram distribution is plotted in the bottom-right
   one. Clearly, neighboring stars define a narrow sequence
 with $\Delta$ `abscissa' $\sim$$-$0.15.
 Their median $\Delta$ `abscissa', which is indicated by the continuous
red line, is assumed to be the best estimate of the differential reddening
suffered by the target star.
\end{enumerate}
After the  median  $\Delta$`abscissa' have  been  subtracted 
to the `abscissa' of  each star in the  rotated CMD,
we obtain an improved  CMD which has been used
to derive a more accurate selection of the sample of MS reference
stars and derive a more precise fiducial line.
        After step 4, we have a newly corrected CMD.  We
        re-run the procedure to see if the fiducial sequence needs
        to be changed (slightly) in response to the adjustments
        made and iterated.  
Typically, the procedure converges after about four
        iterations.
Finally, the corrected `abscissa' and
`ordinate' are converted  to $m_{\rm  F606W}$  and
$m_{\rm F814W}$ magnitudes.

\begin{figure}[ht!]
   \centering
   \includegraphics[width=9.00cm]{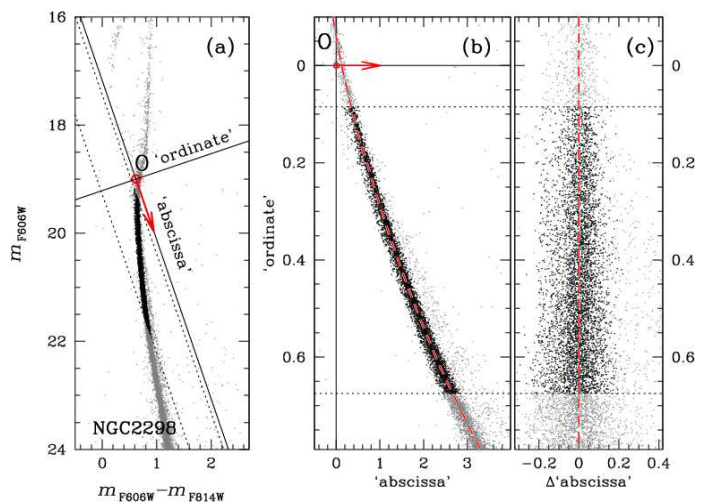}
      \caption{
	\textit{Bottom-left}: Map  of  differential reddening in the  NGC 2298
        field of view.  The gray levels
      correspond to the magnitude  of the variation  in local
        reddening as indicated in the  \textit{upper-right} panel.
	We divided the field of view into 8 horizontal slices
	and 8 vertical slices. \textit{Upper-left} and
        \textit{lower-right} panels plot $\Delta~E(B-V)$ as a function
        of the Y and X coordinate.} 
         \label{redmap}
   \end{figure}

From star-to-star comparison of the original and the corrected magnitudes
we can estimate star to star variations in $E(B-V)$
and derive the reddening map in the direction of our target GCs.
As an example, in Fig.~\ref{redmap},  we divide the field of view into 8 horizontal slices
and 8 vertical slices and plot $\Delta~E(B-V)$ as a function of the Y
(upper panels) and X coordinate (right panels).
We have also divided the whole field of view into 32$\times$32 boxes of 128$\times$128 ACS/WFC pixels
and calculated the average $\Delta~E(B-V)$ within each of them.
The resulting reddening map is shown in the lower-left panel
where each box is represented as a gray square. The  levels of gray
are indicative of the amount of differential reddening as shown in the upper right plot.
The analysis of the intricate reddening structures in our GC fields is beyond the purposes of the
present work and will be presented in a separate paper (King et al., in preparation).

\begin{figure*}[ht!]
   \centering
   \includegraphics[width=14cm]{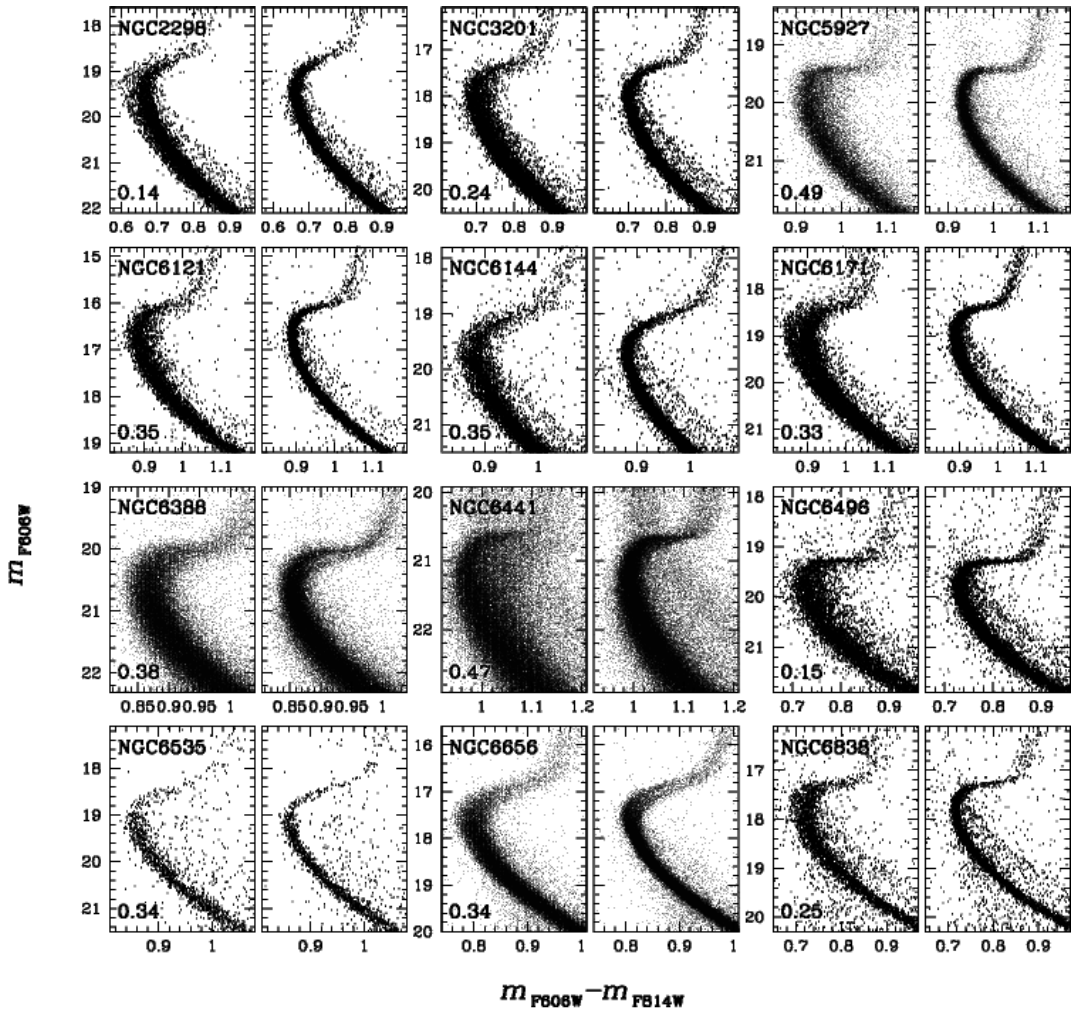}
      \caption{  CMDs of twelve GCs before (left) and after
        (right)
	the correction for differential reddening.
 For each cluster we give the average reddening  from the Harris (1996) catalog.
}
         \label{REDDENINGS}
   \end{figure*}

Fig.~\ref{REDDENINGS} shows the
CMDs of twelve of the GCs studied in this paper including NGC
  2298. These are the clusters that  revealed the largest
differential reddening $\Delta E({\it m}_{\rm F606W}-{\it m}_{\rm F814W})>0.05$.

%
\subsection{ PSF Variations}
\label{zeropoint}
Some GCs have  a  reddening that is close  to  zero  and  therefore we
expect negligible variations of reddening within  their field of view.
In these cases, we need to apply only a correction for the photometric  zero
point  spatial variation due to small, unmodelable  PSF
variations. Usually,   these PSF variations  affect   each filter in a
different way, so their  most evident manifestation  is a slight shift
in the color of the cluster sequence as  a function of the location in
the field (Anderson et al.\ 2008).  For this  reason, 
when the average reddening of the cluster (from Harris 1996) is lower than 0.10
mag, we  did not follow the recipes for the correction of
  differential reddening described in the previous section, but
corrected our photometry
for the effects of the variations of the photometric zero point
along the chip. We used a procedure  that slightly differs from the one
 of Sect.~\ref{reddening}.  
 The only difference from what done in GCs with high reddening is
  that we did not rotated the CMD and so we did not apply
the correction along the reddening line, but along the color direction.

%
   \begin{figure*}[ht!]
   \centering
   \includegraphics[width=12cm]{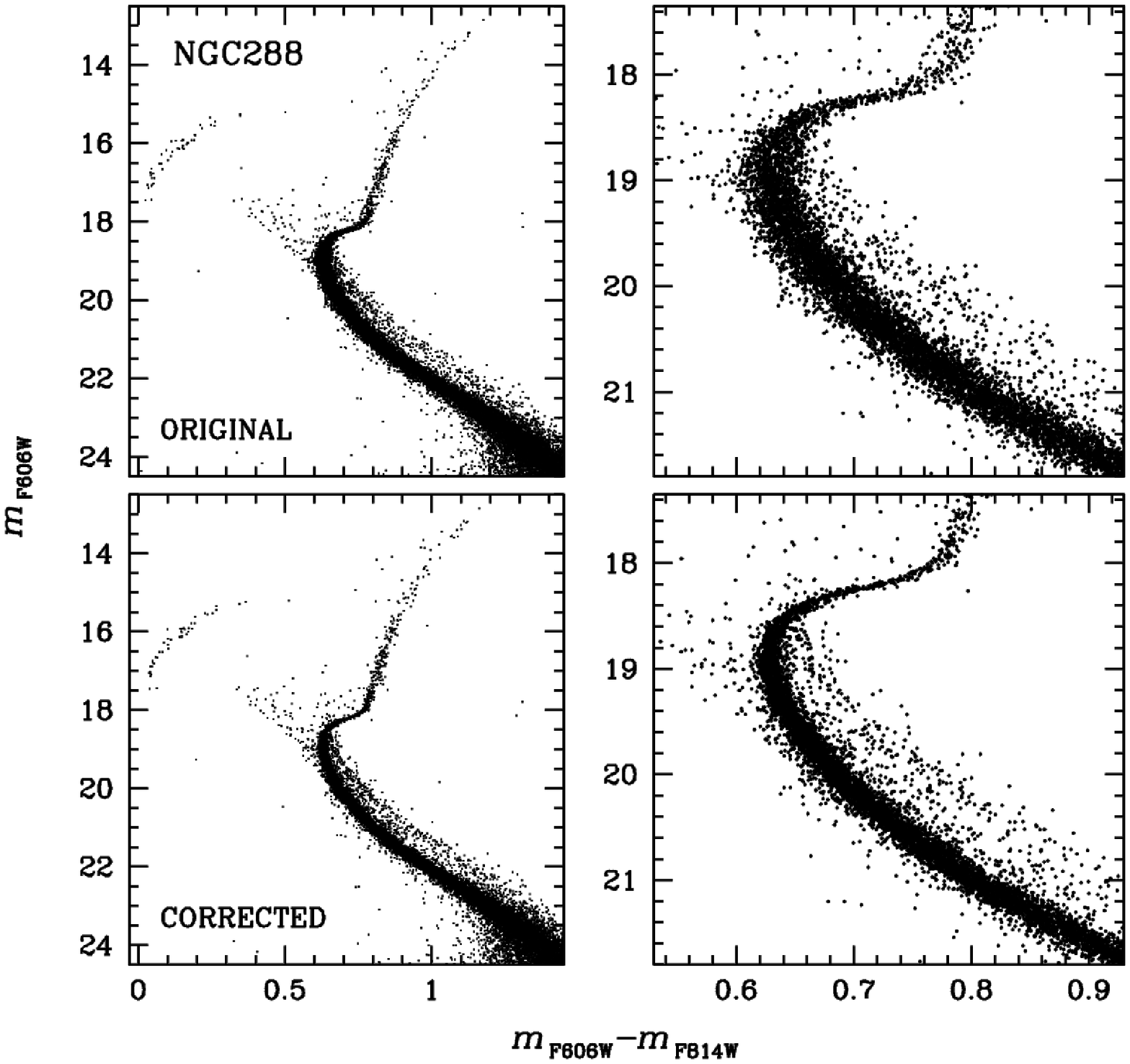}
      \caption{CMDs of NGC 288 before ({\it top}) and after ({\it bottom}) correction for
	photometric zero points variations.
	Right panels show a zoom of CMD around the SGB and upper-MS region.}
         \label{cor288}
   \end{figure*}

%
   \begin{figure*}[ht!]
   \centering
   \includegraphics[width=14cm]{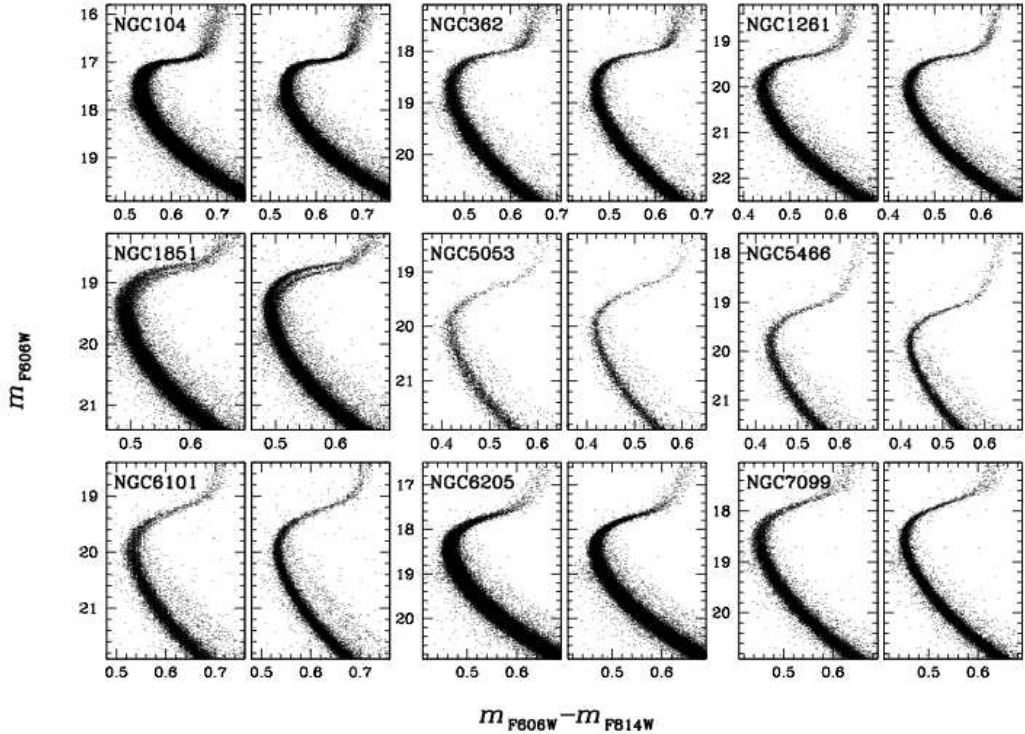}
      \caption{ Comparison of the CMD of nine GCs studied in this paper
	before (\textit{left}) and after (\textit{right})
     the correction for photometric zero points variations. }
         \label{ZP}
   \end{figure*}

The results of this procedure are illustrated in Fig.~\ref{cor288} where
we compare the original and the correct CMD of NGC 288.
The improvement in the quality of our CMD
is exemplified by the comparison in right panels figures that show a zoom
of the SGB and the upper-MS.

Other examples of the improvement in the photometry coming from
this procedure are shown in Fig.~\ref{ZP} where we plot
the nine GCs studied in this
paper for which
we measured the largest color variations.
 The average color variations are typically around 0.005 mag for
  each cluster with $E(B-V)<$0.10 studied in this paper and never exceed
  0.035 mag. 

\section{The measure of the fraction of binaries with high mass ratio }
\label{sec:highq}
Binaries with large mass ratios  have a large offset in luminosity
from the MSRL and are relatively easy to detect.  On the contrary, a
small mass ratio doesn't pull them very far off of the MSRL, making
them hard to distinguish from single MS stars.
Finally, the  low signal to noise photometry of faint
stars  limits  the range  where  binaries can  be detected and studied.

In practice,  the limited photometric   precision and accuracy makes
impossible  the
direct measure of the overall  population of binaries
without assuming a specific distribution of mass ratios {\it f}({\rm q}).
For this reason, in this paper, we followed two different approaches
to study the binary population in our target GCs.\\
{\it 1}) We isolated different samples of high
mass-ratio binaries (i.\ e.\ the  binary systems with $q>$ 0.5, 0.6
and 0.7).
For them, we obtained a direct measure of their fraction with respect
to  the total number of MS stars, and studied the properties of each group
(Sect.~\ref{sec:highq}). \\
{\it 2}) We determined the total fraction of binaries by assuming a
given {\it f}({\rm q}) (Sect.~\ref{totfraction}).

In  each cluster, we estimated the  fractions
 of high q binary stars in the
F814W magnitude interval ranging from 0.75 (${\it m}_{\rm F814W,
  bright}$) to 3.75 (${\it m}_{\rm F814W, faint}$) 
magnitudes below the MSTO.
\footnote{In the cases of NGC 6388 and NGC 6441 we used a smaller
  magnitude interval between 0.75 and 2.25
  magnitudes below the  MSTO. This exception is due to the fact that,
  as we will see in  Sect.~\ref{PM}, we do not have reliable
  proper motions to estimate the numbers of faint field stars in the
  CMDs of these two GCs.}
In this work we used the MSTO magnitudes from Mar\'\i n-Franch et
al.\ (2009), who used our same photometric data base.
The choice of  this  magnitude  interval  represents  a   compromise between  the
necessity of a large set of stars and the need  to avoid faint stars
 to be able to measure the  binary fraction also in clusters with
poorer photometry (because of crowding).

\begin{figure*}[ht!]
   \centering
   \includegraphics[width=14 cm]{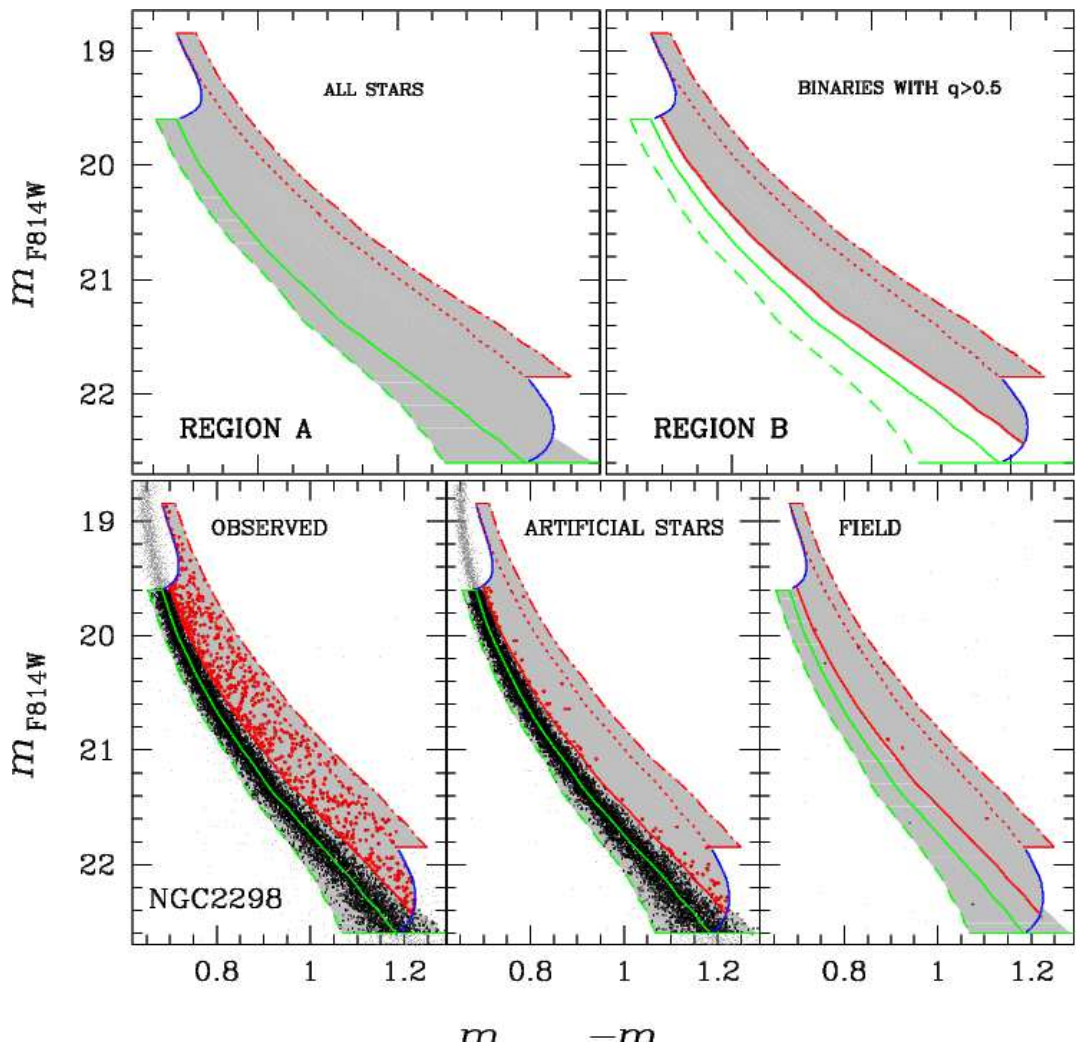}
      \caption{
Gray areas in the upper panels are the regions {\it A}, and {\it B} of the NGC 2298 CMD adopted
      to select all the (single and  binary) cluster stars ({\it left}) and the  candidate  binaries with  {\it q}$>$0.5
      ({\it right}), in a   range of 3 $m_{\rm F814W}$ magnitudes.
	In all panels, the MSRL is represented as a green continuous line, while the
green dashed line is blue shifted from the MSRL by three times the
color error. The red continuous line is the locus of MS-MS binaries
with mass ratio {\it q}=0.5, 
while the red dotted line is the locus of MS-MS binaries
whose components have equal mass.
The red dashed dotted line is displaced
to the red from the dotted line by 3 times the color error.
Lower panels show the observed CMD of NGC 2298 ($left$), the artificial stars CMD ($middle$), and the CMD of field stars ($right$).
       }
         \label{reg1}
   \end{figure*}

To illustrate our setup, Fig.~\ref{reg1} shows
the various regions we studied in the CMD of
NGC 2298 in order to measure the fraction of binaries with mass ratio
${\it q}>0.5$ for this cluster.
The upper half of the figure displays two regions of the
CMD: Region {\it A} (upper left) and region {\it B} (upper right).

Region {\it A} includes all the stars that we can consider to be cluster
members. It includes: all the single MS stars and the MS+MS binaries
with  a primary  star that have
${\it m}_{\rm F814W, bright} < {\it m}_{\rm F814W} < {\it m}_{\rm F814W, faint}$.
The green continuous line is the MS fiducial line, drawn as
described in Sect.~\ref{ZPvariations}. To include stars that have migrated
to the blue due to measuring error, we extend region {\it A} up to the
green dashed line, which is displaced to the blue from the MSRL
by 3 times the
the average color error for a star at that magnitude.
The red dotted line is the locus of MS-MS binaries
whose components have equal mass; we set the limit of
region {\it A} by drawing the red dot-dashed line, displaced
to the red from the dotted line by 3 times the rms color error.
The upper-right panel of Fig.~\ref{reg1} shows Region {\it B}, which is
chosen in such a way that it contains all the binaries with ${\it q}>0.5$.
It starts at the locus of binaries
with mass ratio,
{\it q}=0.5, marked by the continuous red line
and ends at the dotted-dashed red line, which is
the same line defined in the upper-left panel.

The lower half of Fig.~\ref{reg1} shows where observed stars and ASs fall within these
two regions. The left-lower panel plots the observed stars and the middle panel
shows ASs.
We note that a significant number of ASs fall in region {\it B}.
Only a fraction of them can be explained by photometric errors; in many cases
two stars fell at positions so close
together that a pair of stars has blended into a single object,
which would simulate a binary.
Obviously,  in the real CMD, regions {\it A} and {\it B} are
also populated by field stars, as shown in the right panel for NGC 2298.
We will explain how the field star CMD is built in Sect.~\ref{field}.

To determine the fraction  of binaries with {\it q}$>$0.5
we started  by  measuring the number of
stars, corrected for completeness,
 in regions {\it A}
($N_{\rm  REAL}^{\rm A}$) and {\it B}  ($N_{\rm  REAL}^{\rm B}$).
They    are calculated as   $N_{\rm REAL}^{\rm A (B)}=\sum_{1}^{N_{\rm
    OBS}^{A, (B)}}{1/c_{\rm
    i}}$,   where ${\it N}_{\rm OBS}^{\rm A, (B)}$ is the number of stars
observed in the region {\it A} ({\it B}) and ${\it C}_{\rm i}$ is  the
completeness (coming from AS tests). Then, we evaluated
the  corresponding numbers of  artificial stars ($N_{\rm ART}^{\rm A}$ and
$N_{\rm  ART}^{\rm B}$) and field stars  ($N_{\rm  FIELD}^{\rm A}$ and $N_{\rm
FIELD}^{\rm B}$).
In the following Sects.~\ref{field} and~\ref{blends} we will describe
the methods that we used to estimate
$N_{\rm  FIELD}^{\rm A}$ and $N_{\rm FIELD}^{\rm B}$
and $N_{\rm ART}^{\rm A}$ and $N_{\rm ART}^{\rm B}$.

The fraction of binaries with {\it q}$>$0.5 is calculated as
\footnote {
The first term on
the right-hand side of the equation gives the fraction of cluster stars (both
binaries and blends) observed in Region {\it B}, with respect to the number of cluster
stars observed in Region {\it A}. The second right-hand term is the fraction of blends and
is calculated as the ratio of the number of ASs in Regions {\it B} and
{\it A}.
}

\begin{equation}
\label{eq:1}
f_{\rm bin}^{\rm q>0.5}=\frac {N_{\rm REAL}^{\rm B}-N_{\rm FIELD}^{\rm B}}
      {N_{\rm REAL}^{\rm A}-N_{\rm FIELD}^{\rm A}} - \frac {N_{\rm
          ART}^{\rm B}}{N_{\rm ART}^{\rm A}}.
\end{equation}

Similarly, we have calculated the fraction of binaries with {\it q}
$>$0.6 and {\it q}$>$0.7.  
 To do this it is necessary to move redward the left-hand side (red
solid line) of Region {\it B}, according to what is shown in Fig.~\ref{f1}.

The error  associated to each quantity of eq.~\ref{eq:1} is the Poisson
error and the error on the obtained binary fraction
is calculated by following the standard errors propagation.
Therefore it represents a lower limit for  the uncertainty  of
the binary fraction.
We note that the binary fractions strongly
differ from one cluster to another  with $f_{\rm bin}^{\rm q>0.5}$ ranging
from $\sim$0.01 to $\sim$0.40.

In order to analyze the radial distribution of binary stars in GCs and
provide information useful for dynamical  models   of  our target clusters,  we   have
calculated both the total binary fraction and the fraction of binaries with $q>0.5$ at different radial distances from the
cluster center. More specifically, we defined three different regions:

\begin{itemize}
\item{ a circle with a radius of one core radius (${\it r}_{\rm C}$ sample);}
\item{an annulus between the core and the half-mass radius (${\it r}_{\rm C-HM}$ sample);}
\item{a region outside the half-mass radius (${\it r}_{\rm oHM}$ sample).}
\end{itemize}

 The values of the core radius and the half-mass radius are from
  the Harris (1996) catalog.
It should be noted that, even if our data are homogeneous, in the sense that they came from the
same instrument  ({\it ACS/WFC/HST})
and   have   been  reduced  adopting  the   same
techniques, their  photometric quality vary from
cluster to cluster, mainly because of the different stellar densities
(which affects the crowding).
For this reason, for some GCs that have poor photometry in their
central regions, we have measured the fraction of binaries only outside
a minimum radius ( $R_{\rm MIN}$)
where it is possible to distinguish binaries with $q>0.5$ from single
MS stars. The adopted values of
 $R_{\rm MIN}$ are listed in
Table~2.
The fractions of binaries with {\it q}$>$0.5, {\it q}$>$0.6, {\it q}$>$0.7
($f_{\rm bin}^{\rm q>0.5}$, $f_{\rm
  bin}^{\rm q>0.6}$ and $f_{\rm bin}^{\rm q>0.7}$) for the clusters in
our sample are listed in Cols.~3, 4, 5 of Table~2,
respectively. In column 6 there is also our best-estimate of the total
binary fraction (i.\ e.\ the fraction of binaries in the whole range
$0<q<1$) that will be estimated in Sect.~\ref{totfraction}.    
We give both the fractions of binaries calculated over the ACS/WFC
field and those in each of the three regions defined above. 

Following these considerations, it  was possible to  include in the
${\it r}_{\rm C}$ sample
only 43  out of the  original 59 GCs.  In  addition, the  limited ACS
field of view   reduced the number  of   GCs with ${\it r}_{\rm C-HM}$  and   ${\it r}_{\rm oHM}$
samples to 51 and 45 clusters, respectively.

\subsection{Field contamination}
\label{field}
The best ways to quantify foreground/background
contamination of regions {\it A} and {\it B} consists in
identifying field stars on the basis of their proper motion, which usually
differs from the cluster motion.
For several clusters of the sample  considered in this paper there are
previous epoch {\it HST} images with a sufficiently long temporal
baseline and precision to
allow  the  measurement of  proper    motions.  We used  archive
material to determine the proper motions of  20 GCs that
are critically contaminated by field stars:
ARP 2, NGC 104,
NGC 362, NGC 5286, NGC 5927, NGC 6121, NGC 6218,
NGC 6352, NGC 6388, NGC 6441, NGC 6397, NGC 6496, NGC 6535, NGC 6626,
NGC 6637, NGC 6652, NGC 6656, NGC 6681, NGC 6838, and TERZAN 7. The procedure
to measure  proper motions is outlined in Sect.~\ref{PM}

In order to determine field objects contamination in the CMDs of the
remaining clusters, we run a program developed by Girardi et al.\ (2005),
which uses a model to predict star numbers in any
Galactic field. Details of this procedure are given in
Sect.~\ref{galmodel}.
\subsubsection{Proper motions}
\label{PM}
Proper motions are measured by comparing the positions of stars
measured at two or more different epochs. 
 For the majority of the clusters only two epochs were available and we
followed a method that has been widely described in many other papers
(e.\ g.\ see Bedin et al.\ 2008, Anderson \& van der Marel 2010).
In the cases  of NGC 104,
NGC 362, NGC 5927, NGC 6121,  NGC 6397, and  NGC 6656 we  used a
sample of images taken  at three or   even more different  epochs
 and determined proper motions with the procedure given by McLaughlin et al.\ (2006).
 We refer the interested reader to these paper for a detailed description.   

\begin{figure*}[!ht]
   \centering
   \includegraphics[width=15cm]{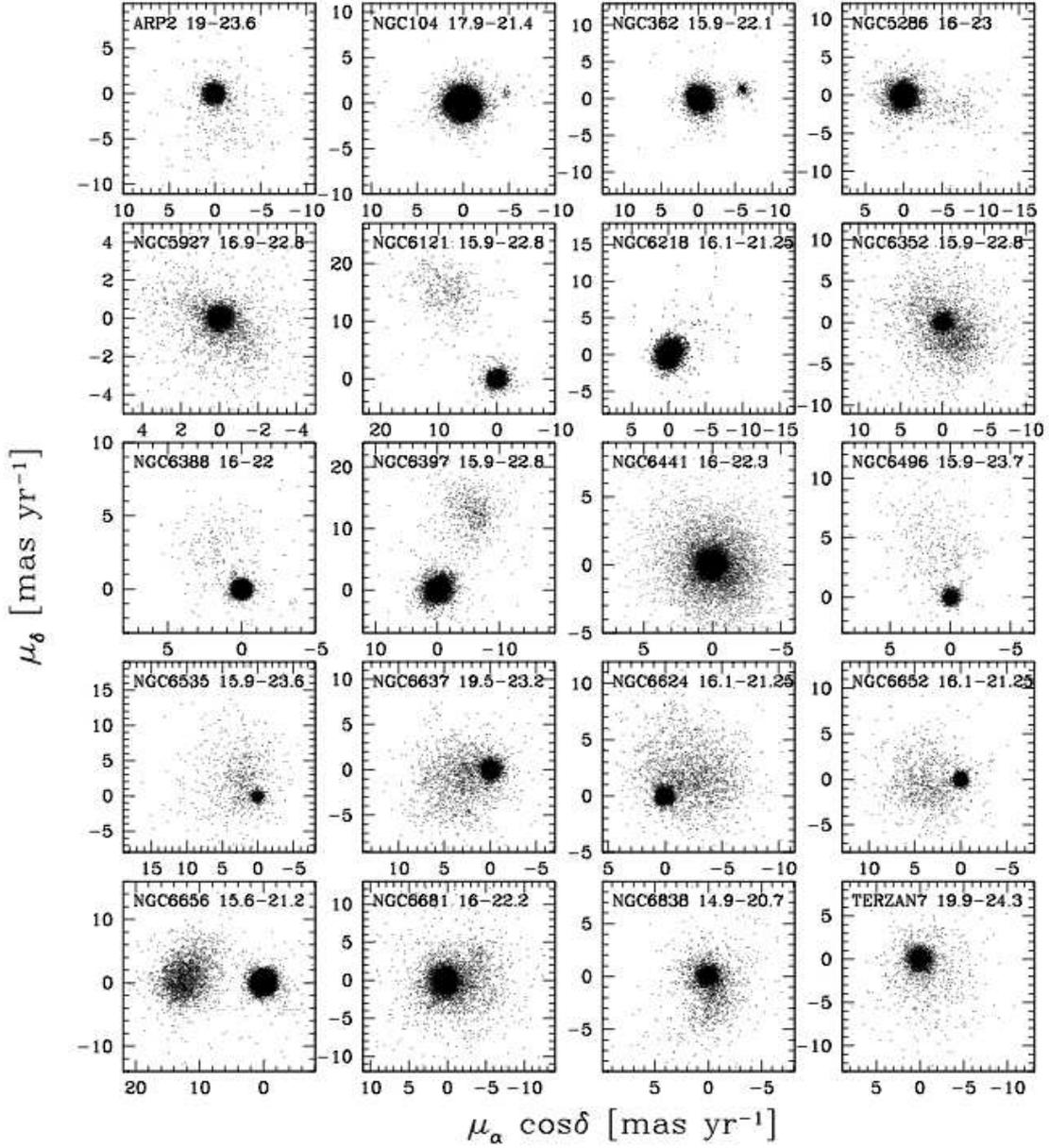}
      \caption{ Vector-point diagram of proper motions,
	in equatorial coordinates, for twenty GCs.
	Only stars in the  ${\it m}_{\rm F814W}$ interval
	indicated in each panel are shown.
       }
        \label{PMALL}
   \end{figure*}

Results are shown in Fig.~\ref{PMALL} which plots  proper motions
for twenty GCs. We plotted only stars in the F814W magnitude range indicated by
the numbers quoted in the insets\footnote{Note that the magnitude range used to create Fig.~\ref{PMALL}  is
 larger than that used  to estimate the field star contamination which enters
into eq.~1 for the the binary fraction calculation.}
 Since  we measured proper  motions
relative to a sample of cluster members, the zero point of the motion
is the mean motion of the cluster.
Therefore, the bulk of stars clustered around the origin of the vector-point diagrams (VPD)
consists mostly of cluster members, while field stars are distributed over a larger
range of proper motions.

\begin{figure*}[!ht]
   \centering
   \includegraphics[width=13cm]{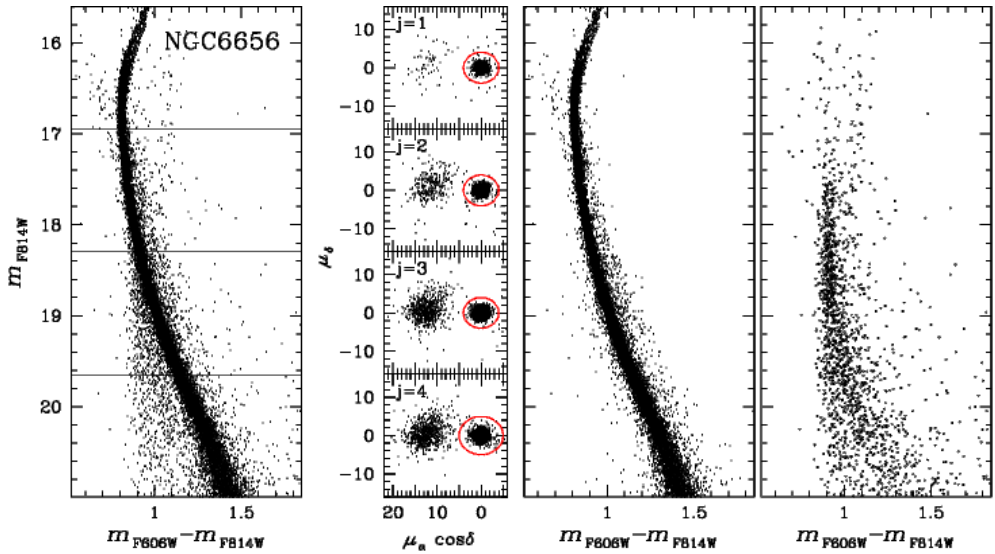}
%
      \caption{\textit{Leftmost column}: ${\it m}_{\rm F814W}$ vs. ${\it
          m}_{\rm F606W}-{\it m}_{\rm F814W}$ CMD for all the stars of
        NGC 6656 with  available measures of proper motions.
	\textit{Second Column}: Proper motion diagrams of the stars in
        the left panels  in $mas yr^{-1}$,
	in intervals of 1.4 mag. \textit{Third column}: The proper motion selected CMD of cluster members.
	  \textit{Rightmost column}: The CMD of field stars.
       }
         \label{PM6656}
   \end{figure*}

\begin{figure*}[!ht]
   \centering
   \includegraphics[width=13cm]{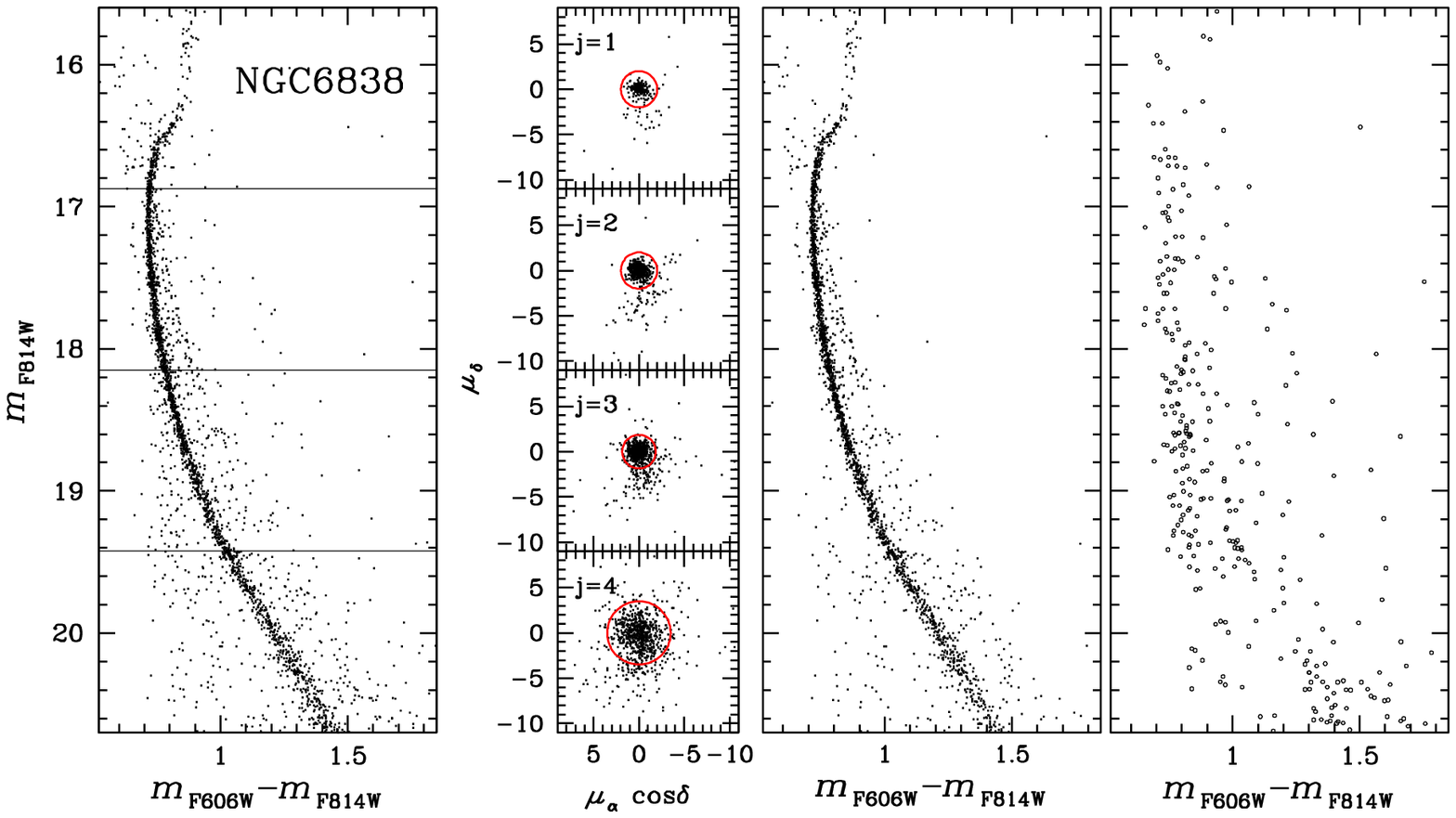}
      \caption{As in Fig.~\ref{PM6656}, but for NGC 6838.       }
         \label{PM6838}
   \end{figure*}

Proper motions offer a unique opportunity to estimate the number of
field stars that populate the regions {\it A} and {\it B} of the CMD.
In order to identify field objects, we began to isolate stars whose proper motions
clearly differ from the cluster mean motion by using the
procedure that is illustrated in Fig.~\ref{PM6656}
 for NGC 6656 (where cluster and field stars
are well separated in the VPD), and in Fig.~\ref{PM6838} for NGC 6838 (where the separation is less evident).

In the left panel of Figs.~\ref{PM6656} and ~\ref{PM6838}
 we show the  CMD for all the stars
for which  proper  motions  measurements are  available.   The second
column of the two figures shows the VPD of the stars in
four different magnitude intervals. The red circle
is drawn to identify the stars that have member-like motions.
In the following, we will indicate as $R_{\rm CL}$ and $R_{\rm OUT}$
the VPD regions within and outside the red circles.
We fixed  the radius of the circles
at  3.25 $\sigma$, where $\sigma$  is the average proper-motion dispersion in
the two  dimensions.  If we assume that  proper  motions  of cluster stars
follow a bivariate Gaussian distribution, the circle should include 99.5
\% of the members in each magnitude interval.
The third panel shows the CMD of stars with
cluster-like proper motion, while selected field objects
 are plotted on the right panel.

%
We 
emphasize
here that, as we will see in detail in the following, proper
motions are used to
evaluate the numbers of field stars that randomly fall within the CMD
regions {\it A} and {\it B} ($N_{\rm FIELD}^{\rm A, B}$) and not to
isolate a sample of cluster stars.
This approach will allow us to determine the binary fraction by means
of eq.~\ref{eq:1}  in the whole ACS/WFC field of view and not only in
the spatial regions covered by multi-epochs images where proper
motions are available.
To determine the values of $N_{\rm FIELD}^{\rm A, B}$ we have to
account for three factors:
\begin{enumerate}
\item 
To accurately measure $N_{\rm FIELD}^{\rm A, B}$
we need a correct estimate of the fraction of field stars that share
cluster-like proper motions.

\item Proper motions are not available for  the whole
  ACS/WFC field of view because, usually, there is only a partial
  overlap between the images at different epochs. As a consequence of
  this we need an accurate measurement of the area of the overlapping
  region.

\item Proper motions may not be available for a fraction of stars in
  the ACS/WFC catalogs even if they are in the overlapping region because these
  stars are not measured in the second-epoch images (that in many cases
  come from WFPC2),
 because they either are too faint or in a too crowded
region.
\end{enumerate}

Specifically the number of field stars in the region A has been evaluated as
\begin{equation}
\label{eq:field}
N_{\rm FIELD}^{\rm A}= \sum_{\rm j=1}^{4} \frac {1+n^{\rm R_{\rm
      CL}}_{\rm FIELD, j}/n_{\rm FIELD, j}}  {F_{\rm AREA}}
\sum_{\rm i=1}^{n_{\rm FIELD, j}^{\rm A}} \frac {1} {f_{\rm PM}^{\rm i} \ c_{\rm i}}
\end{equation}

 where:
\begin{itemize}
\item $n_{\rm FIELD,j}$ and  $n_{\rm FIELD,j}^{\rm A}$ 
 are the total number of field objects and the
number of field objects within Region {\it A} for which we have measured proper motions,
in the magnitude interval j (see the rightmost column of
Figs.~\ref{PM6656} and ~\ref{PM6838}), respectively.
\item $n^{\rm R_{\rm CL}}_{\rm FIELD, j}$ is the fraction of field objects that share proper motions similar
to the cluster;
\item $F_{\rm AREA}$ is the fraction of the ACS/WFC field of view with multi-epoch observations;
\item $c_{\rm i}$ is the completeness of the ACS/WFC catalog calculated in Sect.~\ref{sec:artsample};
\item $f_{\rm PM}^{\rm i}$ is a factor that accounts for the
 availability of proper motions (as in point 3 above).
\end{itemize}
 And the same is done to evaluate the number of field stars in the Region B.
In the following, we describe the procedure used to determine $n^{\rm R_{\rm CL}}$, $F_{\rm AREA}$, and  $f_{\rm PM}^{\rm i}$.

{\bf Field stars with cluster-like proper motions} \\
\label{fieldpm}
   The VPDs of Fig.~\ref{PMALL} show that almost all the clusters
   have some field stars that share the mean cluster motion.
The fraction of these sources with respect to the cluster stars depends on
several factors, such as the astrometric quality of the data,
the temporal baseline, the line of sight, and the motion of the cluster with respect to the field.
Their fraction is almost negligible in NGC 6656 and other cases,  but
makes a significant contribution to the binary fraction
in most of the GCs of Fig.~\ref{PMALL}.
We now describe a method to determine the fraction of field stars with
cluster-like proper motion in order to accurately infer ${\it N}_{\rm
  FIELD}^{\rm A}$ and ${\it N}_{\rm FIELD}^{\rm B}$ in equation~\ref{eq:1}.

We note that, for the purposes of this paper, we do not need to
isolate these intruders. It is sufficient to estimate their total
amount, and, more specifically, the amount of field stars with cluster-like
motions that populate the CMD region associated with MS-MS
binaries or MS single stars.

\begin{figure}[!ht]
   \centering
   \includegraphics[width=8.7cm]{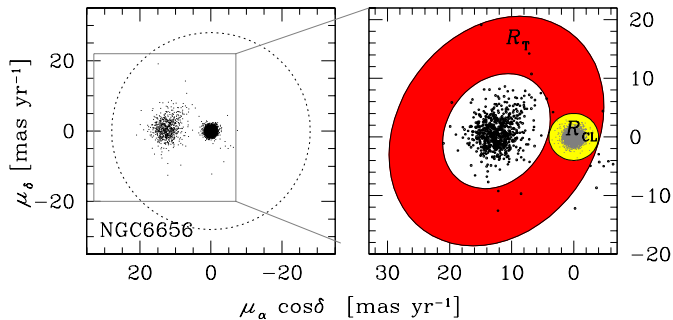}
      \caption{ Setup to estimate the fraction of field stars with
        cluster-like proper motions in NGC 6656.
       }
         \label{PM6656SET}
   \end{figure}

We independently calculated, for the GCs with reliable proper motions,
the number of field stars with cluster-like proper motions
 for each of the four magnitude intervals
of Figs.~\ref{PM6656}, and  ~\ref{PM6838}.
In the cases of  GCs where cluster and field stars
are clearly separated in the proper motion diagram (ARP 2, NGC 104,
NGC 362, NGC 5286, NGC 6121, NGC 6218, NGC 6388, NGC 6397, NGC 6496,
NGC 6535, NGC 6637, NGC 6624, NGC 6652, NGC 6656, and Terzan 7)
we used the method that is illustrated in Fig.~\ref{PM6656SET}
 for NGC 6656.
All the field and cluster stars with reliable proper motions are located
within the dotted circle
of the left panel VPD 
We considered as probable cluster members all the objects
that are plotted as thin gray dots in the yellow area (region ${\it R_{\rm CL}})$
of the zoomed VPD in the right panel, while
 remaining objects are flagged as field stars and are represented
 as heavier points.

\begin{figure}[!ht]
   \centering
   \includegraphics[width=8.7cm]{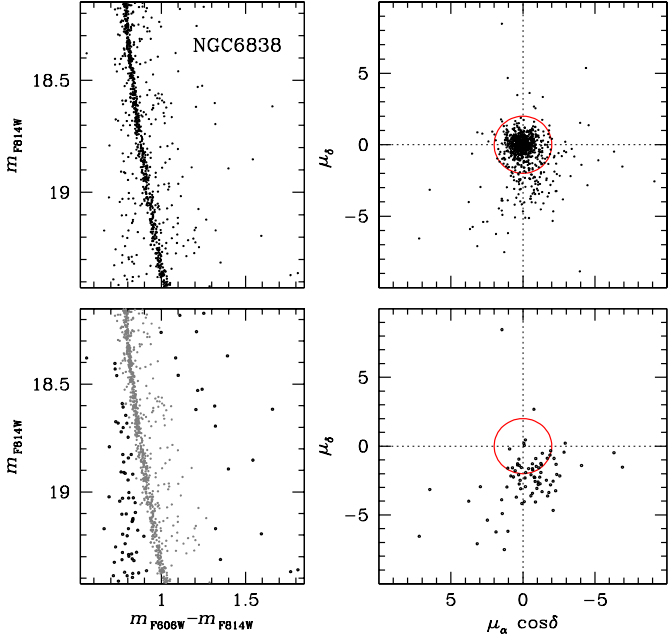}
      \caption{Estimate of the contamination of field stars that
      share cluster proper motions in NGC 6838.
      Upper panels show the CMD ({\it left}) and the VPD ({\it right}) for
      stars with 18.15$<$${\it m}_{\rm F814W}$$<$19.45.
      Lower panels display the CMD ({\it left}) and the VPD ({\it right})
      for those objects that, on
      the basis of their position on the CMD, are probable field stars.
       }
         \label{PM6838SET}
   \end{figure}

The distribution of field stars in the VPD is clearly elongated
and the isodensity contours can be approximately described by ellipses.
In  Fig.~\ref{PM6656SET} we show the two isodensity contours that are tangent
to the region ${\it R}_{\rm CL}$ and define the red region (${\it
  R}_{\rm T}$).
The number of field stars within ${\it R}_{\rm CL}$ is assumed to be:\\
$n_{\rm FIELD}^{\rm R_{\rm CL}}=n_{\rm R_{\rm T}} S_{\rm R_{\rm
    CL}}/S_{\rm R_{\rm T}}$\\
where $S_{\rm R_{\rm CL}}$ and $S_{\rm R_{\rm T}}$ are the areas of regions
${\it R}_{\rm CL}$ and ${\it R}_{\rm T}$ and $n_{\rm R_{\rm T}}$ is the number of stars
within  ${\it R}_{\rm T}$. \\
In the cases of NGC 5927, NGC 6352, NGC 6441, NGC 6681, and NGC 6838, where the separation
of field and cluster stars is less evident,
we followed a different recipe, which is illustrated in
in Fig.~\ref{PM6838SET} for NGC 6838.
The upper panels show the CMD ({\it left}) and the VPD ({\it right}) for stars
in the third interval of magnitudes (j=3) of Fig.~\ref{PM6838}.
We selected, on the CMD, a sample of stars that, on the basis of their
color and magnitude, are probable background/foreground objects.
These stars are marked as
heavy black points in the lower CMD of
Figure~\ref{PM6838SET}, while in the right-lower panel we show their
position in the VPD.

If we assume that the fraction of selected objects
within $R_{\rm CL}$  with respect to the
 total number of selected field objects ($f^{\rm R_{CL}}_{\rm FIELD}$)
is representative of the overall fraction of field
stars that share cluster proper motions we can impose:
$n_{\rm FIELD}^{\rm R_{\rm CL}} = n_{\rm FIELD} f^{\rm R_{CL}}_{\rm
  FIELD}$.  The contribution of $n_{\rm FIELD}^{\rm R_{\rm CL}}$
  to the measure of the binary fraction is, for all the clusters
  smaller than 0.01. \\

 In order to investigate the reliability of this approach, we applied it
also to the 15 GCs for which proper motions allow us to almost completely separate cluster stars from field ones. In all cases, we found full consistency between the two approaches, with the the fraction of binaries with {\it
  q}$>$0.5  listed in Table~2 differing by less than 0.01.

\begin{figure}[ht!]
   \centering
   \includegraphics[width=8.5 cm]{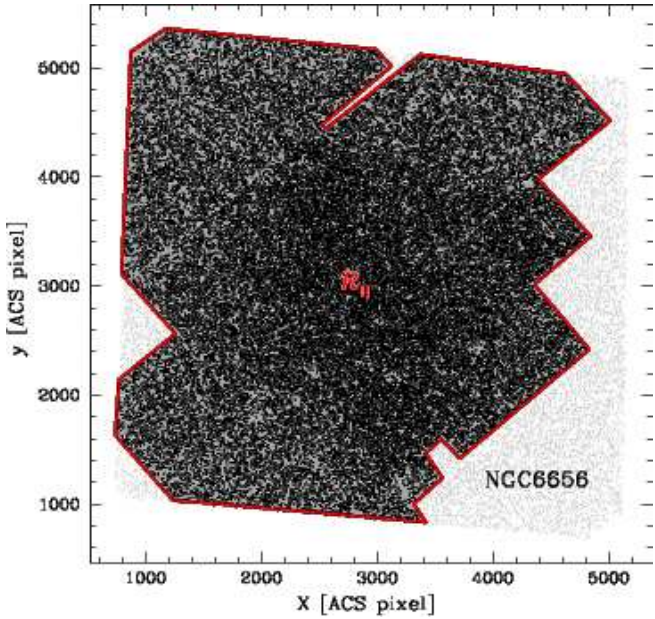}
      \caption{  Map of all  the stars  NGC 6656 (light gray
        points)  with ${\it m}_{\rm F814W}<$18.25. Black points
        mark   all the stars  with   available proper
      motions. The contour of region `${\it R}_{\rm II}$' is colored in red.
       }
         \label{fieldPM}
   \end{figure}

{\bf The spatial coverage of multi-epoch images} \\
For most clusters, there is only a partial overlap among the different
epoch images.  In the following we will  refer to the region that has been
observed in at least two epochs as `${\it R}_{\rm II}$'.
Fig.~\ref{fieldPM} shows the  example for NGC 6656,
where we indicate as light gray points all the stars for which we have
only photometry, and  mark  with black points  the stars  with  both
photometric and proper motion measurements.
As our field is just a few square arcmins, we   can  assume that    the 
background/foreground population is uniformly distributed within it, 
and therefore we estimate the total number of
field  stars in our field of view as the product of the number of
field  stars in the region $R_{\rm II}$  and the ratio  between
the area of the total field of view and the area of $R_{\rm II}$. In
this paper, we will refer to this ratio as: $F_{\rm AREA}$.

\begin{figure}[ht!]
   \centering
   \includegraphics[width=8.5 cm]{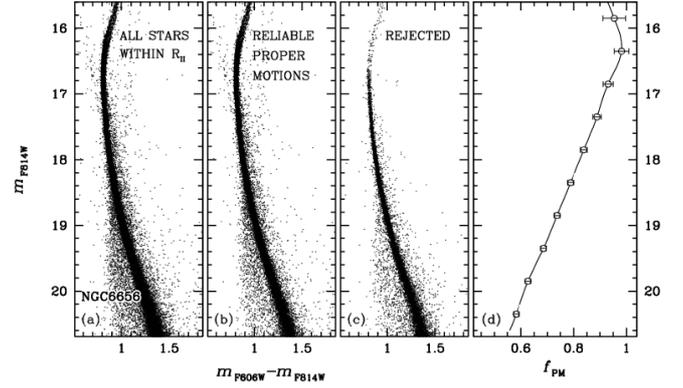}
      \caption{
	\textit{Panel a}: CMD for all the stars within the region `$R_{\rm II}$';
	\textit{Panel b}: CMD for stars with a reliable estimate of proper motions;
	\textit{Panel c}: CMD for stars within `$R_{\rm II}$' for which there are not reliable measurements of proper motions;
	\textit{Panel d}: Fraction of stars with good proper motions with respect to the total number of stars within `$R_{\rm II}$' as a function of ${\it m}_{\rm F814W}$.
       }
         \label{6656compPM}
   \end{figure}

{\bf Completeness correction for field stars }\\
In the  procedure  that  we have  applied   to determine the   cluster
membership using  proper  motions, we have  automatically excluded all
the stars that  might be members  but have poor astrometry.
An accurate estimate of the fraction of these stars is necessary to infer
the  correct values of ${\it N}_{\rm
  FIELD}^{\rm A}$ and ${\it N}_{\rm FIELD}^{\rm B}$.   To estimate  the fraction of cluster
stars lost  by applying   the  proper motion selection   criteria,
we applied the procedure illustrated in Fig.~\ref{6656compPM} for NGC 6656.
In panel $a$ we show the ${\it m}_{\rm F814W}$ vs.\ ${\it m}_{\rm F606W}-{\it m}_{\rm F814W}$ CMD for all the stars in the region `$R_{\rm II}$'.
 Proper motion measurements are available only for a fraction ($f_{\rm
   PM}$) of these stars. Their CMD is shown in panel $b$, while the CMD
for stars with no available proper motions is plotted in panel ($c$).

To determine $f_{\rm PM}$ we started by dividing the CMD into bins of
0.5 ${\it m}_{\rm  F814W}$ magnitudes. In each of them, we counted
the total number  of observed stars (${\it N}_{\rm OBS}$) and the
number of star with a reliable estimate of
proper motions  (${\it N}_{\rm  PM}$).  The fraction  of  stars  with a proper
motions in that bin is:
${\it f}_{\rm PM}={\it N}_{\rm PM}/{\it N}_{\rm OBS}$.

We then calculated the median ${\it m}_{\rm F814W}$
magnitude of the observed stars (${\it m}_{\rm MED}$) in each bin.
We associated to each bin the corresponding value of ${\it f}_{\rm PM}$ and
${\it m}_{\rm MED}$.
The (${\it f}_{\rm PM}^{\rm i}$) for each i-star is calculated by
interpolation with a spline. 
In panel ($d$) of Fig.~\ref{6656compPM} we show the final ${\it f}_{\rm
  PM}$ as a function of  ${\it m}_{\rm F814W}$.
 For the GCs studied here always we have ${\it f}_{\rm  PM}>0.4$ at the
  level of 3.75 F814W magnitudes below the MSTO.

%
\subsubsection{Galactic model}\label{galmodel}
\begin{figure}[ht!]
   \centering
   \includegraphics[width=8.5 cm]{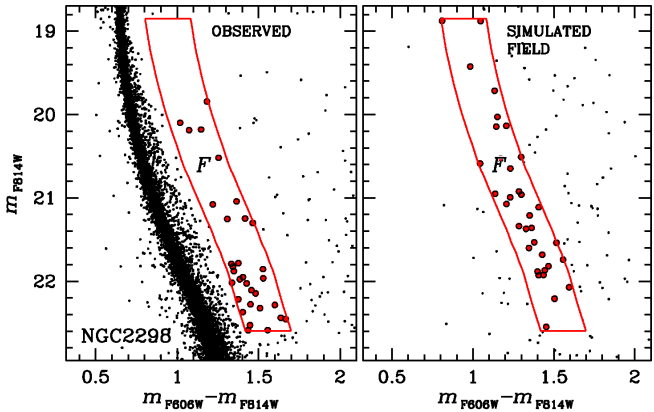}
      \caption { Observed CMD of NGC 2298 ({\it left}) and simulated
        CMD of field stars in the NGC 2298 field of view ({\it right}).  }
   \label{FIELDSET}
   \end{figure}

In order to estimate the number of background/foreground stars in the
field of view of the GCs studied in this paper, and for which we do
not have reliable measurements of proper motions, we   used the
theoretical   Galactic model described  by  Girardi et al.\ (2005).
This model was used to generate  a synthetic  CMD (in
the ACS/WFC F606W and F814W bands) containing  the expected field stars in
the  cluster area  that  we are  studying.
The synthetic CMDs were used to count the number of field stars
in the CMD regions {\it A}, and {\it B}
(${\it N}_{\rm SIM}^{\rm A}$, ${\it N}_{\rm SIM}^{\rm B} $) defined in
Fig~\ref{reg1}.
Obviously, the number of stars in simulated CMDs may differ from
that of observed field stars. To minimize the effect of such
uncertainties on the measure of the fraction of binaries in GCs, we
defined in the CMD a region {\it F} on the red side of 
equal-mass binaries fiducial sequence, that is delimited on the blue side
by the red dashed-dotted  line of Fig.~\ref{reg1} 
 and is likely not populated by cluster stars,
as illustrated in  Fig.~\ref{FIELDSET} for NGC 2298.
We determined the numbers  of stars within {\it F} in the observed and
in the simulated CMDs  (${\it  N}_{\rm OBS}^{\rm  F}$ and ${\it
  N}_{\rm SIM}^{\rm F}$ respectively). 

The number of field stars in the CMD regions {\it A} is
then calculated as:
\begin{equation}
{\it N}_{\rm FIELD}^{\rm A}={\it N}_{\rm SIM}^{\rm A} {\it  N}_{\rm  OBS}^{\rm F}/{\it  N}_{\rm SIM}^{\rm F}
\end{equation}
 and a similar equation is used to estimate the number of
field stars in the region {\it B}.

 As anticipated in Sect.~2, we removed from our list all clusters for
which we had no proper motion (two epochs data) and for which
Girardi et al.\ (2005) model was prediction a field star contamination
larger than 1\%, 
with the only exception of E3 (a 2.4\% expected contamination) and  NGC 6144 (1.3\%). Therefore, for clusters for which we have to rely on a Galactic model to 
estimate the foreground//background stars, the contamination is 
expected to be minimal. On the other hand, we kept into the sample all cluster
for which we could use proper motion to estimate field stars, independently
from the level of contamination.
%
In order to investigate whether the estimate of field stars from
Galactic models is reliable, we applied the synthetic  CMDs method 
also in the 15 GCs for which we have reliable proper motion measurements.
We found that, in the cases of GCs with a small field-star
contamination, the fraction of binaries with {\it
  q}$>$0.5 derived following the two approaches is identical within
the uncertainties, 
 with differences smaller than 0.01. For some GCs with a significant
 background/foreground population, namely NGC 5927, NGC 6352, NGC 6388,
 NGC 6441, NGC 6637, and NGC 6681, the fractions of binaries derived
using a Galactic model differ from those derived using proper motions 
by 0.01 to 0.03 (for  NGC 6441).

\subsection{ Estimate of the fraction of apparent binaries}
\label{blends}
Chance superpositions of two physically unrelated stars that happen to
lie   nearly  along  the  line   of  sight    (apparent binaries)  and
superposition of a faint star  and a  positive  background fluctuation  may
reproduce the  color and luminosity  of a  genuine binary  system,  and
populate the CMD region occupied by binaries.  In a crowded stellar
field, like the core of a GC, a reliable measure of the
binary fraction requires  good accuracy  in deriving the number  of
chance superpositions.

We can identify and  reject a significant fraction
of these objects by analyzing   the stellar profile, and the  PSF-fit
errors.  For  this reason, in  this work, we limited  our study to the
objects that pass  the selection criteria described in  Sect.~\ref{sec:selections}.

In order to account for the blends that have not  been
rejected,  a statistical estimate of their number and distribution in the CMD
is  necessary.    In this paper, we   used  extensive
artificial-star test experiments to evaluate directly the effects of blends.

 Specifically, in this subsection, we illustrate the procedure
  adopted to determine the relative numbers of artificial stars in the
  regions {\it A} and {\it B} of the CMD of Fig.~\ref{reg1} ($N_{\rm
    ART}^{\rm A}$ and $N_{\rm  ART}^{\rm B}$) that are used to
  calculate the last term of eq.~\ref{eq:1}.

This analysis  requires that the  artificial star sample  that we will
compare to  observed  data  reproduce as  much  as  possible all  the
details of  real stars. In particular we  need the best possible match
between the luminosities,  the radial distribution and the photometric
errors of observed and simulated stars.

      The data set described in Anderson et al.\ (2008) includes
      an extensive set of artificial-star tests for each cluster.
      The same quality parameters were determined for
      the artificial stars as for the real stars, so we apply the
      same selection criteria to them as we did to the real stars
      in Sect.~\ref{sec:selections}.

      To apply these generic artificial-star tests to the real
      cluster distribution, for each real star observed, we took
      a set of the artificial stars within $\pm 0.10$ magnitude
      and with radial distances within $100$ pixels of that of
      the star.  These are the stars that were used to estimate
      the measurement errors (random and systematic) of the stars
      in the cluster.

The result of this procedure is a catalog  of simulated stars that reproduces
both the  radial and the luminosity distributions of real stars.
Several effects contribute to the observed width of the main sequence.
In addition to photon noise, we have the contribution
of spatial variations of  the  PSF  and residual differential  reddening that
are beyond the  sensitivity  of the  method   that we used  to  correct
them, as well as scattered light, possible  star-to
star metallicity variations, etc.
However, for  the purposes of this  work, it is not necessary to
distinguish the  contributions of the single  sources of the broadening
and we can include them in the photometric errors ($\sigma$).

Since MS-MS binary systems  and apparent binaries both lie on the red
side of the MS, we can use the  MS scatter to the  blue side of the MS
as an estimate of the photometric error. We note that the blue portion
of the  MS may be contaminated by   MS-white dwarf binaries   but  their
influence on $\sigma$ is expected to be negligible, and further reduced
by the applied "kappa--sigma" rejection algorithm, as described below.

In order   to   estimate $\sigma$,  we  used   the following iterative
procedure, which has been applied to both  the observed and
artificial-star CMD.  First of all, we subtracted the color  of the
MSRL from the color  of each star.  Then  we divided  this CMD into
several intervals  of magnitude, each one containing the  same  number of
stars, and constructed a histogram of the color distribution  for each
magnitude interval. 
The  size of  each  interval is a  compromise between
maximizing the number of stars to reduce the statistical errors
and minimizing the magnitude  intervals  to  account  for the
 variations of   the photometric error as  a  function  of the
luminosity.
For  these reasons, the  size of the adopted interval varies from
one cluster to another, depending on the number of sampled stars.

We used least-squares to fit each histogram with a
    Gaussian that had three fitting parameters:  its
    center, its amplitude, and its dispersion $\sigma$.
Then, we rejected all stars for which color is
far more than $3\sigma$ from  the fiducial  line, because most  of
these  objects must be field  stars  or binaries. Finally  we used the
remaining sample for a new Gaussian fit.

All the stars with negative color in the rectified  CMD (i.\ e., those
on the blue side of  the MS) are used  for a new  Gaussian fit,
but, this time, we fixed the center and  the amplitude of the Gaussian
and considered  $\sigma$  as the  only   free parameter.   The  best
fitting $\sigma$ is  adopted as the  average photometric  error in
that magnitude interval. The errors corresponding to a given
magnitude in the CMD are obtained by interpolations.

As expected, the artificial star color distribution is narrower than
the real star one. We need to properly estimate  the    difference
between  the  artificial-star photometric  error and the photometric
error  of real stars, since, as it will be clearer in next section, we
need  an artificial-star CMD with   the  correct   photometric error
in order  to   estimate the photometric outliers which contaminate the
binary region.

The smaller color dispersion of the artificial star CMD comes from the
fact that the measurement errors of  artificial stars are smaller than
the corresponding  error  of real stars. This difference is due to the
fact that, in fitting artificial
stars, we use exactly the same
PSF that was used to originate them, while we cannot
expect the  same perfect match  of the PSF with  the  real PSF of real
stars.  In addition, and for the same reason, artificial-star
photometry is not affected  by zero 
point photometric errors, and errors associated with the differential
reddening  correction.   

 The difference between the MS color spread  of observed and
  simulated stars might be also due to multiple stellar
  populations. Indeed, nearly all the GCs studied so far host two or more 
  generations of stars with a different light-elements. 
  In few GCs, there are also star-to-stars iron variations (see
  Milone et al.\ 2010b for a recent review).   

Among the clusters studied
  in this paper, multiple MSs associated to helium variation have been
  identified in 47 Tuc, NGC 6752, and NGC 6397 where the  
  $m_{\rm F606W}-m_{\rm F814W}$ color
  difference between the He-rich and He-poor MS is about 0.01 mag 
 (Anderson et al.\ 2009, Milone et al.\  2010a, 2011b,c) i.\ e.\ has
  the same order of magnitude as the color errors of the best measured
 MS stars.
  NGC 6656 (M22) is the only cluster of this paper where two groups of
  stars with a different iron content have been 
  identified. In this case theoretical isochrones show  that the
  measured  [Fe/H] difference of $\sim$0.15 dex do not produce any
  appreciable $m_{\rm  F606W}-m_{\rm F814W}$ color  bimodality among MS
  stars (Marino et al.\  2009, 2011). 
In general the MSs corresponding to the different stellar populations
observed in the majority of GCs (and hence formed by stars that could
have different overall CNO abundance, and light elements variations)
are almost overimposed when observed in the $m_{\rm F606W}-m_{\rm
  F814W}$ color (Sbordone et al.\ 2011).   

\begin{figure*}[ht!]
   \centering
   \includegraphics[width=12 cm]{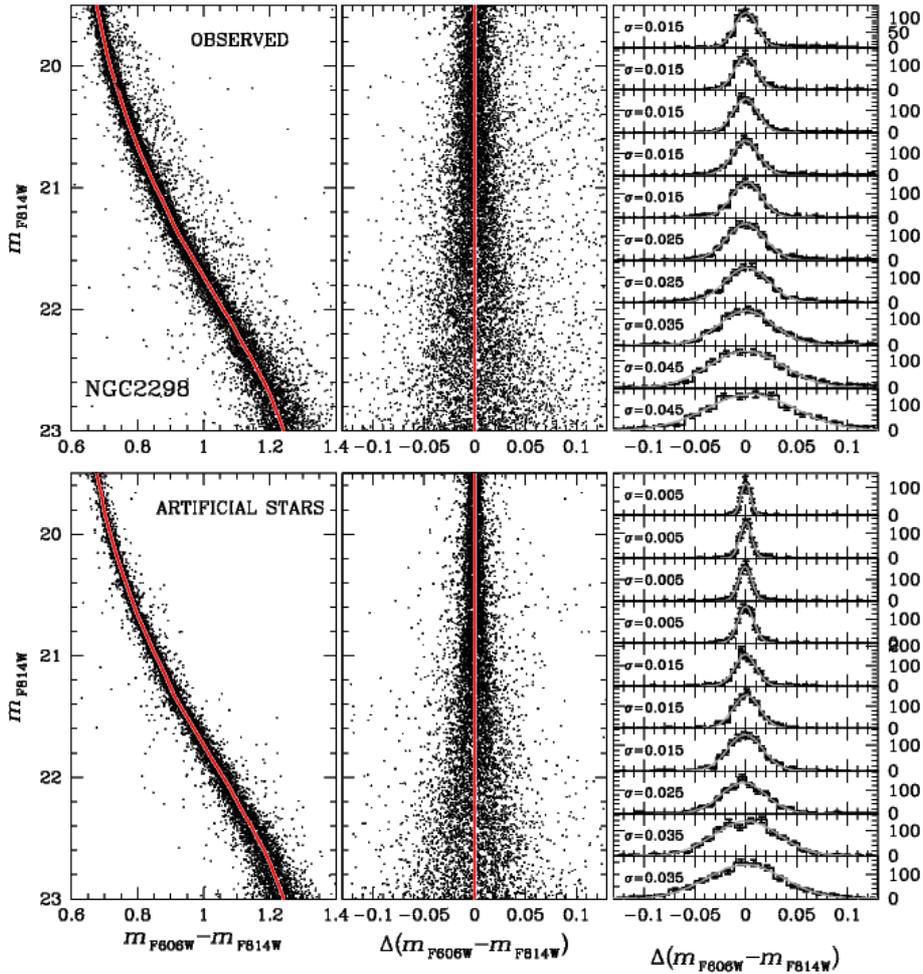}
      \caption { \textit{Left}: The observed (top) and simulated (bottom)
        CMD  of  NGC 2298   with  the  fiducial  line
      overplotted; \textit{Middle}: The  CMDs rectified by subtraction  of the
      fiducial line; \textit{Right}: Color distribution of  the rectified CMDs. The
      $\sigma$ in the inset are those of the best-fitting Gaussian. }
         \label{sp1}
   \end{figure*}

\begin{figure}[ht!]
   \centering
   \includegraphics[width=8.9 cm]{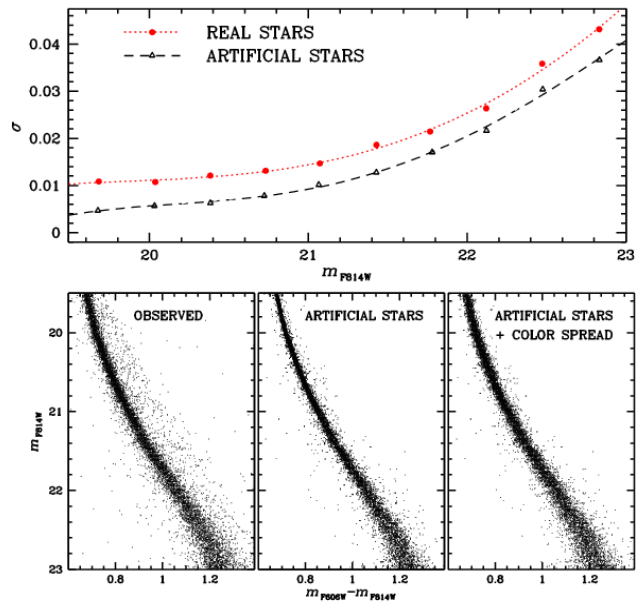}
      \caption{ \textit{Upper panel}:
        MS  dispersion  as   a function of    ${\it m}_{\rm  F814W}$
      magnitude for NGC 2298. Circles and triangles correspond to real
      and artificial stars, respectively.
      The corresponding best fitting $4^{\rm th}$ order
      polynomials are represented by dotted and dashed lines.
      \textit{Lower panels}: comparison of the observed CMD of NGC 2298
      (left) and the CMDs of artificial stars before (middle)
      and after (right) the application of color spread.
       }
         \label{disp}
   \end{figure}

As  an example,    the  difference in   color
dispersion between the real and the artificial star CMDs of NGC 2298
are shown in Fig.~\ref{sp1}.
%
In order to compare the real  and the artificial star color
distribution it is necessary to appropriately re-scale the latter.
For this, we considered the measured  dispersions as a function of the
$m_{\rm  F814W}$ magnitude  for both observed   and simulated  MSs, and
calculated by least squares the $4^{\rm th}$ order polynomials (${\it
  P}_{\rm REAL}$
and  ${\it P}_{\rm ARTS}$) that   best  fit  each   of  them.
As   an  example, Fig.~\ref{disp} (upper panel)
shows the measured   dispersions and the  best fitting
functions for the case of NGC 2298.  In this  paper, we considered the
spread of  the  MS stars as  a  reliable indicator of the  photometric
errors to  be  associated  to  color  measures.   We  believe  that it
represents   a much more accurate  estimate  for
the observed MS breadth
than the one  given by the rms value  obtained from magnitude measures
of the single AS MS stars. In fact it  also accounts for residuals photometric
zero  point errors, errors associated to  the reddening correction
method and possible intrinsic spread due to the presence of multiple
stellar populations. 

The difference   between the observed and   simulated MS dispersion is
expressed as: $\Delta{\sigma}_{\rm
VI}=\sqrt{P_{\rm REAL}^{2}-P_{\rm ARTS}^{2}}$. Assuming that  any  spread of MS
stars around the MS fiducial  line comes only from photometric errors,
$\Delta{\sigma}_{\rm VI}$  indicates how the   artificial-star color errors
underestimate our real-star photometric error.  As a final,
fundamental  step for the  following discussion, we  made the
artificial-star CMD similar to  the observed  one by adding  to each
artificial star  additional random noise   in  color, extracted from a
Gaussian distribution with dispersion $\Delta{\sigma}_{\rm VI}$.
   In the bottom panels of Fig.~\ref{disp} we compare the
     observed CMD of NGC 2298 and the CMDs of artificial stars before
     and after the application of the color spread. The latter CMD has
     been used to calculate the ratio between $N_{\rm ART}^{\rm A}$ and
     $N_{\rm  ART}^{\rm B}$ used in Eq.~\ref{eq:1}. 
\section{Results}
\label{results}
In this Section we illustrate and discuss the main results of this work.
Specifically:
\begin{itemize}
\item In Sect.~\ref{sec:qdist} we analyze the mass-ratio distribution of binaries in each of the 59 GCs studied in this paper in the range
0.5$<q<$1. Results from individual clusters are used to estimate the average mass-ratio distribution of binaries;

\item  Attempt to calculate the total fraction of MS-MS binaries is proposed in Sect.~\ref{totfraction};

\item Sect.~\ref{letteratura} gives a summary of the literature
  measurements of the binary fraction in GCs and
compares them with ours.

\item In Sect.~\ref{magdistr} we investigate the distribution of
  binaries as a function of the primary star mass (magnitude); 

\item The radial distribution of binaries in each GC is studied in Sect.~\ref{sec:RD};

\item Finally, monovariate relations between the binary fraction and the main parent cluster parameters (absolute luminosity, central velocity dispersion, metallicity, age, central density, ellipticity, core and half mass relaxation time, HB morphology, collisional parameter) are discussed in Sect.~\ref{monovariate}.

\end{itemize}
\subsection{Mass-ratio distribution}
\label{sec:qdist}
\begin{figure}[!ht]
   \centering
   \includegraphics[width=8.25cm]{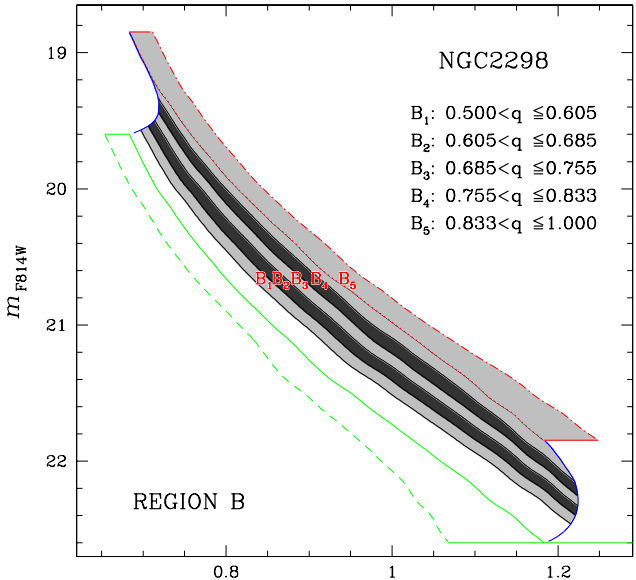}
      \caption{ As an example, we show with gray and black shaded
        areas the five CMD regions
	(${\it B}_{\rm 1, 2, ..., 5}$) used to determine the mass-ratio
	distribution of binary stars in NGC 2298.
       }
         \label{qsetup}
   \end{figure}

This section, presents the mass-ratio
distribution of the binary  population for our target GCs
in the  range of 0.5$<${\it q}$<$1.
To do this, we  have divided Region {\it B} of the CMD
into five intervals of mass ratio (${\it B_{1,2,...,5}}$)
as shown in Fig.~\ref{qsetup} for NGC 2298.
We chose the size of these regions in
such a way that each of them covers almost the same area in the
portion of the CMD populated by binary systems  with $q>0.5$.
The sub-region ${\it B_{5}}$ includes also the gray area on the right side of
equal-mass binaries fiducial that is populated by binary systems with $q \sim 1$
but large photometric errors.

The fraction of binaries in each sub-region ${\it B_{\rm i}}$ is
calculated  over the entire WFC field of view 
following the procedures described in Sect.~\ref{sec:highq}.
Each sub-region includes binary stars within a given mass-ratio interval ($\Delta q_{\rm i}$)
as labeled in Fig.~\ref{qsetup}.
To account for the different mass-ratio values of each sub-region,
and analyze the mass-ratio distribution, we derived the normalized fraction of binaries:
\begin{center}
$ \nu_{\rm bin, i}=f_{\rm bin, i}/ \Delta q_{\rm i}$.
\footnote{
    If we assume that:
    \begin{itemize}
    \item $\phi$({\it q}) is the continuous function that describes the
      distribution of the number of binaries as a function of the mass ratio. 
    \item {\it N}    is the total number of stars (both binaries and
      single stars)
    \item $N_{\rm bin}^{\rm B1}$, $N_{\rm bin}^{\rm B2}$, ..., $N_{\rm
      bin}^{\rm B5}$ the number of binaries in each region ${\it B}_{1,2,...,5}$.
    \end{itemize} 

    Obviously\\
    $\int_{\rm 0.5}^{\rm 1} \phi(q) dq$=$\int_{\rm q1}^{\rm q2} \phi(q) dq$+$\int_{\rm q2}^{\rm q3} \phi(q) dq$+...+$\int_{\rm q5}^{\rm q6} \phi(q) dq$    \\
    where [q1:q2], [q2:q3], ..., [q5:q6] are the mass-ratio intervals
    corresponding to the CMD regions of Fig.~\ref{qsetup}. We have:\\  
    $\int_{\rm q(i)}^{\rm q(i+1)} \phi(q) dq=N_{\rm bin}^{\rm B(i)}$; 
    i=1,2,...,5. \\
    At this point, the best we can do to gather information on $\phi(q)$
    is to use the approximation:\\
    $\int_{\rm q(i)}^{\rm q(i+1)} \phi(q) dq= \phi^{*}_{\rm i}(q) (q(i+1)-q(i))=
    \phi^{*}_{\rm i}(q) \Delta q_{\rm i}$ \\    
    and calculate:\\
    $\phi^{*}_{\rm i}(q)={\big (}\int_{\rm q(i)}^{\rm q(i+1)} \phi(q) dq {\big
      )}/ \Delta q_{\rm i} = N_{\rm bin}^{\rm B(i)}/\Delta q_{\rm i}$.\\
    If we normalize $\phi^{*}_{\rm i}(q)$ by the total number of stars we
    find that the normalized fraction of binaries differs from
    $\phi^{*}_{\rm i}$ by a factor 1/N:
    $\phi^{*}_{\rm i}(q)/N=N_{bin}^{B(i)}/(N \Delta q_{i})=f_{bin, i}/\Delta q_{i}=\nu_{\rm bin, i}$.
    
    Since the total number of stars changes from one cluster to each
    other, we use here
    $\nu_{\rm bin, i}$ as the best approximation of the mass-ratio
    distribution in each $q$ interval. 
    \\  
}
\end{center}

\begin{figure*}[!ht]
   \centering
   \includegraphics[width=15.0cm]{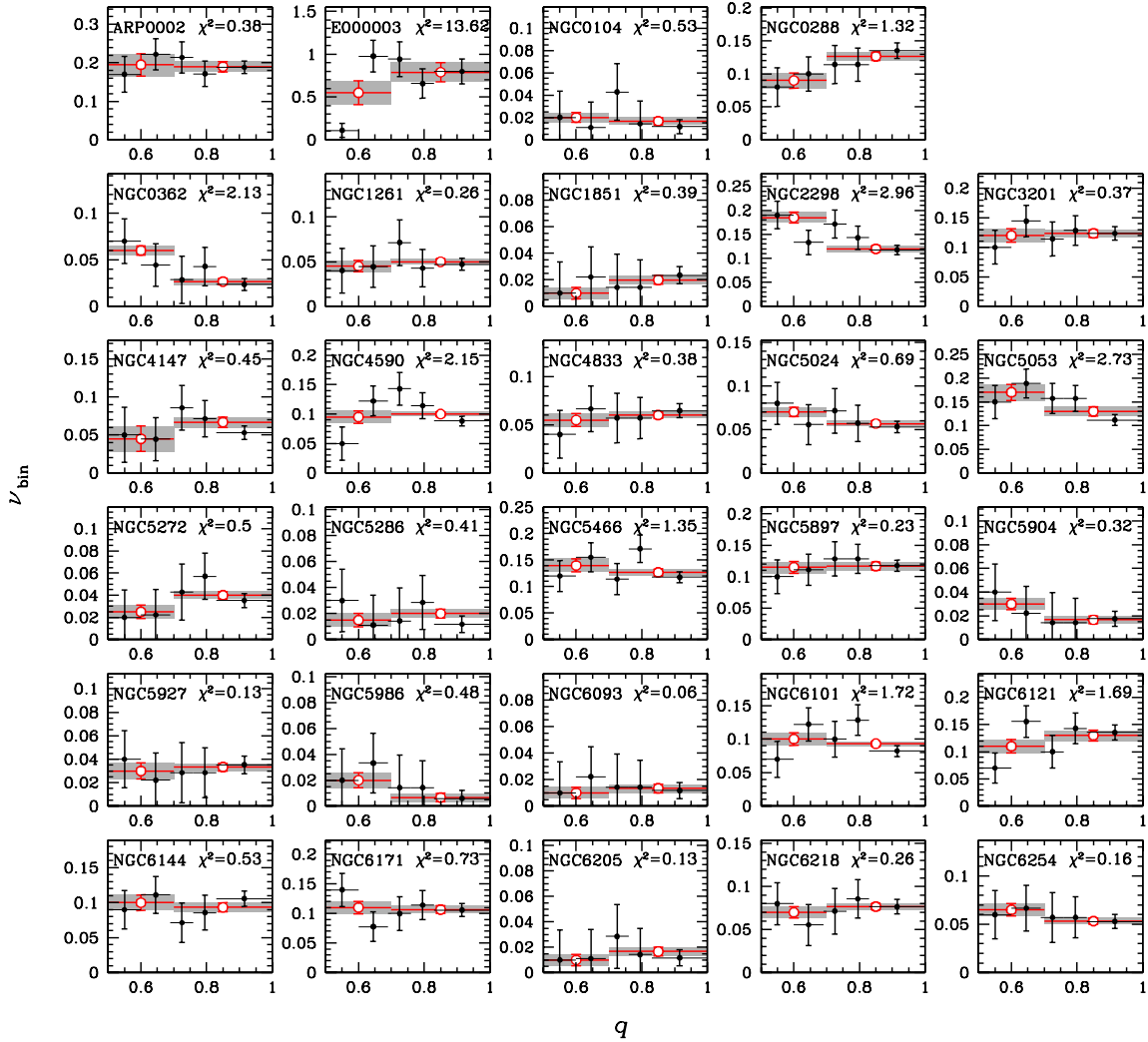}
      \caption{ Mass-ratio distribution for the binary population
	in the ACS field of 29 GCs.
        Black filled-circles show the normalized fraction of
        binaries in five mass-ratio intervals, while red
        open-circles indicate the $\nu_{\rm bin, i}$ values
        obtained by using only two bins with 0.5$<${\it q}$<$0.7, and
        0.7$<${\it q}$<$1. Horizontal  segments mark
        the adopted mass-ratio interval, while observational errors
        are plotted as the vertical lines and
        shadowed areas.
       }
         \label{QD1}
   \end{figure*}

\begin{figure*}[!ht]
   \centering
   \includegraphics[width=15.0cm]{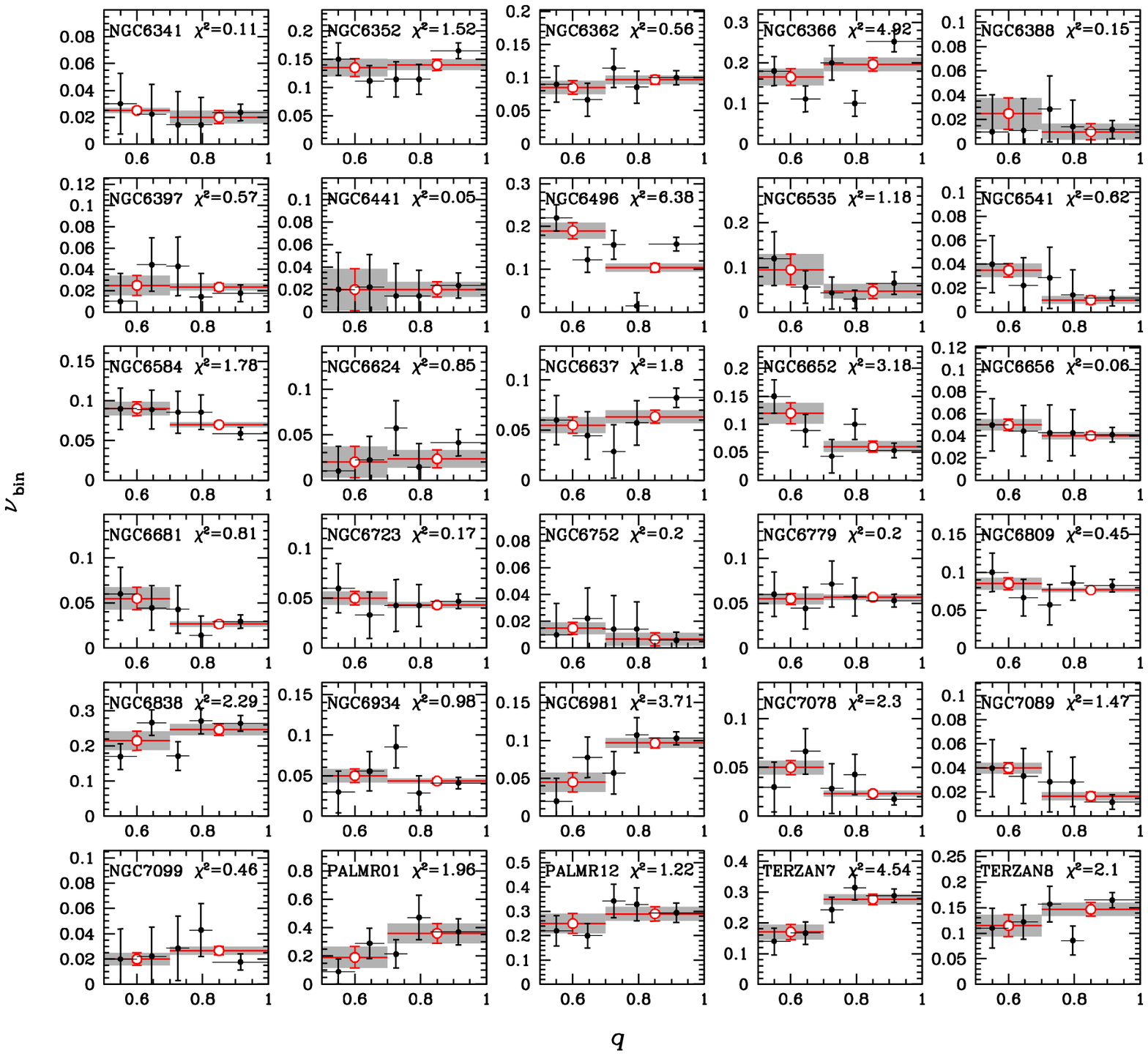}
      \caption{ As in Fig.~\ref{QD1} for the remaining 30 GCs.
       }
         \label{QD2}
   \end{figure*}

Results for all
clusters are shown with black symbols in Figs.~\ref{QD1} and~\ref{QD2}.
To increase the statistics, we have also divided Region {\it B} into two large mass
ratio intervals with $0.5<q<0.7$ and $0.7<q<1$ and calculated  $
\nu_{\rm bin}$ in each of them. The results we obtained by
using these $q$ bins are marked with red open circles in Figs.~\ref{QD1}
and~\ref{QD2}.

 The mass-ratio distribution is almost flat for most of the GCs of our
sample but in few cases we cannot exclude possible deviations from
this general trend. To investigate this statement we compared the
  observations with   a flat distribution, calculated for each
cluster the reduced $\chi^{2}$  and quoted it in
Figs.~\ref{QD1} and~\ref{QD2}.  
Montecarlo simulations demonstrate that in the case of a flat
distribution we expect the 50\% of the total number of clusters
having $\chi^{2}<$1.1  and the 99\%
 $\chi^{2}<$3.8. 
 We found $\chi^{2}$ values higher
than 3.8 in four GCs namely NGC 6366 ( $\chi^{2}$=4.92), NGC 6496
($\chi^{2}$=6.38),  TERZAN 7 ( $\chi^{2}$=4.45) and E 3 ($\chi^{2}$=13.62).
 
To compare the trend of the fraction of binaries as a function of {\it q} 
for different GCs we divided $\nu_{\rm bin, i}$  by
two times the fraction of binaries with {\it
  q}$>$0.5 \footnote{since $\nu_{\rm bin, i}$ depends on the fraction of
  binaries, which changes from one cluster to each other, to compare
  results from different clusters, we have to normalize it by the
  total fraction of binaries. Due to the lack of information on
  binaries with q$<$0.5, we normalized $\nu_{\rm bin, i}$ by $f_{\rm
    bin}^{q>0.5}$. We also multiplied the latter by a factor of two to
  normalize to one. 
(Note that, by chance, $2 f_{\rm bin}^{q>0.5}$ corresponds to the
  total fraction of binaries for the case of flat mass-ratio distribution). }. 

\begin{figure}[!ht]
   \centering
   \includegraphics[width=8.5cm]{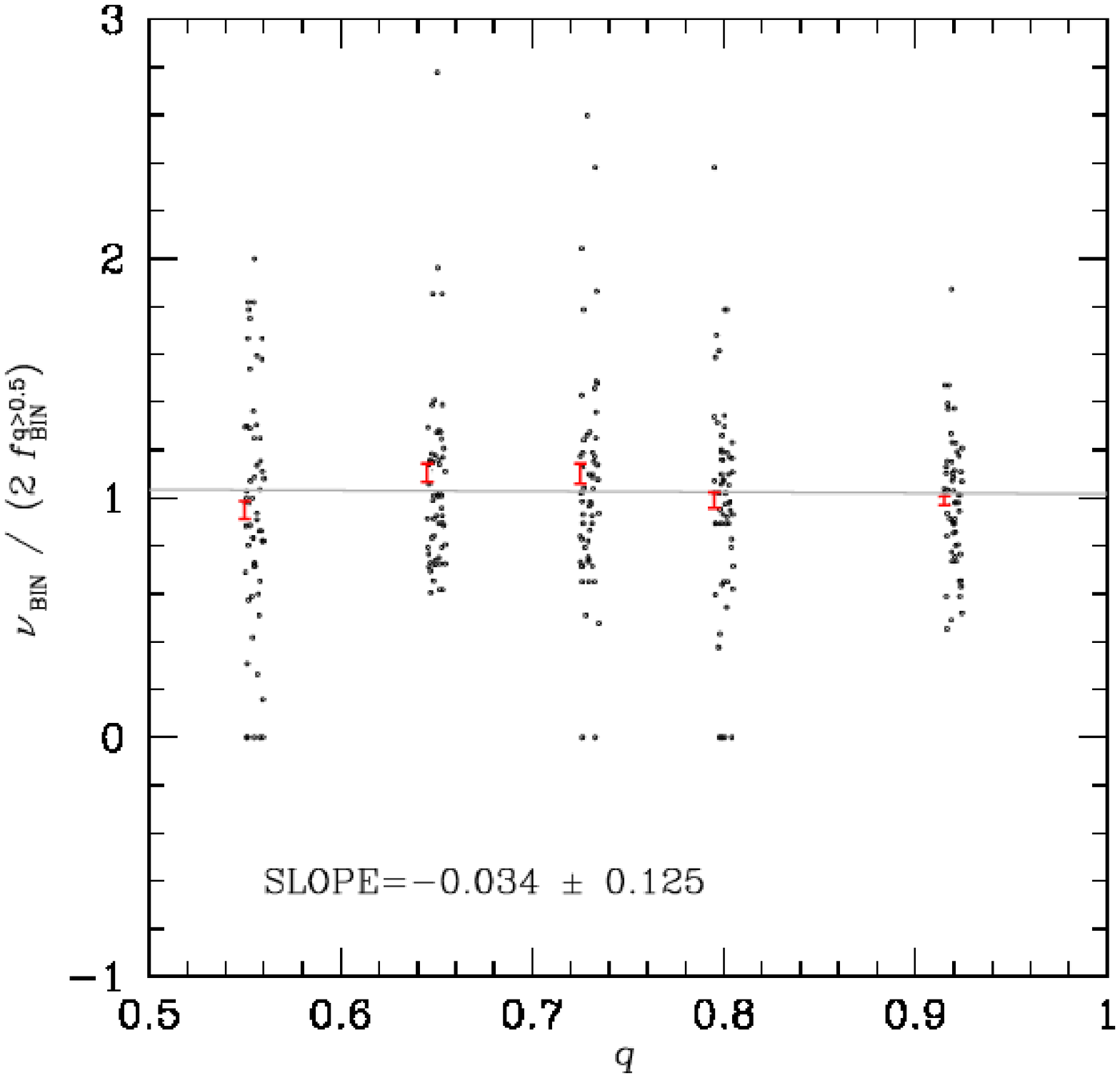}
      \caption{
	Black points show the  normalized fractions of binaries,
         $\nu_{\rm bin}$ 
	measured in five mass-ratio intervals as a function of {\it q} for all the GCs studied in this paper.
	To compare the fraction of binaries  in different clusters we divided
        $\nu_{\rm bin}$ by two times  the fraction of binaries with {\it q}$>$0.5.
	For clarity, black points have been randomly scattered around the corresponding  {\it q} value.
	Red points with error bars are the means in each mass-ratio bin, while the
	gray line is the best fitting line, whose slope is quoted in the inset.
       }
         \label{QDALL}
   \end{figure}

Results are in Fig.~\ref{QDALL}.
Black points indicate the measurements for all the GCs, while
red points with error bars are the averages in each mass-ratio bin.
The gray line is the best fitting line. Its slope is indicated in the
figure and suggests that the mass-ratio distribution is nearly  flat
for {\it q}$>$0.5. In the Appendix we will demonstrate that this
result is not affected by any significant systematic error.

 Since we have determined the mass-ratio
  distribution over the entire ACS/WFC field of view, our conclusions
  should indicate the general behavior of the binaries in
  GCs. Unfortunately, due to the relatively small numbers of binaries,
  we could not extend
this analysis to each sample of ${\it r}_{\rm C}$, the ${\it r}_{\rm C-HM}$, and the ${\it r}_{\rm oHM}$ stars.
  In these regions, due to  mass-segregation effects,  the mass-ratio
  distribution could differ from that shown in Fig.~\ref{QDALL}. 

\begin{figure}[ht!]
   \centering
   \includegraphics[width=8.5cm]{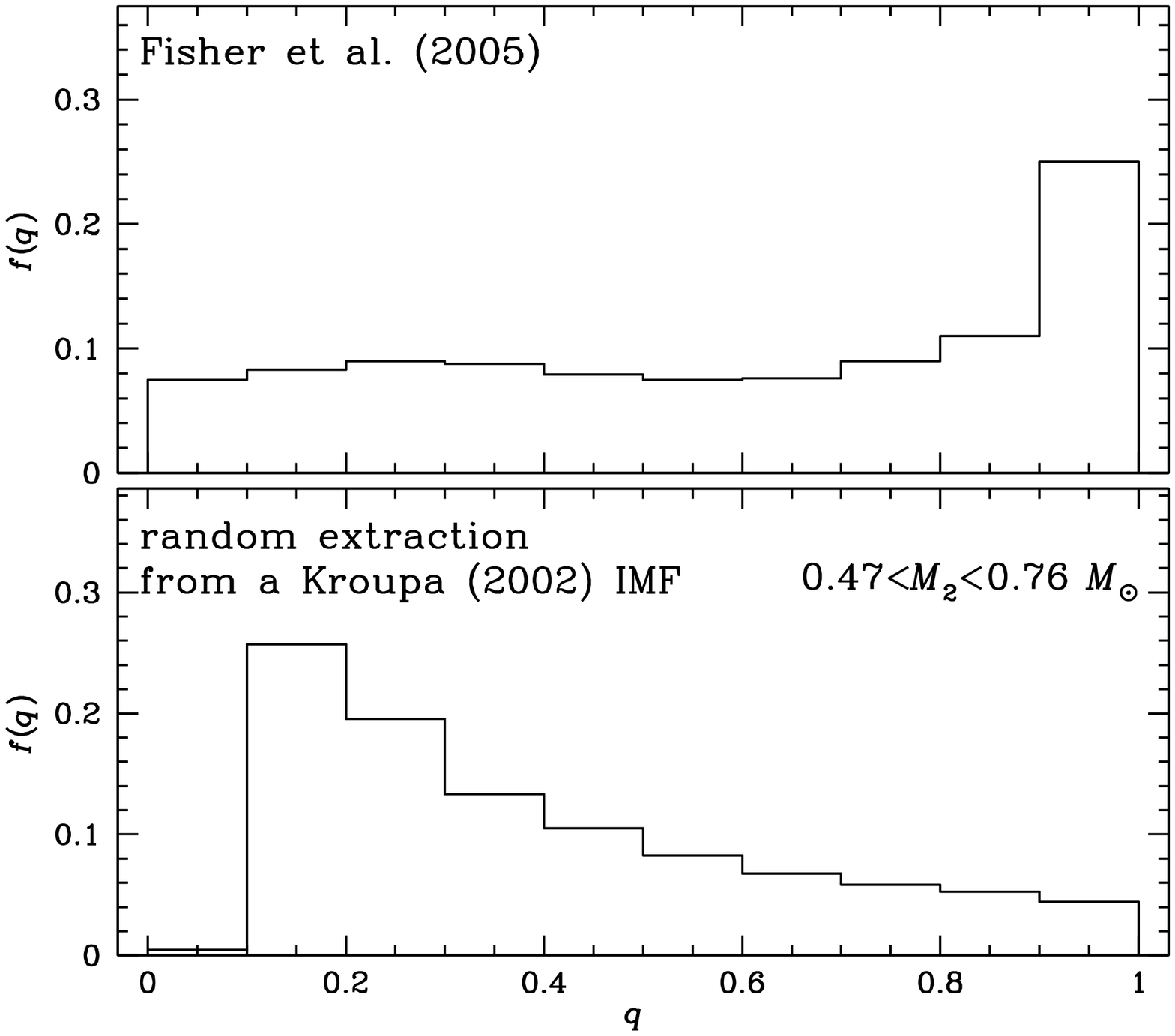}
      \caption{\textit{ Upper panel}: Mass-ratio distribution derived by Fisher et al.\ (2005).
		\textit{ Lower panel}: Mass-ratio distribution simulated from random extraction from a Kroupa\ (2002) IMF
	in the primary star mass intervals quoted in the inset.
	}
         \label{qdisFISHER}
   \end{figure}

 Up  to now, there  are few observational  constraints
on the overall mass-ratio distribution  of the binary  population in GCs.
One of the few measures of {\it f}(q) for binary systems, available
in the literature, comes  from Fisher et   al.\ (2005) who estimated   the
 the mass-ratio distribution function from
spectroscopic observations of field binaries within 100 parsecs
 from the Sun.
The {\it f}(q) derived by Fisher et al.\ (2005) is shown in the
    upper panel of
Fig.~\ref{qdisFISHER}. Binaries  with {\it q}$<$$\sim$0.9  have a
nearly flat distribution while there is a large concentration of
binaries formed by two components of similar mass. 
Tout (1991) studied the binary systems located in the local field and
suggests that {\it f}(q) can   be  derived  by randomly  extracting  secondary
stars from  the  observed  initial mass function (IMF).
The mass-ratio distribution that we obtain by randomly extracting pairs
of stars from a Kroupa (2002) IMF is displayed in
the  upper panel of Fig.~\ref{qdisFISHER}
for MS binaries with a primary with $0.47<M<0.76 M_{\odot}$ which is
the typical mass interval corresponding to the magnitude interval we
analyzed in the present work. 
In this case, the {\it f}({\rm q}) shape rapidly decreases from low
to high mass-ratio values with only the  24\% of binaries having
{\it q}$>$0.5.

In order to investigate whether
the observations of Fig.~\ref{QDALL} are
consistent with any of the two mass-ratio distributions described
above, we calculated the
normalized fraction of binaries we expect in the CMD of a GCs where
binary stars follow the distribution by Fisher et al.\ (2005) and the
distribution obtained from
random extraction of secondary stars from a Kroupa (2002) IMF
($\nu_{\rm bin, F}$, $\nu_{\rm bin, R}$). We also divided each of
these quantity by two times the 
fraction of binaries with {\it q}$>$0.5 of the corresponding CMD
($f_{\rm bin, F}^{\rm q>0.5}$, $f_{\rm bin, R}^{\rm q>0.5}$) in close
analogy to what done for real stars. 

\begin{figure}[ht!]
   \centering
   \includegraphics[width=8.5cm]{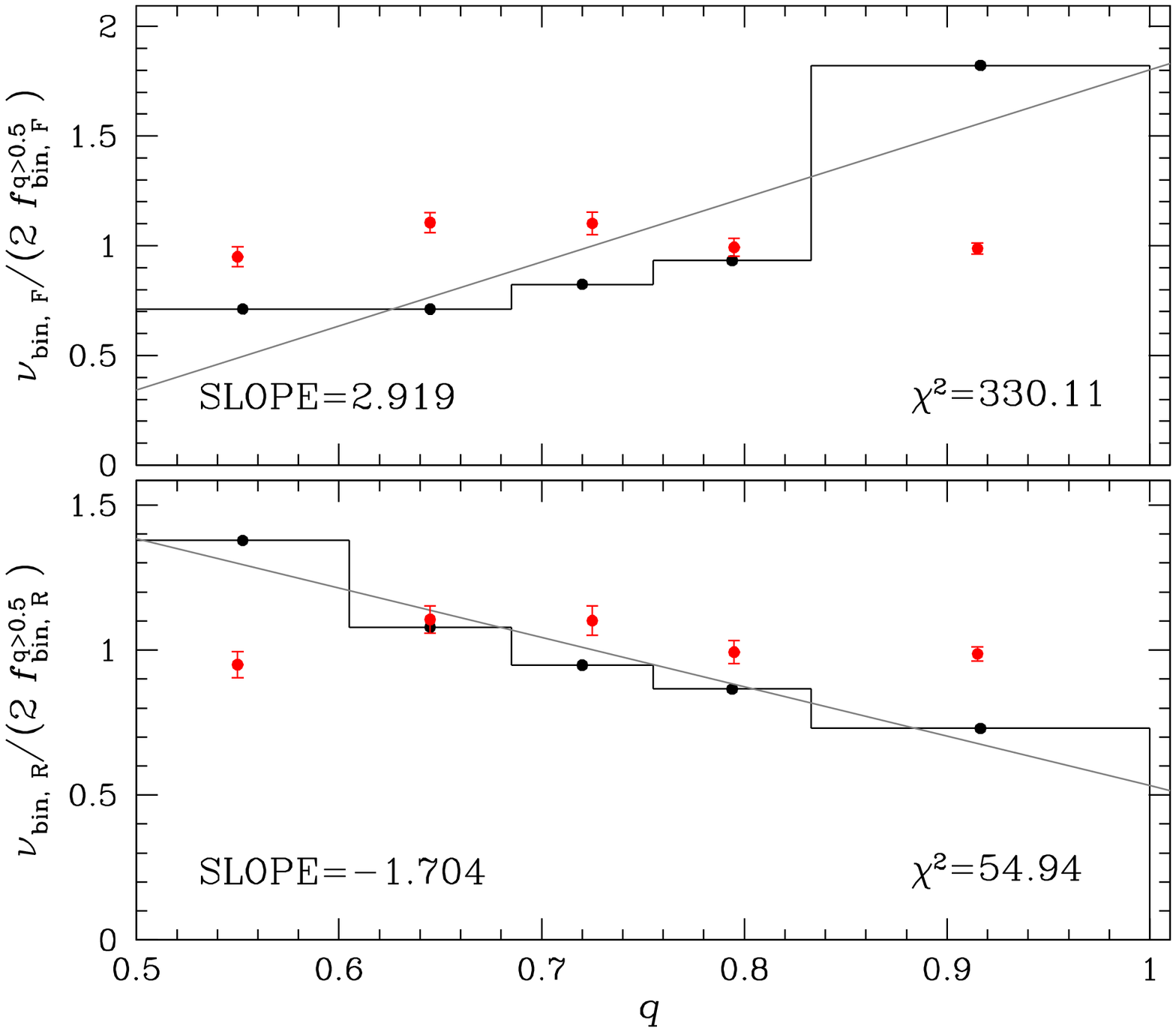}
      \caption{Frequency of binaries divided by two times the fraction
        of binaries with {\it q}$>$0.5 expected for the cases of Fisher
        et al.\ (2005) mass-ratio distribution (upper panel) and from
        the distribution obtained by randomly extracting secondary
        stars from a Kroupa (2002) IMF (bottom panels). Red points
        with the error bars are the mean values of the observed
        binary frequency normalized by two times ${\it f}_{\rm
          bin}^{\rm q>0.5}$ and have been already plotted in
        Fig.~\ref{QDALL}. The slope of the best-fitting least-squares
        gray straight lines  and the reduced-$\chi^{2}$ obtained from
        the comparison of the observed and theoretical distribution
        are quoted in the figure.
	}
         \label{vuSIMU}
   \end{figure}

Results are in Fig.~\ref{vuSIMU} where the values of 
$\nu_{\rm bin, F}/(2~f_{\rm bin, F}^{\rm q>0.5})$ and $\nu_{\rm bin,
  R}/(2~f_{\rm bin, R}^{\rm q>0.5})$ are plotted as a function of {\it
q}. The best-fitting least-squares lines are colored gray and their
slopes are quoted in the inset. Red points are the observed average
binary frequencies of Fig.~\ref{QDALL}. 
The large reduced-$\chi$ square values obtained from the comparison of
the theoretical and the observed points, and quoted in the
figure, indicate that neither the Fisher et al.\ (2005) nor the Tout
(1991) distribution properly matches the distribution we observe in GCs.

\subsection{The total binary fraction}
\label{totfraction}
The procedure described in the previous section allowed us to directly
measure the fraction of binaries with {\it q}$>$0.5
without any assumptions regarding  {\it f}({\rm q}).
On the other hand, because of the photometric errors, binaries with
small mass ratios ({\it q}$<$0.5) are  indistinguishable from
single MS stars in this dataset, therefore, any attempt to
determine the total fraction of MS-MS binaries without
assumption on the mass-ratio distribution is impossible with this
approach. 

The approach we follow to estimate the total fraction of binaries
is similar to that used by Sollima et al.\ (2007) and consists of assuming a
form for {\it f}({\rm q}). 
Since none of the two mass-ratio distributions available from
literature properly matches the observed distribution
in order to estimate the total fraction of binaries ($f_{\rm
  bin}^{\rm TOT}$), we extrapolated 
the results of Sect.~\ref{sec:qdist} adopting
a  flat {\it f(q)} also for binary systems with {\it q}$<$0.5;
i.\ e.\ , we assumed a constant mass-ratio distribution for all {\it q} values.
In this case as $f_{\rm bin}^{\rm q>0.5} \equiv f_{\rm bin}^{\rm q<0.5}$ the total fraction of binaries is simply\\
$f_{\rm bin}^{\rm TOT}=2.0 ~ f_{\rm bin}^{\rm q>0.5}$.\\
The final
$f_{\rm bin}^{\rm TOT}$ are listed in the fifth column of Table~2 for
the $r_{\rm C}$, the $r_{\rm C-HM}$, the $r_{\rm oHM}$ sample, and the
WFC field.

For completeness, we note that, according to Fisher et al.\ (2005),
66.5\% of binary systems have mass ratio larger than 0.5.
Hence, assuming a Fisher et al.\ (2005) mass ration distribution,
the total fraction of binaries should be: \\
$f_{\rm bin}^{\rm TOT, F}=1.504 ~ f_{\rm bin}^{\rm q>0.5}$.\\
If we assume that binary stars are formed by random associations
between stars of different masses, only 24\% of binaries have {\it
q}$>$0.5, and the total fraction of binaries becomes: \\
$f_{\rm bin}^{\rm TOT, R}=4.167 ~ f_{\rm bin}^{\rm q>0.5}$. \\
 
\begin{table*}
\centering
\scriptsize{
\caption{Fraction of binaries with mass ratio $q>0.5$, $q>0.6$ and
  $q>0.7$, and total fraction of binaries measured for ${\it r}_{\rm
    C}$, ${\it r}_{\rm C-HM}$, and ${\it r}_{\rm oHM}$ sample and in
  the whole ACS/WFC field of view for the  target GCs. We
  have analyzed only the ACS/WFC region with radial distance from the
  cluster center larger than a minimum radius ${\it R}_{\rm min}$
  whose value, in arcminutes, is listed in the first column. }
\begin{tabular}{cccccc}
\hline
\hline
\noalign{\smallskip}
\multicolumn{1}{c}{ID} &  
\multicolumn{1}{c}{REGION} &  
\multicolumn{1}{c}{${\it f}_{\rm bin}^{\rm q>0.5}$} & 
\multicolumn{1}{c}{${\it f}_{\rm bin}^{\rm q>0.6}$} &    
\multicolumn{1}{c}{${\it f}_{\rm bin}^{\rm q>0.7}$} &     
\multicolumn{1}{c}{${\it f}_{\rm bin}^{\rm TOT}$} \\ 
\\
\hline
\noalign{\smallskip}
ARP 2    & ${\it r}_{\rm C}$ sample     & 0.093$\pm$0.010  &  0.076$\pm$0.007 &  0.055$\pm$0.005 &  0.186$\pm$0.020 \\
           & ${\it r}_{\rm C-HM}$ sample  & 0.119$\pm$0.023  &  0.093$\pm$0.017 &  0.056$\pm$0.012 &  0.238$\pm$0.046 \\
           & ${\it r}_{\rm oHM}$ sample   & 0.091$\pm$0.031  &  0.086$\pm$0.024 &  0.081$\pm$0.017 &  0.182$\pm$0.062 \\
${\it R}_{\rm min}$=0.00 & WFC field                  & 0.096$\pm$0.009  &  0.079$\pm$0.006 &  0.057$\pm$0.004 &  0.192$\pm$0.018 \\
\hline
E 3    & ${\it r}_{\rm C}$ sample     & 0.360$\pm$0.043  &  0.350$\pm$0.042 &  0.247$\pm$0.035 &  0.720$\pm$0.086 \\
           & ${\it r}_{\rm C-HM}$ sample  & 0.317$\pm$0.203  &  0.147$\pm$0.171 &  0.264$\pm$0.171 &  0.634$\pm$0.406 \\
           & ${\it r}_{\rm oHM}$ sample   & 0.082$\pm$0.107  &  0.103$\pm$0.107 &  0.029$\pm$0.075 &  0.164$\pm$0.214 \\
${\it R}_{\rm min}$=0.00 & WFC field                  & 0.347$\pm$0.041  &  0.336$\pm$0.039 &  0.237$\pm$0.033 &  0.694$\pm$0.082 \\
\hline
NGC 104    & ${\it r}_{\rm C}$ sample     &       ---        &        ---       &        ---       &        ---       \\
           & ${\it r}_{\rm C-HM}$ sample  & 0.009$\pm$0.003  &  0.007$\pm$0.003 &  0.005$\pm$0.003 &  0.018$\pm$0.006 \\
           & ${\it r}_{\rm oHM}$ sample   &       ---        &        ---       &        ---       &        ---       \\
${\it R}_{\rm min}$=0.83 & WFC field                  & 0.009$\pm$0.003  &  0.007$\pm$0.003 &  0.005$\pm$0.003 &  0.018$\pm$0.006 \\
\hline
NGC 288    & ${\it r}_{\rm C}$ sample     & 0.056$\pm$0.005  &  0.050$\pm$0.004 &  0.041$\pm$0.003 &  0.112$\pm$0.010 \\
           & ${\it r}_{\rm C-HM}$ sample  & 0.054$\pm$0.007  &  0.045$\pm$0.005 &  0.030$\pm$0.004 &  0.108$\pm$0.014 \\
           & ${\it r}_{\rm oHM}$ sample   & 0.092$\pm$0.040  &  0.032$\pm$0.016 &  0.021$\pm$0.011 &  0.184$\pm$0.080 \\
${\it R}_{\rm min}$=0.00 & WFC field                  & 0.056$\pm$0.004  &  0.048$\pm$0.003 &  0.038$\pm$0.003 &  0.112$\pm$0.008 \\
\hline
NGC 362    & ${\it r}_{\rm C}$ sample     &       ---        &        ---       &        ---       &        ---       \\
           & ${\it r}_{\rm C-HM}$ sample  & 0.025$\pm$0.004  &  0.018$\pm$0.003 &  0.010$\pm$0.003 &  0.050$\pm$0.008 \\
           & ${\it r}_{\rm oHM}$ sample   & 0.016$\pm$0.003  &  0.011$\pm$0.003 &  0.008$\pm$0.003 &  0.032$\pm$0.006 \\
${\it R}_{\rm min}$=0.42 & WFC field                  & 0.020$\pm$0.003  &  0.013$\pm$0.003 &  0.008$\pm$0.003 &  0.040$\pm$0.006 \\
\hline
NGC 1261    & ${\it r}_{\rm C}$ sample     & 0.023$\pm$0.009  &  0.023$\pm$0.006 &  0.021$\pm$0.005 &  0.046$\pm$0.018 \\
           & ${\it r}_{\rm C-HM}$ sample  & 0.032$\pm$0.004  &  0.028$\pm$0.003 &  0.021$\pm$0.003 &  0.064$\pm$0.008 \\
           & ${\it r}_{\rm oHM}$ sample   & 0.020$\pm$0.003  &  0.018$\pm$0.003 &  0.012$\pm$0.003 &  0.040$\pm$0.006 \\
${\it R}_{\rm min}$=0.00 & WFC field                  & 0.024$\pm$0.003  &  0.021$\pm$0.003 &  0.015$\pm$0.003 &  0.048$\pm$0.006 \\
\hline
NGC 1851    & ${\it r}_{\rm C}$ sample     &       ---        &        ---       &        ---       &        ---       \\
           & ${\it r}_{\rm C-HM}$ sample  &       ---        &        ---       &        ---       &        ---       \\
           & ${\it r}_{\rm oHM}$ sample   & 0.008$\pm$0.003  &  0.008$\pm$0.003 &  0.006$\pm$0.003 &  0.016$\pm$0.006 \\
${\it R}_{\rm min}$=0.67 & WFC field                  & 0.008$\pm$0.003  &  0.008$\pm$0.003 &  0.006$\pm$0.003 &  0.016$\pm$0.006 \\
\hline
NGC 2298    & ${\it r}_{\rm C}$ sample     & 0.077$\pm$0.009  &  0.066$\pm$0.006 &  0.041$\pm$0.004 &  0.154$\pm$0.018 \\
           & ${\it r}_{\rm C-HM}$ sample  & 0.056$\pm$0.007  &  0.047$\pm$0.005 &  0.036$\pm$0.004 &  0.112$\pm$0.014 \\
           & ${\it r}_{\rm oHM}$ sample   & 0.047$\pm$0.004  &  0.034$\pm$0.003 &  0.023$\pm$0.003 &  0.094$\pm$0.008 \\
${\it R}_{\rm min}$=0.00 & WFC field                  & 0.073$\pm$0.004  &  0.054$\pm$0.003 &  0.036$\pm$0.003 &  0.146$\pm$0.008 \\
\hline
NGC 3201    & ${\it r}_{\rm C}$ sample     & 0.064$\pm$0.004  &  0.056$\pm$0.003 &  0.042$\pm$0.003 &  0.128$\pm$0.008 \\
           & ${\it r}_{\rm C-HM}$ sample  & 0.054$\pm$0.006  &  0.039$\pm$0.004 &  0.026$\pm$0.003 &  0.108$\pm$0.012 \\
           & ${\it r}_{\rm oHM}$ sample   &       ---        &        ---       &        ---       &        ---       \\
${\it R}_{\rm min}$=0.00 & WFC field                  & 0.061$\pm$0.003  &  0.051$\pm$0.003 &  0.037$\pm$0.003 &  0.122$\pm$0.006 \\
\hline
NGC 4147    & ${\it r}_{\rm C}$ sample     & 0.131$\pm$0.047  &  0.103$\pm$0.036 &  0.044$\pm$0.021 &  0.262$\pm$0.094 \\
           & ${\it r}_{\rm C-HM}$ sample  & 0.017$\pm$0.011  &  0.041$\pm$0.007 &  0.036$\pm$0.005 &  0.034$\pm$0.022 \\
           & ${\it r}_{\rm oHM}$ sample   & 0.019$\pm$0.006  &  0.019$\pm$0.003 &  0.012$\pm$0.003 &  0.038$\pm$0.012 \\
${\it R}_{\rm min}$=0.00 & WFC field                  & 0.029$\pm$0.005  &  0.027$\pm$0.003 &  0.020$\pm$0.003 &  0.058$\pm$0.010 \\
\hline
NGC 4590    & ${\it r}_{\rm C}$ sample     & 0.057$\pm$0.006  &  0.054$\pm$0.004 &  0.040$\pm$0.003 &  0.114$\pm$0.012 \\
           & ${\it r}_{\rm C-HM}$ sample  & 0.040$\pm$0.004  &  0.037$\pm$0.003 &  0.023$\pm$0.003 &  0.080$\pm$0.008 \\
           & ${\it r}_{\rm oHM}$ sample   & 0.053$\pm$0.007  &  0.038$\pm$0.005 &  0.025$\pm$0.003 &  0.106$\pm$0.014 \\
${\it R}_{\rm min}$=0.00 & WFC field                  & 0.049$\pm$0.003  &  0.044$\pm$0.003 &  0.030$\pm$0.003 &  0.098$\pm$0.006 \\
\hline
NGC 4833    & ${\it r}_{\rm C}$ sample     & 0.033$\pm$0.004  &  0.029$\pm$0.003 &  0.021$\pm$0.003 &  0.066$\pm$0.008 \\
           & ${\it r}_{\rm C-HM}$ sample  & 0.020$\pm$0.003  &  0.018$\pm$0.003 &  0.014$\pm$0.003 &  0.040$\pm$0.006 \\
           & ${\it r}_{\rm oHM}$ sample   &       ---        &        ---       &        ---       &        ---       \\
${\it R}_{\rm min}$=0.00 & WFC field                  & 0.029$\pm$0.003  &  0.025$\pm$0.003 &  0.018$\pm$0.003 &  0.058$\pm$0.006 \\
\hline
NGC 5024    & ${\it r}_{\rm C}$ sample     &       ---        &        ---       &        ---       &        ---       \\
           & ${\it r}_{\rm C-HM}$ sample  & 0.028$\pm$0.003  &  0.021$\pm$0.003 &  0.014$\pm$0.003 &  0.056$\pm$0.006 \\
           & ${\it r}_{\rm oHM}$ sample   & 0.033$\pm$0.003  &  0.024$\pm$0.003 &  0.019$\pm$0.003 &  0.066$\pm$0.006 \\
${\it R}_{\rm min}$=0.75 & WFC field                  & 0.031$\pm$0.003  &  0.023$\pm$0.003 &  0.017$\pm$0.003 &  0.062$\pm$0.006 \\
\hline
NGC 5053    & ${\it r}_{\rm C}$ sample     & 0.072$\pm$0.005  &  0.058$\pm$0.004 &  0.038$\pm$0.003 &  0.144$\pm$0.010 \\
           & ${\it r}_{\rm C-HM}$ sample  & 0.093$\pm$0.020  &  0.072$\pm$0.013 &  0.050$\pm$0.010 &  0.186$\pm$0.040 \\
           & ${\it r}_{\rm oHM}$ sample   &       ---        &        ---       &        ---       &        ---       \\
${\it R}_{\rm min}$=0.00 & WFC field                  & 0.073$\pm$0.005  &  0.059$\pm$0.004 &  0.039$\pm$0.003 &  0.146$\pm$0.010 \\
\hline
NGC 5272    & ${\it r}_{\rm C}$ sample     & 0.027$\pm$0.007  &  0.031$\pm$0.004 &  0.024$\pm$0.003 &  0.054$\pm$0.014 \\
           & ${\it r}_{\rm C-HM}$ sample  & 0.012$\pm$0.003  &  0.011$\pm$0.003 &  0.010$\pm$0.003 &  0.024$\pm$0.006 \\
           & ${\it r}_{\rm oHM}$ sample   & 0.019$\pm$0.003  &  0.015$\pm$0.003 &  0.012$\pm$0.003 &  0.038$\pm$0.006 \\
${\it R}_{\rm min}$=0.00 & WFC field                  & 0.017$\pm$0.003  &  0.015$\pm$0.003 &  0.012$\pm$0.003 &  0.034$\pm$0.006 \\
\hline
NGC 5286    & ${\it r}_{\rm C}$ sample     &       ---        &        ---       &        ---       &        ---       \\
           & ${\it r}_{\rm C-HM}$ sample  &       ---        &        ---       &        ---       &        ---       \\
           & ${\it r}_{\rm oHM}$ sample   & 0.011$\pm$0.003  &  0.008$\pm$0.003 &  0.007$\pm$0.003 &  0.022$\pm$0.006 \\
${\it R}_{\rm min}$=0.83 & WFC field                  & 0.009$\pm$0.003  &  0.006$\pm$0.003 &  0.006$\pm$0.003 &  0.018$\pm$0.006 \\
\hline
NGC 5466    & ${\it r}_{\rm C}$ sample     & 0.071$\pm$0.004  &  0.058$\pm$0.003 &  0.041$\pm$0.003 &  0.142$\pm$0.008 \\
           & ${\it r}_{\rm C-HM}$ sample  & 0.055$\pm$0.008  &  0.049$\pm$0.006 &  0.029$\pm$0.004 &  0.110$\pm$0.016 \\
           & ${\it r}_{\rm oHM}$ sample   & 0.016$\pm$0.035  &  0.022$\pm$0.024 &  0.009$\pm$0.010 &  0.032$\pm$0.070 \\
${\it R}_{\rm min}$=0.00 & WFC field                  & 0.066$\pm$0.004  &  0.055$\pm$0.003 &  0.038$\pm$0.003 &  0.132$\pm$0.008 \\
\hline
NGC 5897    & ${\it r}_{\rm C}$ sample     & 0.059$\pm$0.003  &  0.051$\pm$0.003 &  0.037$\pm$0.003 &  0.118$\pm$0.006 \\
           & ${\it r}_{\rm C-HM}$ sample  & 0.025$\pm$0.017  &  0.012$\pm$0.011 &  0.008$\pm$0.008 &  0.050$\pm$0.034 \\
           & ${\it r}_{\rm oHM}$ sample   &       ---        &        ---       &        ---       &        ---       \\
${\it R}_{\rm min}$=0.00 & WFC field                  & 0.058$\pm$0.003  &  0.049$\pm$0.003 &  0.035$\pm$0.003 &  0.116$\pm$0.006 \\
\hline
NGC 5904    & ${\it r}_{\rm C}$ sample     &       ---        &        ---       &        ---       &        ---       \\
           & ${\it r}_{\rm C-HM}$ sample  & 0.012$\pm$0.003  &  0.007$\pm$0.003 &  0.005$\pm$0.003 &  0.024$\pm$0.006 \\
           & ${\it r}_{\rm oHM}$ sample   & 0.006$\pm$0.009  &  0.003$\pm$0.004 &  0.005$\pm$0.003 &  0.012$\pm$0.018 \\
${\it R}_{\rm min}$=0.67 & WFC field                  & 0.011$\pm$0.003  &  0.007$\pm$0.003 &  0.005$\pm$0.003 &  0.022$\pm$0.006 \\
\hline
\hline
\noalign{\smallskip}
\end{tabular}
}
\end{table*}

\begin{table*}
\addtocounter{table}{-1}
\caption{Cont.} 
\centering
\scriptsize{
\begin{tabular}{cccccc}
\hline
\hline
\noalign{\smallskip}
\noalign{\smallskip}
\multicolumn{1}{c}{ID} &  
\multicolumn{1}{c}{REGION} &  
\multicolumn{1}{c}{${\it f}_{\rm bin}^{\rm q>0.5}$} & 
\multicolumn{1}{c}{${\it f}_{\rm bin}^{\rm q>0.6}$} &    
\multicolumn{1}{c}{${\it f}_{\rm bin}^{\rm q>0.7}$} &     
\multicolumn{1}{c}{${\it f}_{\rm bin}^{\rm TOT}$} \\ 
\\
\hline
\noalign{\smallskip}
NGC 5927    & ${\it r}_{\rm C}$ sample     & 0.052$\pm$0.009  &  0.037$\pm$0.007 &  0.030$\pm$0.006 &  0.104$\pm$0.018 \\
           & ${\it r}_{\rm C-HM}$ sample  & 0.026$\pm$0.003  &  0.016$\pm$0.003 &  0.014$\pm$0.003 &  0.052$\pm$0.006 \\
           & ${\it r}_{\rm oHM}$ sample   & 0.006$\pm$0.003  &  0.006$\pm$0.003 &  0.004$\pm$0.003 &  0.012$\pm$0.006 \\
${\it R}_{\rm min}$=0.00 & WFC field                  & 0.016$\pm$0.003  &  0.012$\pm$0.003 &  0.010$\pm$0.003 &  0.032$\pm$0.006 \\
\hline
NGC 5986    & ${\it r}_{\rm C}$ sample     &       ---        &        ---       &        ---       &        ---       \\
           & ${\it r}_{\rm C-HM}$ sample  & 0.005$\pm$0.004  &  0.003$\pm$0.003 &  0.003$\pm$0.003 &  0.010$\pm$0.008 \\
           & ${\it r}_{\rm oHM}$ sample   & 0.006$\pm$0.003  &  0.003$\pm$0.003 &  0.001$\pm$0.003 &  0.012$\pm$0.006 \\
${\it R}_{\rm min}$=0.83 & WFC field                  & 0.006$\pm$0.003  &  0.003$\pm$0.003 &  0.002$\pm$0.003 &  0.012$\pm$0.006 \\
\hline
NGC 6093    & ${\it r}_{\rm C}$ sample     &       ---        &        ---       &        ---       &        ---       \\
           & ${\it r}_{\rm C-HM}$ sample  &       ---        &        ---       &        ---       &        ---       \\
           & ${\it r}_{\rm oHM}$ sample   & 0.006$\pm$0.003  &  0.006$\pm$0.003 &  0.004$\pm$0.003 &  0.012$\pm$0.006 \\
${\it R}_{\rm min}$=0.58 & WFC field                  & 0.006$\pm$0.003  &  0.006$\pm$0.003 &  0.004$\pm$0.003 &  0.012$\pm$0.006 \\
\hline
NGC 6101    & ${\it r}_{\rm C}$ sample     & 0.050$\pm$0.004  &  0.043$\pm$0.003 &  0.031$\pm$0.003 &  0.100$\pm$0.008 \\
           & ${\it r}_{\rm C-HM}$ sample  & 0.042$\pm$0.004  &  0.040$\pm$0.003 &  0.026$\pm$0.003 &  0.084$\pm$0.008 \\
           & ${\it r}_{\rm oHM}$ sample   & 0.054$\pm$0.007  &  0.039$\pm$0.005 &  0.021$\pm$0.003 &  0.108$\pm$0.014 \\
${\it R}_{\rm min}$=0.00 & WFC field                  & 0.048$\pm$0.003  &  0.041$\pm$0.003 &  0.028$\pm$0.003 &  0.096$\pm$0.006 \\
\hline
NGC 6121    & ${\it r}_{\rm C}$ sample     & 0.074$\pm$0.007  &  0.073$\pm$0.006 &  0.052$\pm$0.005 &  0.148$\pm$0.014 \\
           & ${\it r}_{\rm C-HM}$ sample  & 0.051$\pm$0.005  &  0.042$\pm$0.004 &  0.030$\pm$0.003 &  0.102$\pm$0.010 \\
           & ${\it r}_{\rm oHM}$ sample   &       ---        &        ---       &        ---       &        ---       \\
${\it R}_{\rm min}$=0.00 & WFC field                  & 0.061$\pm$0.004  &  0.055$\pm$0.004 &  0.039$\pm$0.003 &  0.122$\pm$0.008 \\
\hline
NGC 6144    & ${\it r}_{\rm C}$ sample     & 0.066$\pm$0.006  &  0.059$\pm$0.005 &  0.046$\pm$0.004 &  0.132$\pm$0.012 \\
           & ${\it r}_{\rm C-HM}$ sample  & 0.039$\pm$0.005  &  0.029$\pm$0.004 &  0.017$\pm$0.003 &  0.078$\pm$0.010 \\
           & ${\it r}_{\rm oHM}$ sample   & 0.030$\pm$0.007  &  0.021$\pm$0.005 &  0.010$\pm$0.004 &  0.060$\pm$0.014 \\
${\it R}_{\rm min}$=0.00 & WFC field                  & 0.048$\pm$0.003  &  0.040$\pm$0.003 &  0.028$\pm$0.003 &  0.096$\pm$0.006 \\
\hline
NGC 6171    & ${\it r}_{\rm C}$ sample     & 0.093$\pm$0.011  &  0.071$\pm$0.008 &  0.052$\pm$0.007 &  0.186$\pm$0.022 \\
           & ${\it r}_{\rm C-HM}$ sample  & 0.046$\pm$0.003  &  0.035$\pm$0.003 &  0.027$\pm$0.003 &  0.092$\pm$0.006 \\
           & ${\it r}_{\rm oHM}$ sample   &       ---        &        ---       &        ---       &        ---       \\
${\it R}_{\rm min}$=0.00 & WFC field                  & 0.054$\pm$0.003  &  0.042$\pm$0.003 &  0.032$\pm$0.003 &  0.108$\pm$0.006 \\
\hline
NGC 6205    & ${\it r}_{\rm C}$ sample     & 0.005$\pm$0.003  &  0.010$\pm$0.003 &  0.007$\pm$0.003 &  0.010$\pm$0.006 \\
           & ${\it r}_{\rm C-HM}$ sample  & 0.006$\pm$0.003  &  0.004$\pm$0.003 &  0.004$\pm$0.003 &  0.012$\pm$0.006 \\
           & ${\it r}_{\rm oHM}$ sample   & 0.012$\pm$0.003  &  0.006$\pm$0.003 &  0.004$\pm$0.003 &  0.024$\pm$0.006 \\
${\it R}_{\rm min}$=0.00 & WFC field                  & 0.007$\pm$0.003  &  0.006$\pm$0.003 &  0.005$\pm$0.003 &  0.014$\pm$0.006 \\
\hline
NGC 6218    & ${\it r}_{\rm C}$ sample     & 0.057$\pm$0.005  &  0.046$\pm$0.004 &  0.034$\pm$0.004 &  0.114$\pm$0.010 \\
           & ${\it r}_{\rm C-HM}$ sample  & 0.032$\pm$0.003  &  0.025$\pm$0.003 &  0.019$\pm$0.003 &  0.064$\pm$0.006 \\
           & ${\it r}_{\rm oHM}$ sample   & 0.011$\pm$0.013  &  0.007$\pm$0.009 &  0.004$\pm$0.007 &  0.022$\pm$0.026 \\
${\it R}_{\rm min}$=0.00 & WFC field                  & 0.037$\pm$0.003  &  0.030$\pm$0.003 &  0.023$\pm$0.003 &  0.074$\pm$0.006 \\
\hline
NGC 6254    & ${\it r}_{\rm C}$ sample     & 0.039$\pm$0.004  &  0.032$\pm$0.003 &  0.023$\pm$0.003 &  0.078$\pm$0.008 \\
           & ${\it r}_{\rm C-HM}$ sample  & 0.022$\pm$0.003  &  0.017$\pm$0.003 &  0.012$\pm$0.003 &  0.044$\pm$0.006 \\
           & ${\it r}_{\rm oHM}$ sample   & 0.027$\pm$0.007  &  0.018$\pm$0.005 &  0.012$\pm$0.003 &  0.054$\pm$0.014 \\
${\it R}_{\rm min}$=0.00 & WFC field                  & 0.029$\pm$0.003  &  0.023$\pm$0.003 &  0.016$\pm$0.003 &  0.058$\pm$0.006 \\
\hline
NGC 6341    & ${\it r}_{\rm C}$ sample     &       ---        &        ---       &        ---       &        ---       \\
           & ${\it r}_{\rm C-HM}$ sample  & 0.010$\pm$0.003  &  0.007$\pm$0.003 &  0.005$\pm$0.003 &  0.020$\pm$0.006 \\
           & ${\it r}_{\rm oHM}$ sample   & 0.009$\pm$0.003  &  0.007$\pm$0.003 &  0.004$\pm$0.003 &  0.018$\pm$0.006 \\
${\it R}_{\rm min}$=0.42 & WFC field                  & 0.011$\pm$0.003  &  0.008$\pm$0.003 &  0.006$\pm$0.003 &  0.022$\pm$0.006 \\
\hline
NGC 6352    & ${\it r}_{\rm C}$ sample     & 0.092$\pm$0.008  &  0.078$\pm$0.007 &  0.054$\pm$0.005 &  0.184$\pm$0.016 \\
           & ${\it r}_{\rm C-HM}$ sample  & 0.053$\pm$0.005  &  0.041$\pm$0.004 &  0.034$\pm$0.003 &  0.106$\pm$0.010 \\
           & ${\it r}_{\rm oHM}$ sample   & 0.039$\pm$0.017  &  0.026$\pm$0.014 &  0.015$\pm$0.011 &  0.078$\pm$0.034 \\
${\it R}_{\rm min}$=0.00 & WFC field                  & 0.069$\pm$0.004  &  0.055$\pm$0.003 &  0.042$\pm$0.003 &  0.138$\pm$0.008 \\
\hline
NGC 6362    & ${\it r}_{\rm C}$ sample     & 0.060$\pm$0.004  &  0.044$\pm$0.003 &  0.034$\pm$0.003 &  0.120$\pm$0.008 \\
           & ${\it r}_{\rm C-HM}$ sample  & 0.021$\pm$0.005  &  0.020$\pm$0.004 &  0.016$\pm$0.003 &  0.042$\pm$0.010 \\
           & ${\it r}_{\rm oHM}$ sample   & 0.032$\pm$0.037  &  0.023$\pm$0.026 &  0.043$\pm$0.024 &  0.064$\pm$0.074 \\
${\it R}_{\rm min}$=0.00 & WFC field                  & 0.046$\pm$0.003  &  0.037$\pm$0.003 &  0.029$\pm$0.003 &  0.092$\pm$0.006 \\
\hline
NGC 6366    & ${\it r}_{\rm C}$ sample     & 0.099$\pm$0.007  &  0.082$\pm$0.006 &  0.064$\pm$0.006 &  0.198$\pm$0.014 \\
           & ${\it r}_{\rm C-HM}$ sample  & 0.057$\pm$0.015  &  0.035$\pm$0.012 &  0.042$\pm$0.012 &  0.114$\pm$0.030 \\
           & ${\it r}_{\rm oHM}$ sample   &       ---        &        ---       &        ---       &        ---       \\
${\it R}_{\rm min}$=0.00 & WFC field                  & 0.092$\pm$0.007  &  0.074$\pm$0.006 &  0.059$\pm$0.005 &  0.184$\pm$0.014 \\
\hline
NGC 6388    & ${\it r}_{\rm C}$ sample     &       ---        &        ---       &        ---       &        ---       \\
           & ${\it r}_{\rm C-HM}$ sample  &       ---        &        ---       &        ---       &        ---       \\
           & ${\it r}_{\rm oHM}$ sample   & 0.004$\pm$0.004  &  0.006$\pm$0.003 &  0.003$\pm$0.003 &  0.008$\pm$0.008 \\
${\it R}_{\rm min}$=0.83 & WFC field                  & 0.008$\pm$0.004  &  0.006$\pm$0.003 &  0.003$\pm$0.003 &  0.016$\pm$0.008 \\
\hline
NGC 6397    & ${\it r}_{\rm C}$ sample     & 0.035$\pm$0.018  &  0.037$\pm$0.015 &  0.037$\pm$0.013 &  0.070$\pm$0.036 \\
           & ${\it r}_{\rm C-HM}$ sample  & 0.012$\pm$0.003  &  0.010$\pm$0.003 &  0.005$\pm$0.003 &  0.024$\pm$0.006 \\
           & ${\it r}_{\rm oHM}$ sample   & 0.014$\pm$0.026  &  0.005$\pm$0.003 &  0.002$\pm$0.003 &  0.028$\pm$0.052 \\
${\it R}_{\rm min}$=0.00 & WFC field                  & 0.012$\pm$0.003  &  0.011$\pm$0.003 &  0.007$\pm$0.003 &  0.024$\pm$0.006 \\
\hline
NGC 6441    & ${\it r}_{\rm C}$ sample     &       ---        &        ---       &        ---       &        ---       \\
           & ${\it r}_{\rm C-HM}$ sample  &       ---        &        ---       &        ---       &        ---       \\
           & ${\it r}_{\rm oHM}$ sample   & 0.010$\pm$0.005  &  0.008$\pm$0.004 &  0.006$\pm$0.003 &  0.020$\pm$0.010 \\
${\it R}_{\rm min}$=1.00 & WFC field                  & 0.010$\pm$0.005  &  0.008$\pm$0.004 &  0.006$\pm$0.003 &  0.020$\pm$0.010 \\
\hline
NGC 6496    & ${\it r}_{\rm C}$ sample     & 0.089$\pm$0.006  &  0.073$\pm$0.005 &  0.051$\pm$0.004 &  0.178$\pm$0.012 \\
           & ${\it r}_{\rm C-HM}$ sample  & 0.077$\pm$0.008  &  0.053$\pm$0.007 &  0.036$\pm$0.006 &  0.154$\pm$0.016 \\
           & ${\it r}_{\rm oHM}$ sample   & 0.046$\pm$0.024  &  0.021$\pm$0.018 &  0.015$\pm$0.015 &  0.092$\pm$0.048 \\
${\it R}_{\rm min}$=0.00 & WFC field                  & 0.069$\pm$0.005  &  0.049$\pm$0.004 &  0.031$\pm$0.003 &  0.138$\pm$0.010 \\
\hline
NGC 6535    & ${\it r}_{\rm C}$ sample     & 0.046$\pm$0.016  &  0.027$\pm$0.012 &  0.014$\pm$0.008 &  0.092$\pm$0.032 \\
           & ${\it r}_{\rm C-HM}$ sample  & 0.026$\pm$0.013  &  0.018$\pm$0.009 &  0.018$\pm$0.009 &  0.052$\pm$0.026 \\
           & ${\it r}_{\rm oHM}$ sample   & 0.028$\pm$0.010  &  0.016$\pm$0.007 &  0.012$\pm$0.006 &  0.056$\pm$0.020 \\
${\it R}_{\rm min}$=0.00 & WFC field                  & 0.033$\pm$0.009  &  0.021$\pm$0.006 &  0.014$\pm$0.005 &  0.066$\pm$0.018 \\
\hline
\hline
\noalign{\smallskip}
\end{tabular}
}
\end{table*}

\begin{table*}
\addtocounter{table}{-1}
\caption{Cont.} 
\centering
\scriptsize{
\begin{tabular}{cccccc}
\hline
\hline
\noalign{\smallskip}
\noalign{\smallskip}
\multicolumn{1}{c}{ID} &  
\multicolumn{1}{c}{REGION} &  
\multicolumn{1}{c}{${\it f}_{\rm bin}^{\rm q>0.5}$} & 
\multicolumn{1}{c}{${\it f}_{\rm bin}^{\rm q>0.6}$} &    
\multicolumn{1}{c}{${\it f}_{\rm bin}^{\rm q>0.7}$} &     
\multicolumn{1}{c}{${\it f}_{\rm bin}^{\rm TOT}$} \\ 
\\
\hline
\noalign{\smallskip}
NGC 6541    & ${\it r}_{\rm C}$ sample     &       ---        &        ---       &        ---       &        ---       \\
           & ${\it r}_{\rm C-HM}$ sample  & 0.014$\pm$0.003  &  0.010$\pm$0.003 &  0.005$\pm$0.003 &  0.028$\pm$0.006 \\
           & ${\it r}_{\rm oHM}$ sample   & 0.010$\pm$0.003  &  0.005$\pm$0.003 &  0.001$\pm$0.003 &  0.020$\pm$0.006 \\
${\it R}_{\rm min}$=0.42 & WFC field                  & 0.010$\pm$0.003  &  0.007$\pm$0.003 &  0.003$\pm$0.003 &  0.020$\pm$0.006 \\
\hline
NGC 6584    & ${\it r}_{\rm C}$ sample     & 0.045$\pm$0.006  &  0.045$\pm$0.004 &  0.034$\pm$0.003 &  0.090$\pm$0.012 \\
           & ${\it r}_{\rm C-HM}$ sample  & 0.036$\pm$0.007  &  0.025$\pm$0.005 &  0.020$\pm$0.003 &  0.072$\pm$0.014 \\
           & ${\it r}_{\rm oHM}$ sample   & 0.025$\pm$0.003  &  0.016$\pm$0.003 &  0.009$\pm$0.003 &  0.050$\pm$0.006 \\
${\it R}_{\rm min}$=0.00 & WFC field                  & 0.039$\pm$0.003  &  0.030$\pm$0.003 &  0.021$\pm$0.003 &  0.078$\pm$0.006 \\
\hline
NGC 6624    & ${\it r}_{\rm C}$ sample     &       ---        &        ---       &        ---       &        ---       \\
           & ${\it r}_{\rm C-HM}$ sample  & 0.013$\pm$0.004  &  0.002$\pm$0.003 &  0.001$\pm$0.003 &  0.026$\pm$0.008 \\
           & ${\it r}_{\rm oHM}$ sample   & 0.013$\pm$0.005  &  0.018$\pm$0.005 &  0.010$\pm$0.004 &  0.026$\pm$0.010 \\
${\it R}_{\rm min}$=0.42 & WFC field                  & 0.011$\pm$0.004  &  0.012$\pm$0.003 &  0.007$\pm$0.003 &  0.022$\pm$0.008 \\
\hline
NGC 6637    & ${\it r}_{\rm C}$ sample     & 0.062$\pm$0.010  &  0.060$\pm$0.007 &  0.057$\pm$0.006 &  0.124$\pm$0.020 \\
           & ${\it r}_{\rm C-HM}$ sample  & 0.029$\pm$0.004  &  0.028$\pm$0.003 &  0.020$\pm$0.003 &  0.058$\pm$0.008 \\
           & ${\it r}_{\rm oHM}$ sample   & 0.013$\pm$0.003  &  0.008$\pm$0.003 &  0.005$\pm$0.003 &  0.026$\pm$0.006 \\
${\it R}_{\rm min}$=0.00 & WFC field                  & 0.030$\pm$0.003  &  0.024$\pm$0.003 &  0.019$\pm$0.003 &  0.060$\pm$0.006 \\
\hline
NGC 6652    & ${\it r}_{\rm C}$ sample     & 0.172$\pm$0.055  &  0.091$\pm$0.038 &  0.059$\pm$0.029 &  0.344$\pm$0.110 \\
           & ${\it r}_{\rm C-HM}$ sample  & 0.052$\pm$0.006  &  0.032$\pm$0.004 &  0.018$\pm$0.003 &  0.104$\pm$0.012 \\
           & ${\it r}_{\rm oHM}$ sample   & 0.027$\pm$0.006  &  0.021$\pm$0.005 &  0.016$\pm$0.004 &  0.054$\pm$0.012 \\
${\it R}_{\rm min}$=0.00 & WFC field                  & 0.042$\pm$0.004  &  0.027$\pm$0.003 &  0.018$\pm$0.003 &  0.084$\pm$0.008 \\
\hline
NGC 6656    & ${\it r}_{\rm C}$ sample     & 0.023$\pm$0.003  &  0.018$\pm$0.003 &  0.013$\pm$0.003 &  0.046$\pm$0.006 \\
           & ${\it r}_{\rm C-HM}$ sample  & 0.020$\pm$0.003  &  0.015$\pm$0.003 &  0.010$\pm$0.003 &  0.040$\pm$0.006 \\
           & ${\it r}_{\rm oHM}$ sample   &       ---        &        ---       &        ---       &        ---       \\
${\it R}_{\rm min}$=0.00 & WFC field                  & 0.022$\pm$0.003  &  0.017$\pm$0.003 &  0.012$\pm$0.003 &  0.044$\pm$0.006 \\
\hline
NGC 6681    & ${\it r}_{\rm C}$ sample     &       ---        &        ---       &        ---       &        ---       \\
           & ${\it r}_{\rm C-HM}$ sample  & 0.026$\pm$0.005  &  0.013$\pm$0.003 &  0.006$\pm$0.003 &  0.052$\pm$0.010 \\
           & ${\it r}_{\rm oHM}$ sample   & 0.005$\pm$0.004  &  0.011$\pm$0.003 &  0.010$\pm$0.003 &  0.010$\pm$0.008 \\
${\it R}_{\rm min}$=0.10 & WFC field                  & 0.019$\pm$0.003  &  0.013$\pm$0.003 &  0.008$\pm$0.003 &  0.038$\pm$0.006 \\
\hline
NGC 6723    & ${\it r}_{\rm C}$ sample     & 0.031$\pm$0.004  &  0.025$\pm$0.003 &  0.020$\pm$0.003 &  0.062$\pm$0.008 \\
           & ${\it r}_{\rm C-HM}$ sample  & 0.013$\pm$0.003  &  0.011$\pm$0.003 &  0.006$\pm$0.003 &  0.026$\pm$0.006 \\
           & ${\it r}_{\rm oHM}$ sample   & 0.017$\pm$0.004  &  0.008$\pm$0.003 &  0.008$\pm$0.003 &  0.034$\pm$0.008 \\
${\it R}_{\rm min}$=0.00 & WFC field                  & 0.023$\pm$0.003  &  0.017$\pm$0.003 &  0.013$\pm$0.003 &  0.046$\pm$0.006 \\
\hline
NGC 6752    & ${\it r}_{\rm C}$ sample     & 0.017$\pm$0.016  &  0.011$\pm$0.008 &  0.006$\pm$0.004 &  0.034$\pm$0.032 \\
           & ${\it r}_{\rm C-HM}$ sample  & 0.005$\pm$0.003  &  0.004$\pm$0.003 &  0.002$\pm$0.003 &  0.009$\pm$0.006 \\
           & ${\it r}_{\rm oHM}$ sample   &       ---        &        ---       &        ---       &        ---       \\
${\it R}_{\rm min}$=0.00 & WFC field                  & 0.005$\pm$0.003  &  0.004$\pm$0.003 &  0.002$\pm$0.003 &  0.010$\pm$0.006 \\
\hline
NGC 6779    & ${\it r}_{\rm C}$ sample     & 0.050$\pm$0.009  &  0.050$\pm$0.006 &  0.038$\pm$0.005 &  0.100$\pm$0.018 \\
           & ${\it r}_{\rm C-HM}$ sample  & 0.028$\pm$0.003  &  0.022$\pm$0.003 &  0.017$\pm$0.003 &  0.056$\pm$0.006 \\
           & ${\it r}_{\rm oHM}$ sample   & 0.023$\pm$0.003  &  0.016$\pm$0.003 &  0.012$\pm$0.003 &  0.046$\pm$0.006 \\
${\it R}_{\rm min}$=0.00 & WFC field                  & 0.028$\pm$0.003  &  0.022$\pm$0.003 &  0.017$\pm$0.003 &  0.056$\pm$0.006 \\
\hline
NGC 6809    & ${\it r}_{\rm C}$ sample     & 0.040$\pm$0.003  &  0.031$\pm$0.003 &  0.023$\pm$0.003 &  0.080$\pm$0.006 \\
           & ${\it r}_{\rm C-HM}$ sample  &       ---        &        ---       &        ---       &        ---       \\
           & ${\it r}_{\rm oHM}$ sample   &       ---        &        ---       &        ---       &        ---       \\
${\it R}_{\rm min}$=0.00 & WFC field                  & 0.040$\pm$0.003  &  0.031$\pm$0.003 &  0.023$\pm$0.003 &  0.080$\pm$0.006 \\
\hline
NGC 6838    & ${\it r}_{\rm C}$ sample     & 0.152$\pm$0.017  &  0.120$\pm$0.015 &  0.080$\pm$0.012 &  0.304$\pm$0.034 \\
           & ${\it r}_{\rm C-HM}$ sample  & 0.110$\pm$0.008  &  0.100$\pm$0.007 &  0.072$\pm$0.006 &  0.220$\pm$0.016 \\
           & ${\it r}_{\rm oHM}$ sample   & 0.104$\pm$0.014  &  0.084$\pm$0.012 &  0.076$\pm$0.011 &  0.208$\pm$0.028 \\
${\it R}_{\rm min}$=0.00 & WFC field                  & 0.117$\pm$0.007  &  0.101$\pm$0.006 &  0.074$\pm$0.005 &  0.234$\pm$0.014 \\
\hline
NGC 6934    & ${\it r}_{\rm C}$ sample     &       ---        &        ---       &        ---       &        ---       \\
           & ${\it r}_{\rm C-HM}$ sample  & 0.032$\pm$0.003  &  0.027$\pm$0.003 &  0.017$\pm$0.003 &  0.064$\pm$0.006 \\
           & ${\it r}_{\rm oHM}$ sample   & 0.020$\pm$0.004  &  0.019$\pm$0.003 &  0.012$\pm$0.003 &  0.040$\pm$0.008 \\
${\it R}_{\rm min}$=0.42 & WFC field                  & 0.023$\pm$0.003  &  0.021$\pm$0.003 &  0.013$\pm$0.003 &  0.046$\pm$0.006 \\
\hline
NGC 6981    & ${\it r}_{\rm C}$ sample     & 0.049$\pm$0.009  &  0.053$\pm$0.006 &  0.041$\pm$0.005 &  0.098$\pm$0.018 \\
           & ${\it r}_{\rm C-HM}$ sample  & 0.031$\pm$0.008  &  0.035$\pm$0.006 &  0.031$\pm$0.004 &  0.062$\pm$0.016 \\
           & ${\it r}_{\rm oHM}$ sample   & 0.034$\pm$0.006  &  0.028$\pm$0.004 &  0.019$\pm$0.003 &  0.068$\pm$0.012 \\
${\it R}_{\rm min}$=0.00 & WFC field                  & 0.038$\pm$0.004  &  0.037$\pm$0.003 &  0.029$\pm$0.003 &  0.076$\pm$0.008 \\
\hline
NGC 7078    & ${\it r}_{\rm C}$ sample     &       ---        &        ---       &        ---       &        ---       \\
           & ${\it r}_{\rm C-HM}$ sample  & 0.010$\pm$0.005  &  0.012$\pm$0.003 &  0.009$\pm$0.003 &  0.020$\pm$0.010 \\
           & ${\it r}_{\rm oHM}$ sample   & 0.018$\pm$0.003  &  0.014$\pm$0.003 &  0.007$\pm$0.003 &  0.036$\pm$0.006 \\
${\it R}_{\rm min}$=0.83 & WFC field                  & 0.017$\pm$0.003  &  0.014$\pm$0.003 &  0.007$\pm$0.003 &  0.034$\pm$0.006 \\
\hline
NGC 7089    & ${\it r}_{\rm C}$ sample     &       ---        &        ---       &        ---       &        ---       \\
           & ${\it r}_{\rm C-HM}$ sample  & 0.032$\pm$0.006  &  0.018$\pm$0.004 &  0.009$\pm$0.003 &  0.064$\pm$0.012 \\
           & ${\it r}_{\rm oHM}$ sample   & 0.011$\pm$0.003  &  0.009$\pm$0.003 &  0.005$\pm$0.003 &  0.022$\pm$0.006 \\
${\it R}_{\rm min}$=0.83 & WFC field                  & 0.013$\pm$0.003  &  0.009$\pm$0.003 &  0.005$\pm$0.003 &  0.026$\pm$0.006 \\
\hline
NGC 7099    & ${\it r}_{\rm C}$ sample     & 0.035$\pm$0.015  &  0.033$\pm$0.015 &  0.010$\pm$0.003 &  0.070$\pm$0.030 \\
           & ${\it r}_{\rm C-HM}$ sample  & 0.012$\pm$0.003  &  0.010$\pm$0.003 &  0.008$\pm$0.003 &  0.024$\pm$0.006 \\
           & ${\it r}_{\rm oHM}$ sample   & 0.013$\pm$0.003  &  0.009$\pm$0.003 &  0.007$\pm$0.003 &  0.026$\pm$0.006 \\
${\it R}_{\rm min}$=0.00 & WFC field                  & 0.012$\pm$0.003  &  0.010$\pm$0.003 &  0.008$\pm$0.003 &  0.024$\pm$0.006 \\
\hline
PALOMAR 1    & ${\it r}_{\rm C}$ sample     & 0.333$\pm$0.096  &  0.311$\pm$0.092 &  0.244$\pm$0.079 &  0.666$\pm$0.192 \\
           & ${\it r}_{\rm C-HM}$ sample  & 0.130$\pm$0.042  &  0.116$\pm$0.037 &  0.093$\pm$0.033 &  0.260$\pm$0.084 \\
           & ${\it r}_{\rm oHM}$ sample   & 0.095$\pm$0.031  &  0.089$\pm$0.027 &  0.070$\pm$0.023 &  0.190$\pm$0.062 \\
${\it R}_{\rm min}$=0.00 & WFC field                  & 0.146$\pm$0.027  &  0.136$\pm$0.024 &  0.108$\pm$0.021 &  0.292$\pm$0.054 \\
\hline
PALOMAR 12    & ${\it r}_{\rm C}$ sample     & 0.130$\pm$0.057  &  0.130$\pm$0.045 &  0.104$\pm$0.037 &  0.260$\pm$0.114 \\
           & ${\it r}_{\rm C-HM}$ sample  & 0.175$\pm$0.018  &  0.144$\pm$0.015 &  0.108$\pm$0.013 &  0.350$\pm$0.036 \\
           & ${\it r}_{\rm oHM}$ sample   & 0.066$\pm$0.019  &  0.055$\pm$0.014 &  0.044$\pm$0.012 &  0.132$\pm$0.038 \\
${\it R}_{\rm min}$=0.00 & WFC field                  & 0.137$\pm$0.013  &  0.114$\pm$0.011 &  0.087$\pm$0.009 &  0.274$\pm$0.026 \\
\hline
TERZAN 7    & ${\it r}_{\rm C}$ sample     & 0.187$\pm$0.017  &  0.159$\pm$0.013 &  0.140$\pm$0.011 &  0.374$\pm$0.034 \\
           & ${\it r}_{\rm C-HM}$ sample  & 0.084$\pm$0.016  &  0.092$\pm$0.013 &  0.073$\pm$0.010 &  0.168$\pm$0.032 \\
           & ${\it r}_{\rm oHM}$ sample   & 0.088$\pm$0.011  &  0.075$\pm$0.008 &  0.051$\pm$0.006 &  0.176$\pm$0.022 \\
${\it R}_{\rm min}$=0.00 & WFC field                  & 0.117$\pm$0.008  &  0.104$\pm$0.006 &  0.083$\pm$0.005 &  0.234$\pm$0.016 \\
\hline
TERZAN 8    & ${\it r}_{\rm C}$ sample     & 0.083$\pm$0.011  &  0.072$\pm$0.008 &  0.056$\pm$0.006 &  0.166$\pm$0.022 \\
           & ${\it r}_{\rm C-HM}$ sample  &       ---        &        ---       &        ---       &        ---       \\
           & ${\it r}_{\rm oHM}$ sample   & 0.059$\pm$0.009  &  0.047$\pm$0.006 &  0.037$\pm$0.005 &  0.118$\pm$0.018 \\
${\it R}_{\rm min}$=0.00 & WFC field                  & 0.067$\pm$0.007  &  0.056$\pm$0.005 &  0.044$\pm$0.004 &  0.134$\pm$0.014 \\
\hline
\hline
\noalign{\smallskip}
\end{tabular}
}
\end{table*}

\subsection{Comparison with previous measurements of the binary fraction in GCs}
\label{letteratura}
To date, the fraction of binaries has been measured for 30 GCs. 
In Table~3 we list the photometric binary fraction in Galactic GCs from previous 
measurements and available in the literature.
Although for some GCs of our sample the fraction of binaries were already estimated
in previous works, caution must be used to compare the results presented in this paper with literature ones.
In particular, it should be noted that
the inferred values of the total fraction of binaries are tightly
related to the assumed {\it f}({\it q}). 
Many authors just determined lower limits to the binary
  fraction, as they studied binary systems with large {\it q} 
that are clearly separated from single MS stars.
Without any indication on the mass-ratio interval analyzed, a
quantitative comparison of results with these studies is not possible.

 From the comparison between Table~2 and~3 
we note that in some cases the fraction of binaries
measured in the same cluster region by different authors strongly
differs from the results presented here.
As an example, in the case of NGC 6752, Rubenstein \& Bailyn 
  (1997) estimated an high fraction of binaries in the core ($f_{\rm  bin}$=0.27$\pm$0.12), 
in disagreement with the results presented in
this paper ($f_{\rm  bin}^{\rm TOT}$=0.03$\pm$0.03) and in  Milone et
al.\ (2010) ($f_{\rm  bin}^{\rm q>0.5}$=0.03$\pm$0.01).  
 To investigate these different results, Milone et al.\ (2010)
  re-examined  the Rubenstein \& Bailyn (1997) findings first by
  analyzing the same data with the improved photometric techniques that are now
available, and then using the better datasets that have been collected
more recently. 
They concluded that the disagreement comes from the use of
the stellar photometry tools they used, which allow a
better separation of stellar blends. 
Similarly, the large fraction of binaries detected by Albrow et al.\ (2001) in NGC 104,
and Fisher et al.\ (1995) are not confirmed by our study.

Sollima et al.\ (2007) have recently measured the fraction of binaries in the core of 13 low-density GCs
by using the same  images 
studied in this paper. First, they analyzed the color distribution of MS stars to
directly derive the minimum fraction of binary systems required to reproduce the observed CMD morphologies, then they inferred two different estimates
of the total fraction of binaries by assuming the mass-ratio distribution obtained from random extractions from a de Marchi et al.\ (2005) IMF, and from the
distribution measured by Fisher et al.\ (2005). 

\begin{figure}[ht!]
   \centering
   \includegraphics[width=8.5cm]{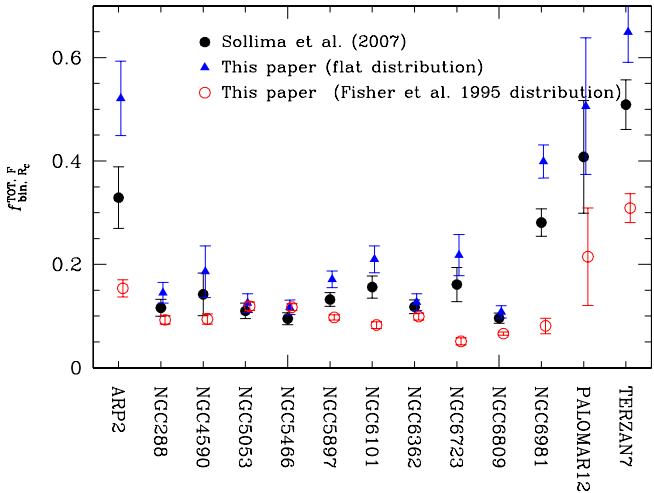}
      \caption{Comparison of the core binary fractions of 13 GCs measured in this paper (red circles)
	and in Sollima et al.\ (2007, black circles). In both cases
        has been assumed the mass-ratio distribution from Fisher et al.\ (2005). 
 Blue triangles indicate the binary fraction estimated in this
 work assuming a flat {\it q} distribution.}
         \label{sollima}
   \end{figure}

Even if we have shown that the Fisher et al.\ (2005)
distribution is not consistent with what found in the present work and
because the images are the same as in
this paper, for a meaningful comparison with Sollima et al.\ (2007), in 
Fig.~\ref{sollima} we compare the total fraction of binaries in the core that we obtained by  assuming the Fisher et al.\ (2005) distribution (red circles) with
 the values from Sollima et al.\ (2007). 
 Blue triangles correspond to the binary fraction estimated in this
paper assuming a flat {\it q} distribution.
For eight out of thirteen GCs, results are in agreement, at the level
of less than three $\sigma$. In the cases of ARP 2, NGC 6101, NGC
6723, NGC 6981, and Terzan 7  the fraction of binaries measured in
this work is systematically smaller than those found by Sollima and
collaborators. 

%
\subsection{The binary fraction as a function of primary-star mass}
\label{magdistr}
\begin{figure*}[ht!]
   \centering
   \includegraphics[width=13 cm]{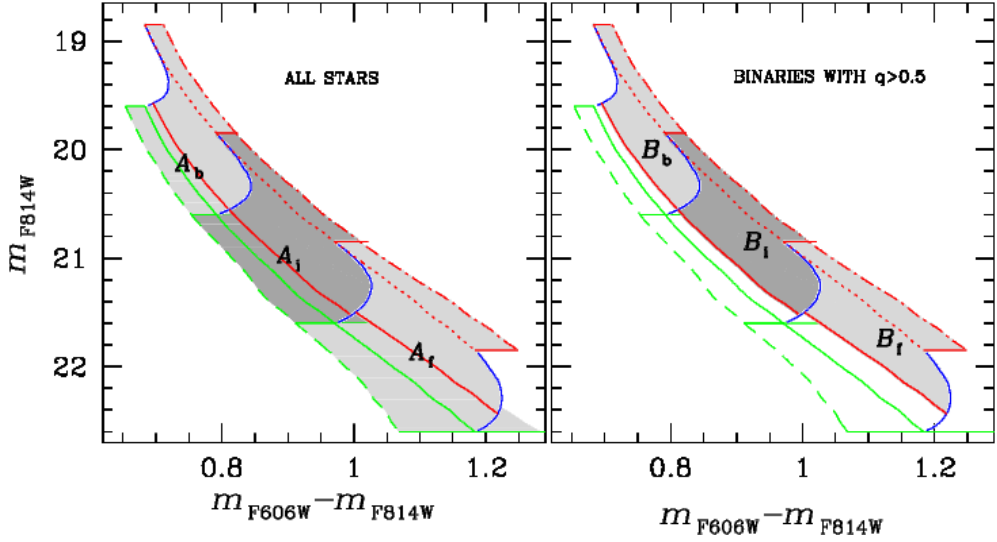}
      \caption{Dark and light gray areas indicate the CMD regions used
        to measure the fraction of binaries in three magnitude
        intervals.}
         \label{3reg}
   \end{figure*}

In this Section we investigate the distribution of binary
systems as a function of the magnitude. To do this, we calculated the
fraction of binaries  over the entire WFC/ACS field of view in
the three magnitude intervals,  containing all the single MS stars
  and the binary systems with a primary star:
[0.75,1.75],  [1.75,2.75],  [2.75,3.75], F814W magnitudes below
the MSTO respectively. 
In the cases of NGC 6388 and NGC 6441 we used smaller magnitudes
intervals of [0.75,1.25],  [1.25,1.75],  [1.75,2.25] F814W magnitudes below
the MSTO.
We divided the CMD regions {\it A} and {\it B} defined in Sect.~\ref{sec:highq}
and illustrated in Fig.~\ref{reg1}
into three subregions (named ${\it A}_{\rm b}$, ${\it A}_{\rm i}$,
${\it A}_{\rm f}$
and ${\it B}_{\rm b}$, ${\it B}_{\rm i}$, ${\it B}_{\rm f}$) as shown in Fig.~\ref{3reg}
and calculated the fraction of binaries in each magnitude interval 
 see eq.~\ref{eq:1}.

\begin{figure*}[ht!]
   \centering
   \includegraphics[width=13 cm]{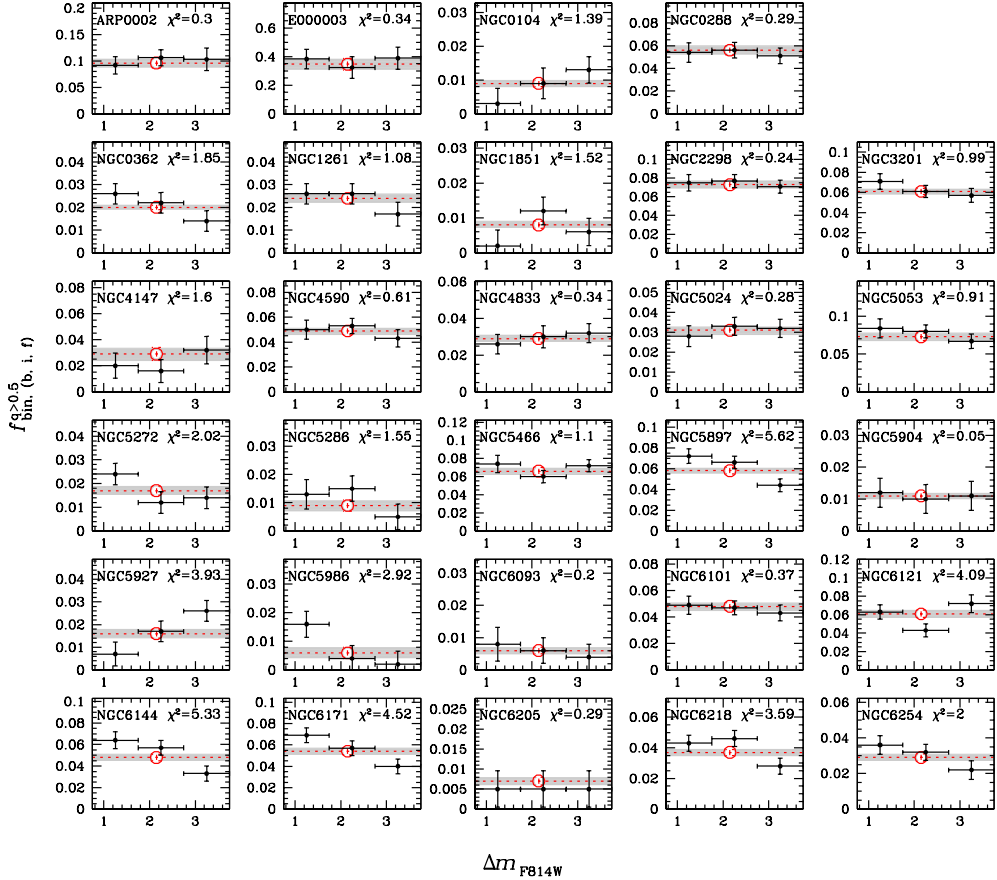}
      \caption{Fraction of binaries with mass ratio {\it q}$>$0.5  for 29 GCs
measured in three magnitude intervals (black points) and  in the
interval between 0.75 and 3.75
F814W  magnitudes below the MS turn off
(red points) as a function of $\Delta {\it m}_{\rm F814W}$. Horizontal
       segments indicate the magnitude coverage
      corresponding to each point.} 
         \label{MD1}
   \end{figure*}

\begin{figure*}[ht!]
   \centering
   \includegraphics[width=13 cm]{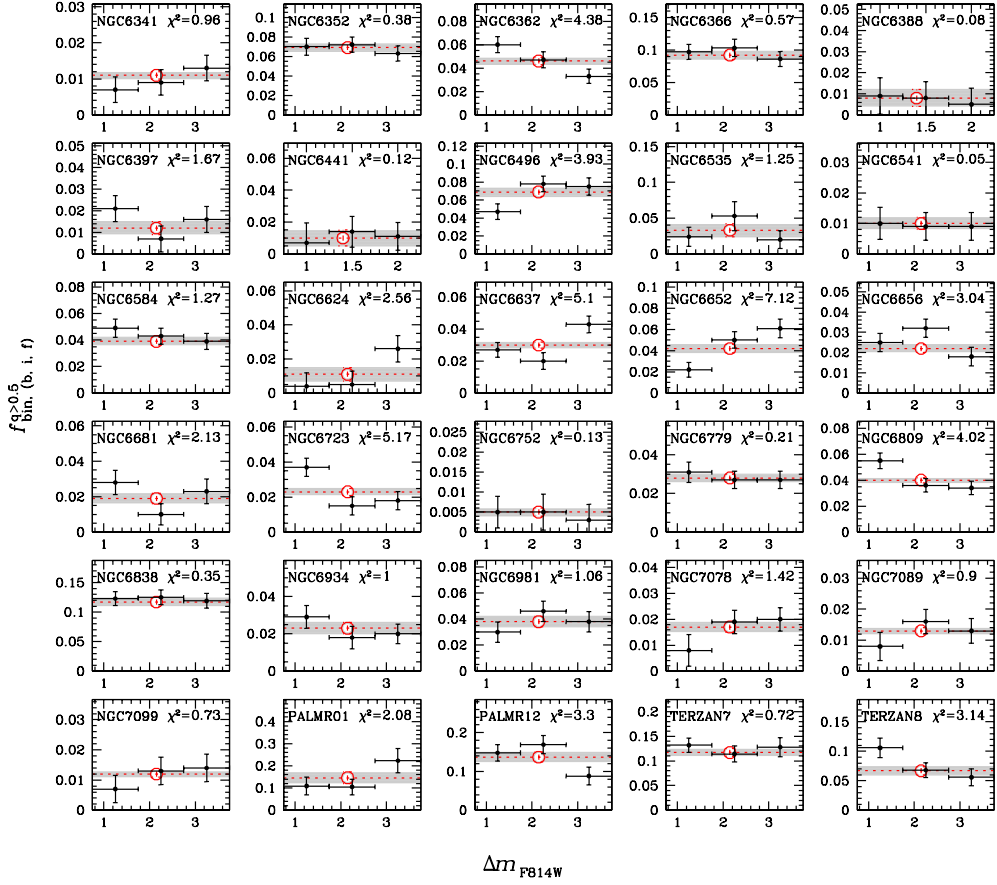}
      \caption{As in Fig.~\ref{MD1} for the remaining 30 GCs.
In the cases of NGC 6388 and NGC 6441 the binary fraction
 has been measured in the interval between 0.75 and 2.25  F814W
magnitudes below the MS turn off.
}
         \label{MD2}
   \end{figure*}

Results are shown in Figs.~\ref{MD1} and ~\ref{MD2} where we plot the
fractions of binaries with mass ratio {\it q}$>$0.5 calculated in
three magnitude bins as a function of the difference between the mean
F814W magnitude of the bin and the   F814W magnitude of the MS turn
off ($\Delta {\it m}_{\rm F814W}$). Red points indicate the fraction
of binaries in the full interval [0.75:3.75]
([0.75:2.15] for NGC 6388 and NGC 6441), while the shadowed area
indicates the error associated to this measure.

In general we find no evidence for a significant trend in the fraction of
binaries with magnitude (which is a proxy for primary mass), as
suggested by the reduced-$\chi^{2}$ values quoted in
Figs.~\ref{MD1} and~\ref{MD2}.
Montecarlo simulations show that in the case of a flat distribution
the 50\% and 99\% of objects have  $\chi^{2}$ values smaller
than 1.1 and 5.5 respectively.
Possible exceptions to this rule of a flat trend are represented by NGC 5897
 and NGC 6652 for which we have estimated $\chi^{2}$ values
 higher than 5.5. And large $\chi^{2}>5.0$ are obtained
 also for NGC 6144, NGC 6637, and NGC 6723.

\begin{figure}[ht!]
   \centering
   \includegraphics[width=8.5 cm]{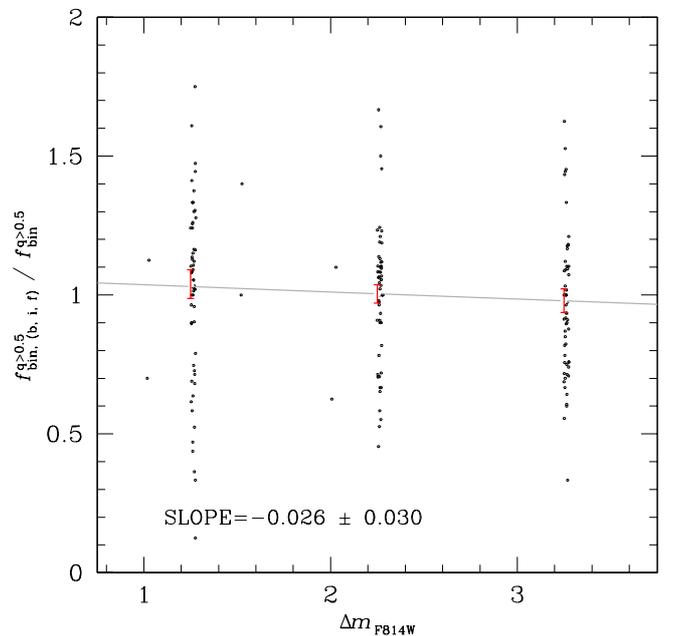}
      \caption{
Fraction of binaries with {\it q}$>$0.5 measured in three magnitude
intervals (black points) as a function of $\Delta {\it m}_{\rm F814W}$
for the 59 GCs studied in this work. 
To compare the fraction of binaries in different clusters we have
divided the fraction of binaries in each bin by the
value of ${\it f}_{\rm bin}^{\rm q>0.5}$ measured in the
interval between 0.75 and 3.75
F814W  magnitudes below the MS turn off.
For clarity black points have
been randomly scattered around the corresponding $\Delta {\it m}_{\rm
  F814W}$ value. Red points with error bars are the average binary
fractions in each interval while the gray line is the best fitting
least-square line whose slope is quoted in the inset.  
}
         \label{MDALL}
   \end{figure}

In order to further analyze the general trend of the binary fraction
with the  magnitude  for all the GCs studied in this paper we divided
the values of ${\it f}_{\rm bin}^{\rm q>0.5}$ measured in each magnitude bin by the fraction of
 binaries with {\it q}$>$0.5 in the interval between 0.75 and 3.75
 magnitudes below the MS turn off.  Results are shown in
 Fig.~\ref{MDALL} and confirm that
 the fraction of binaries is nearly flat in the analyzed magnitude range.

\begin{figure}[ht!]
   \centering
   \includegraphics[width=8.5 cm]{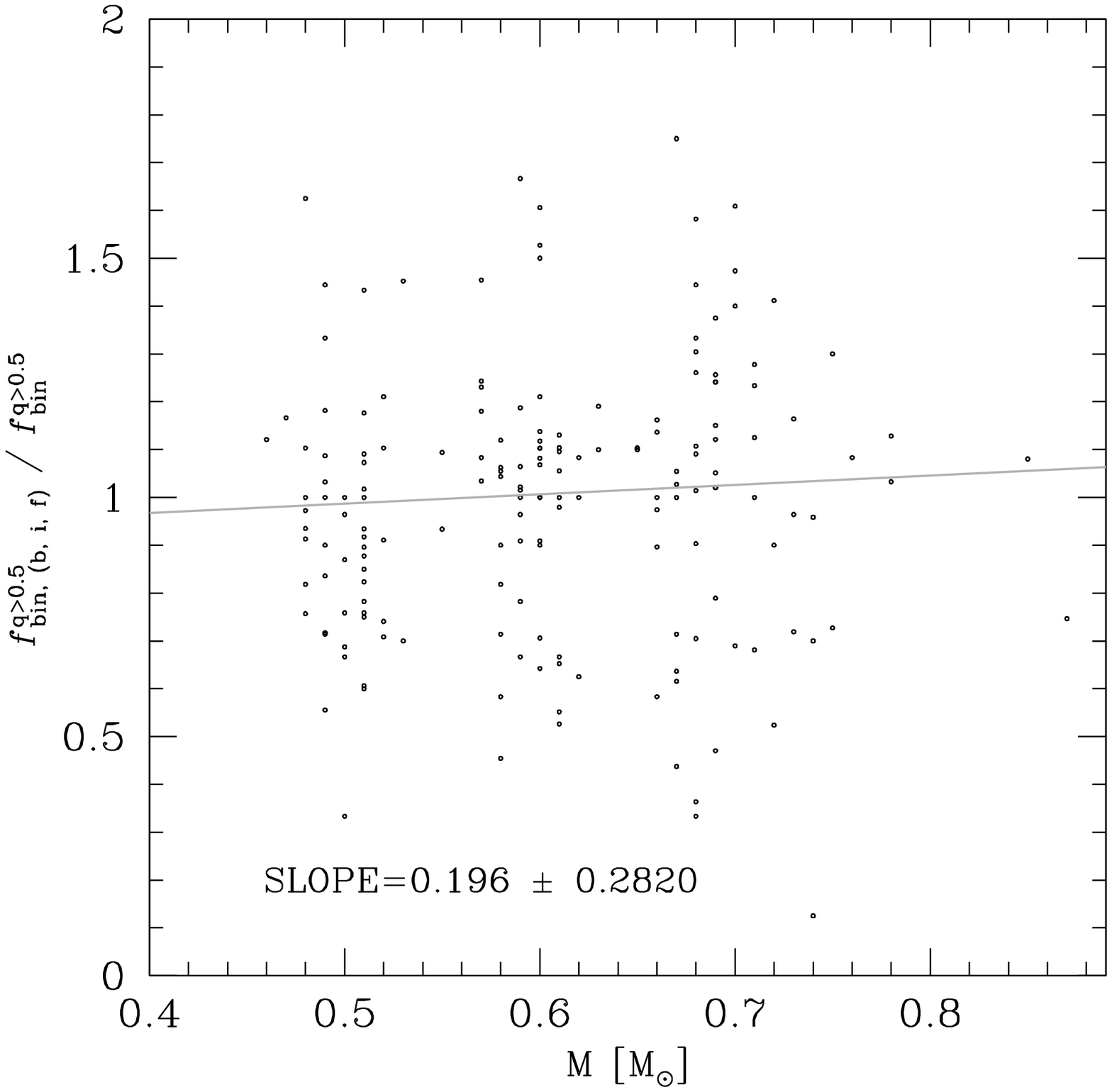}
      \caption{
 Fraction of binaries with {\it q}$>$0.5 measured in three magnitude
intervals and normalized by ${\it f}_{\rm bin}^{\rm q>0.5}$ (black
points) as a function of the mass of the primary star
for the 59 GCs studied in this work. 
}
         \label{MMDALL}
   \end{figure}

Finally,  we used isochrones to estimate the average mass of the
  single stars and the primary component of binary systems in the regions ${\it
  A}_{\rm b}$, ${\it A}_{\rm i}$, and ${\it A}_{\rm f}$.
 To do this we converted the  mean F814W magnitudes of 
  the single stars contained in each of these regions
 into
masses through the Dotter et al.\ (2007)   mass-luminosity relations. 
 Figure~\ref{MMDALL} shows the ratio ${\it f}_{\rm bin, b, i, f}^{\rm
    q>0.5}/{\it f}_{\rm bin}^{\rm q>0.5}$ as a function of
the  average mass estimated above  and suggests that the binary fraction is
nearly flat in the analyzed mass interval.   

We recall here, that the results presented in this subsection come
from the analysis of the binary fractions measured over the entire
ACS/WFC field of view. 
 Due to the relatively small numbers of binaries,
  we did not extended this analysis to each group of ${\it r}_{\rm
    C}$, the ${\it r}_{\rm C-HM}$, and the ${\it r}_{\rm oHM}$ stars.


\subsection{The radial distribution}
\label{sec:RD}

In order to investigate how the fraction of high-mass-ratio binaries
depends on the radial distance, we divided the  ACS field of   view into four concentric
annuli,  and calculated the  fraction  of  binaries by
 following the recipes described in Sects.~\ref{sec:highq} and
\ref{totfraction}. We chose the size
of the annulus  such that the number   of stars that populate the  CMD
region {\it A} is equal in each of them.

\begin{figure*}[ht!]
   \centering
   \includegraphics[width=15cm]{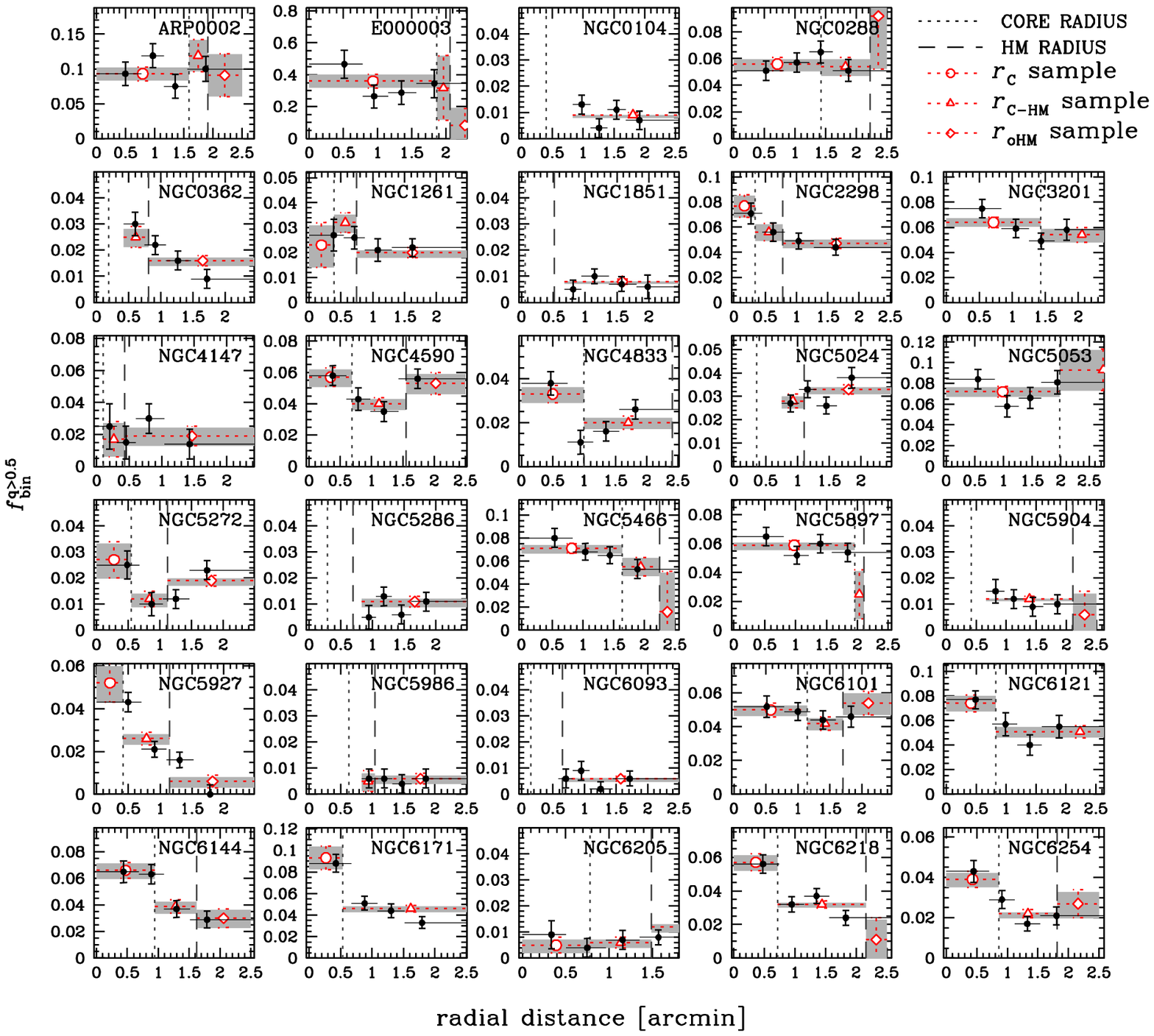}
      \caption{Fraction of binaries with {\it q}$>$0.5
       as a function of the radial distance from the cluster center
	for 29 GCs. The dotted and dashed vertical lines mark
	the core and the half mass radius respectively.
        Black filled-circles show binary fractions in four radial
        intervals while red open-symbols indicate the binary fraction
        for the $r_{\rm C}$, $r_{\rm C-HM}$, and $r_{\rm oHM}$ sample.
	Horizontal black  segments
	indicate the radial coverage corresponding to each point while
         observational errors are plotted as vertical lines and shadowed areas.
       }
         \label{RD1}
   \end{figure*}

\begin{figure*}[ht!]
   \centering
   \includegraphics[width=15cm]{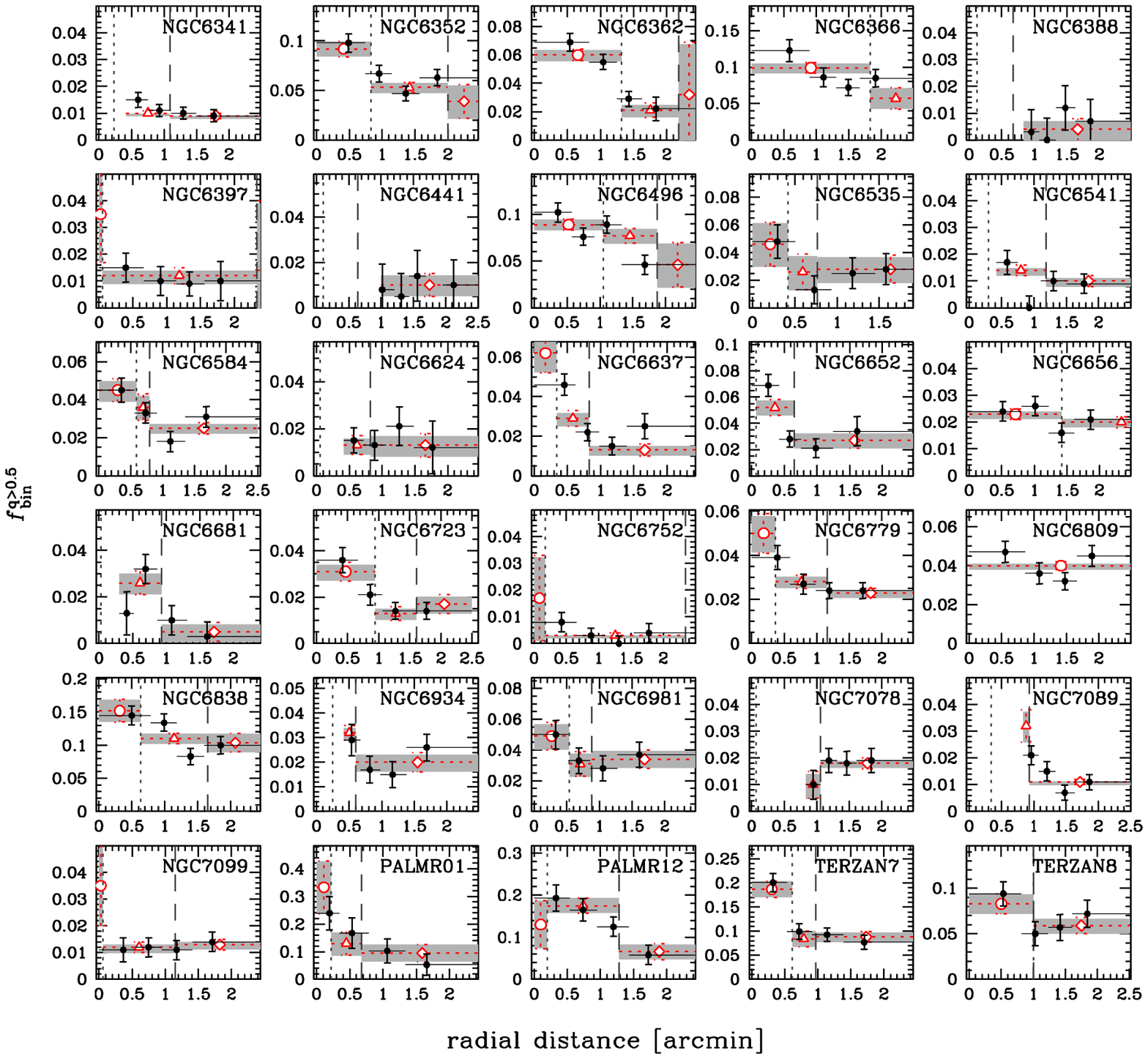}
      \caption{ As in Fig.~\ref{RD1}
	for the remaining  GCs.
       }
         \label{RD2}
   \end{figure*}

Results are shown in Figs.~\ref{RD1} and \ref{RD2} where we plotted
 ${\it f}_{\rm bin}^{\rm  q>0.5}$ as a function of the explored radial distance
for all the GCs studied in this paper. 
and confirm that, in most of the GCs
where the fraction of binaries has been calculated both in the core
and in the outer regions,
 binaries are significantly more centrally concentrated than single MS stars. 

   \begin{figure}[ht!]
   \centering
   \includegraphics[width=8.5cm]{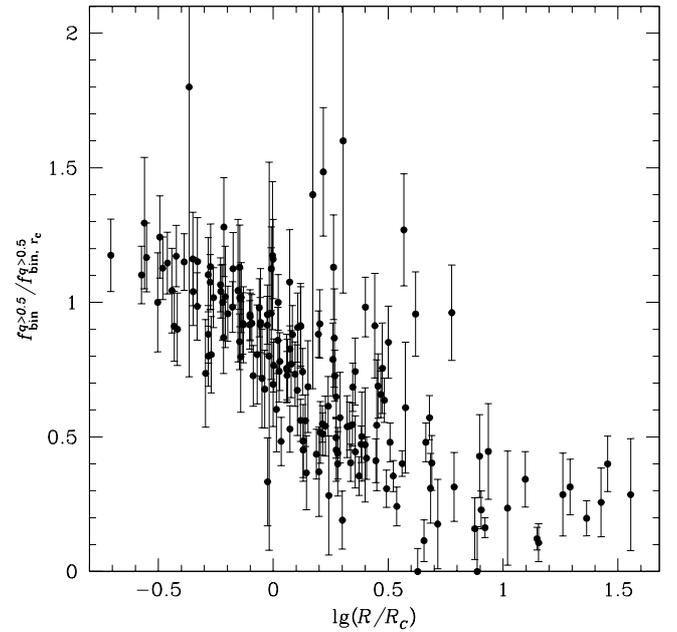}
      \caption{
Binary fraction with {\it q}$>$0.5 (in units of core binaries) as a function of the distance from the cluster center in units of core radii }
	   \label{Rad3}
   \end{figure}
%
In Fig.~\ref{Rad3}, for
 the 43 GCs studied in this paper  for which we
measured the core binary fraction,
we plot 
the fraction of binaries with {\it q}$>$0.5 in units of core-binary fraction as a function
of the radial distance in units of core radii 
While it would be naive to assume a similar radial distribution of binaries
for all Galactic GCs, it seems clear that the fraction of binaries
normalized to the core binary fraction is correlated with radius,
and that the binary fraction typically decreases by a factor of
$\sim$2 at two core radii with respect to the core binary
fraction. The latter  behavior was also suggested by Sollima et
al.\ (2007) on the basis of their analysis of  the radial distribution
of binaries in seven GCs. In the cluster envelope, the binary-fraction
trend with radius tends to flatten. 

\subsection {Correlation between the binary fraction and the parent-cluster parameters}
\label{monovariate}
In this  section  we  investigate  whether the  binary  fraction  is
correlated with any of  the  physical and morphological parameters  of
their host GCs.
In particular, our analysis makes use of the following quantities:\\
{\bf Relative ages.} We used the most recent age measures by
Mar\'\i n-Franch et al.\ (2009). Relative ages were obtained from the
same photometric database used in this paper
by comparing the relative position of the clusters' MS turnoffs,
using MS fitting to cross-compare clusters within the sample.
Typical errors on the 
relative age measurements are between 2 \% and 7 \%. 
 We also used absolute ages from Salaris \& Weiss (2002) and De Angeli et
al.\ (2005). Absolute ages are not available for 15 GCs, namely: E3, NGC 4147, NGC
4833, NGC 5024, NGC 5286, NGC 5927, NGC 5986, NGC 6144, NGC 6388, NGC
6441, NGC 6496, NGC 6541, NGC 7089, PAL 1, and TERZAN 8. \\
{\bf Metallicity.}
 We performed our analysis with both the metallicity scales defined by
 Zinn \& West  (1984) and Carretta \& Gratton \ (1997), which were also used by
Mar\'\i n-Franch et al.\ (2009) to determine relative ages.\\
{\bf BSS Frequency.} We used the counts of BSS
derived by Moretti et al.\ (2008) from the WFPC2 photometric catalogs
published by Piotto et al.\ (2002). In particular, we used
the normalized number of BSS, which is the absolute number of
BSS in a given region divided by the total luminosity coming
from the stars in the same region (in unit of 10$^{4} L_{\odot}$).\\
{\bf Rate of stellar collisions per year.} 

King, Surdin \& Rastorguev (2002) have shown that the rate of stellar
collisions per cluster and per year is $\Gamma_{\rm C}=5 \times
10^{-15} (\Sigma_{0}^{3} r_{\rm C})^{1/2}$, where $\Sigma_{0}$ is the
central surface brightness in units of $L_{\odot, V} pc^{-2}$ and
$r_{\rm C}$ is the core radius in units of parsecs. 
We calculated the probability ($\Gamma_{\rm  *}$) that a given star
will have a collision in 1 yr, by dividing the collision rate by
the total number of stars in the cluster. This is calculated by assuming a
mass-luminosity ratio of 2 and a mean mass for colliding stars of 0.4
${\it M}_{\odot}$.

 We also compared the measured fraction of binaries with the encounter
frequency adopted by Pooley \& Hut (2006) in the 
 approximation used for virialized systems:
$\rho_{0}^{1.5} r_{\rm C}^{2}$  where $r_{\rm C}$ is the core radius
 and $\rho_{0}$ the central stellar density. \\

The other parameters involved in this analysis are the absolute visual
magnitude ${\it M}_{\rm V}$, the ellipticity ($e$),
the central concentration ($c$), the core relaxation timescale, $\tau_{\rm c}$,
the half-mass relaxation timescale $\tau_{\rm hm}$,
 and the  logarithm of the central luminosity density $\rho_{0}$,
and are taken from the Harris (1996) compilation.
We also used three different parameters related to the cluster HB morphologies,
as discussed in Sect.~\ref{HB}. 
 Ellipticity  ($e$) 
measurements are not available for six clusters, namely
  ARP2, E3, NGC 288, PALOMAR 12, Terzan 7 and Terzan 8. 

\begin{figure*}[ht!]
   \centering
   \includegraphics[width=12.0cm]{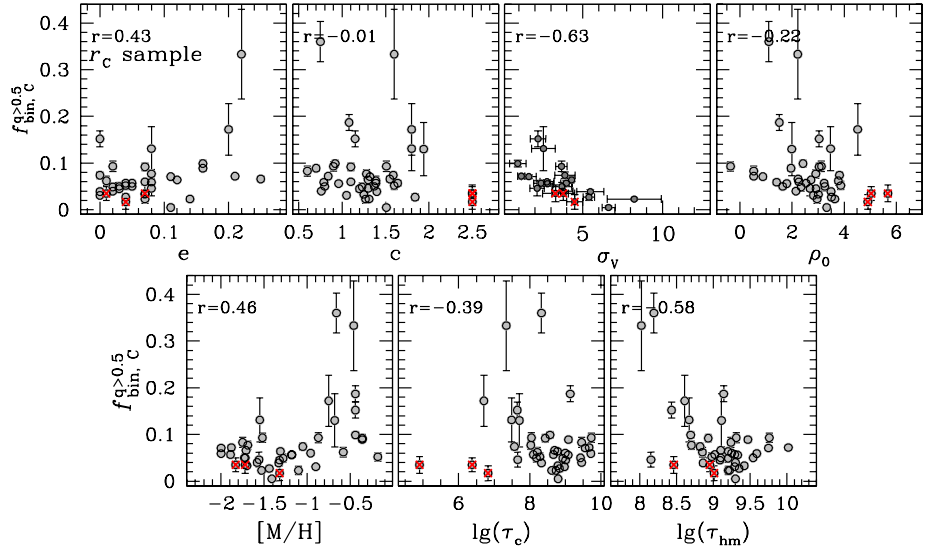}
      \caption{Fraction of binaries with $q> 0.5$ in the core as a function
	of some parameters of their host GCs. Clockwise: ellipticity, central
        concentration, central velocity dispersion,  logarithm of
          the central luminosity
        density, half-mass and core relaxation timescale, and
        metallicity.
         In each panel we quoted the Pearson correlation coefficient (r).
        PCC clusters are marked with red crosses  and are not used
        to calculate r (see text for details).}
         \label{PAR1}
   \end{figure*}

\begin{figure*}[ht!]
   \centering
   \includegraphics[width=12.0cm]{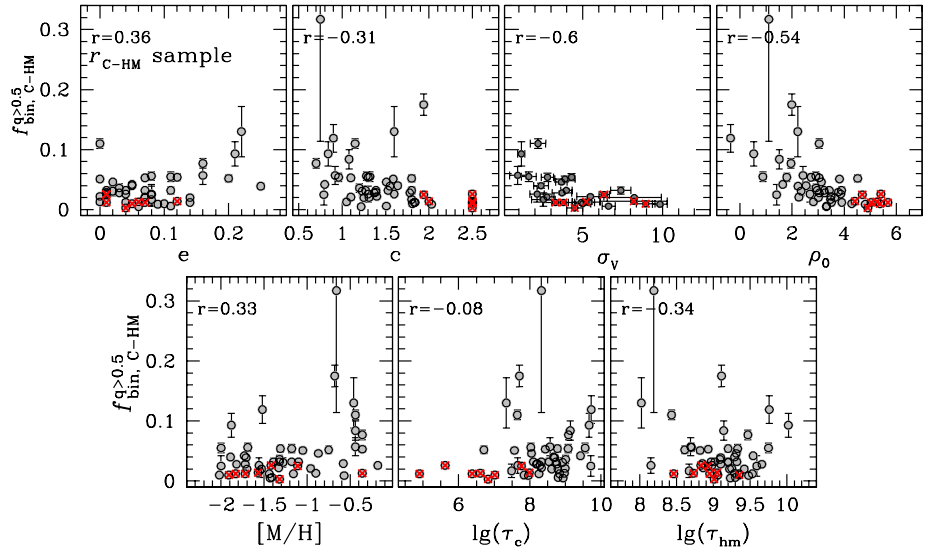}
      \caption{As in Fig.~\ref{PAR1} for the ${\it r}_{\rm C-HM}$ sample.}
         \label{PAR1HM}
   \end{figure*}

\begin{figure*}[ht!]
   \centering
   \includegraphics[width=12.0cm]{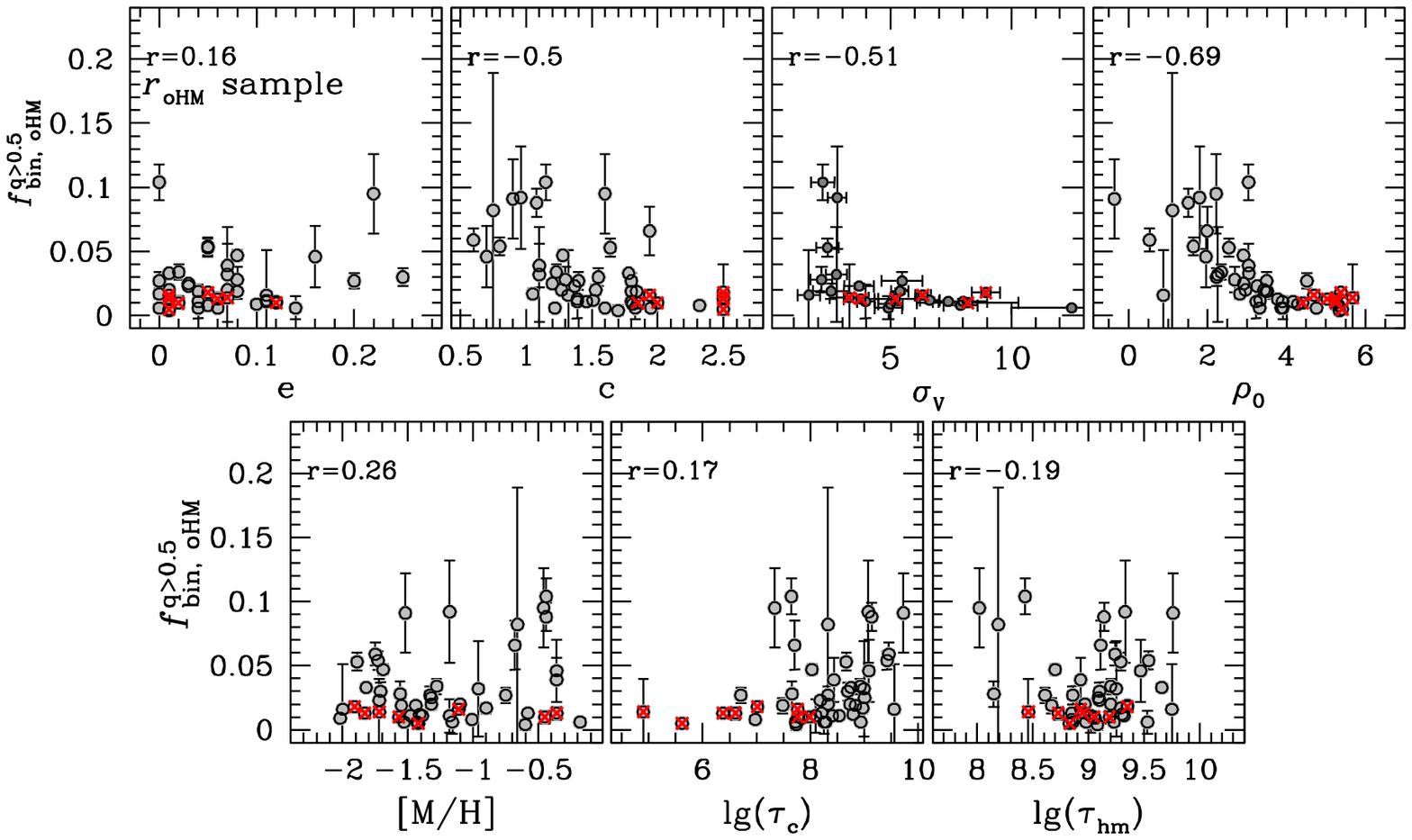}
      \caption{As in Fig.~\ref{PAR1} for the ${\it r}_{\rm oHM}$ sample.}
         \label{PAR1TI}
   \end{figure*}

Figures~\ref{PAR1}--\ref{PARHB} show the monovariate correlations.
 Note that, in our study of the core population of binaries, we did
not include the post-core-collapse (PCC) GCs, because, for
these objects, the definition of core radius is not reliable (Trager
et al.\ 1993). Specifically, PCC clusters are marked with red crosses in these
figures but are not used to study the statistical significance of the
correlations. 
Figures~\ref{PAR1}--\ref{PARHB} show
that there are no significant correlations between
the binary fractions and the cluster ellipticity, core 
and half mass relaxation time, central concentration and metallicity
 as suggested by the small values of the Pearson correlation coefficient.
Some marginal correlation with the central density can not be excluded.

In the following we will discuss some of the relevant relations between
the cluster parameters listed above
and the fraction of binaries calculated in three radial
regions defined in Sect.~\ref{sec:highq}.
A noteworthy correlation of the   binary fraction is with the
central  velocity dispersion (r$\sim$-0.6).
as shown in Fig.~\ref{PAR1},~\ref{PAR1HM}, and ~\ref{PAR1TI}.
 The central velocity dispersions $\sigma_{\rm V}$  come from  Meylan (1989),
 and are available only for a subsample of the GCs that are studied in
 the present work. 

%

\subsubsection{${\it f}_{\rm bin}$ versus ${\it M}_{\rm V}$,
 $\Gamma_{*}$, and BSS frequency}
   \begin{figure*}[ht!]
   \centering
   \includegraphics[width=12.0cm]{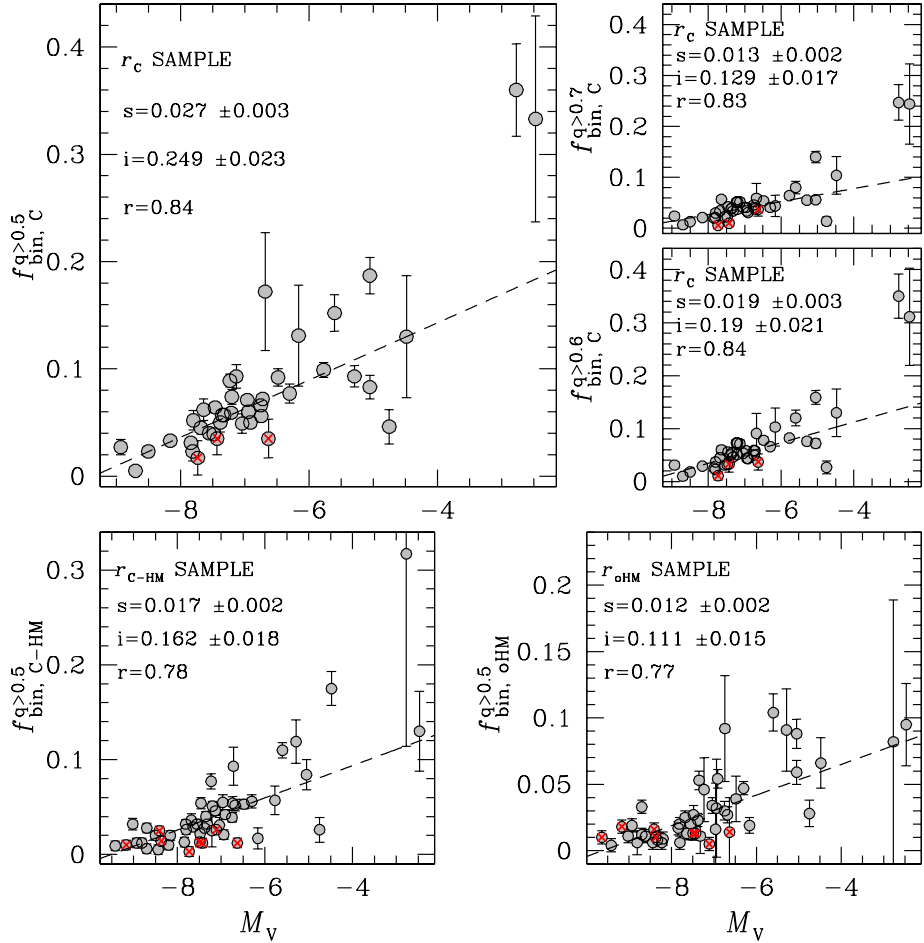}
      \caption{ \textit{Upper-left}: Fraction of binaries with ${\it q}>
        0.5$ in the core as a function of the absolute visual
        magnitude of the host GC. Dashed line is the best fitting
        straight line whose slope (s) and intercept (i) are quoted in the figure 
           together with the Pearson correlation coefficient (r). PCC
           clusters are marked with red crosses and are not used to
           calculate neither the best-fitting line nor r.
         For completeness in the \textit{upper-right}
        panels we show the same plot for the fraction of binaries with
        ${\it q}> 0.6$, and  ${\it q}> 0.7$.  \textit{Lower panels}: Fraction
        of binaries with ${\it q}> 0.5$ in the $r_{\rm C-HM}$ (\textit{
          left}) and $r_{\rm oHM}$ (\textit{right}) sample as a function
        of $M_{\rm V}$.
       }
         \label{MV}
   \end{figure*}
%

The most  significant correlation we found is the correlation  between the
cluster binary fraction and its absolute magnitude.  Clusters
with  fainter absolute luminosity  (smaller mass)
have  higher binary fractions. This correlation is present in all
binary groups, i.\ e.\ for binaries  inside the cluster core,
 for binaries located between  the core and the
half-mass  radius,  for  binaries outside the half-mass
radius, and for binaries with {\it q}$>$0.5 (Fig.~\ref{MV}).
 An anticorrelation between the fraction of binaries and the mass of
  the host GCs is predicted by theoretical models 
(Sollima 2008, see also Fregeau et al.\ 2009).
These authors suggest that this correlation could be the due to the fact that
  cluster mass and the efficiency of binary destruction have the same
  dependence on the cluster density and velocity dispersion.

 This  anticorrelation might extend to open cluster masses.
In fact, Sollima et   al.\ (2010) 
found a dependence of the fraction
  of binaries and the cluster mass in a sample of five open clusters. 
Sollima et al.\ (2010) suggests  that the binary
  disruption within the cluster core is the dominant process that
  determine the fraction of binaries in star clusters.

\begin{figure}[ht!]
   \centering
   \includegraphics[width=8.5cm]{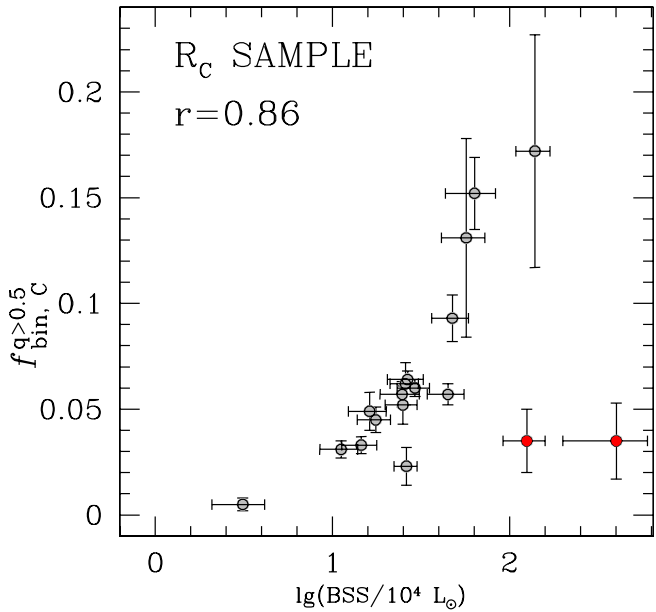}
      \caption{ Fraction of binaries with ${\it q}> 0.5$ as a function of the BSS frequency in the core.
	PCC GCs are marked with red points.
       }
         \label{BINvsBSS}
   \end{figure}

\begin{figure*}[ht!]
   \centering
   \includegraphics[width=12.0cm]{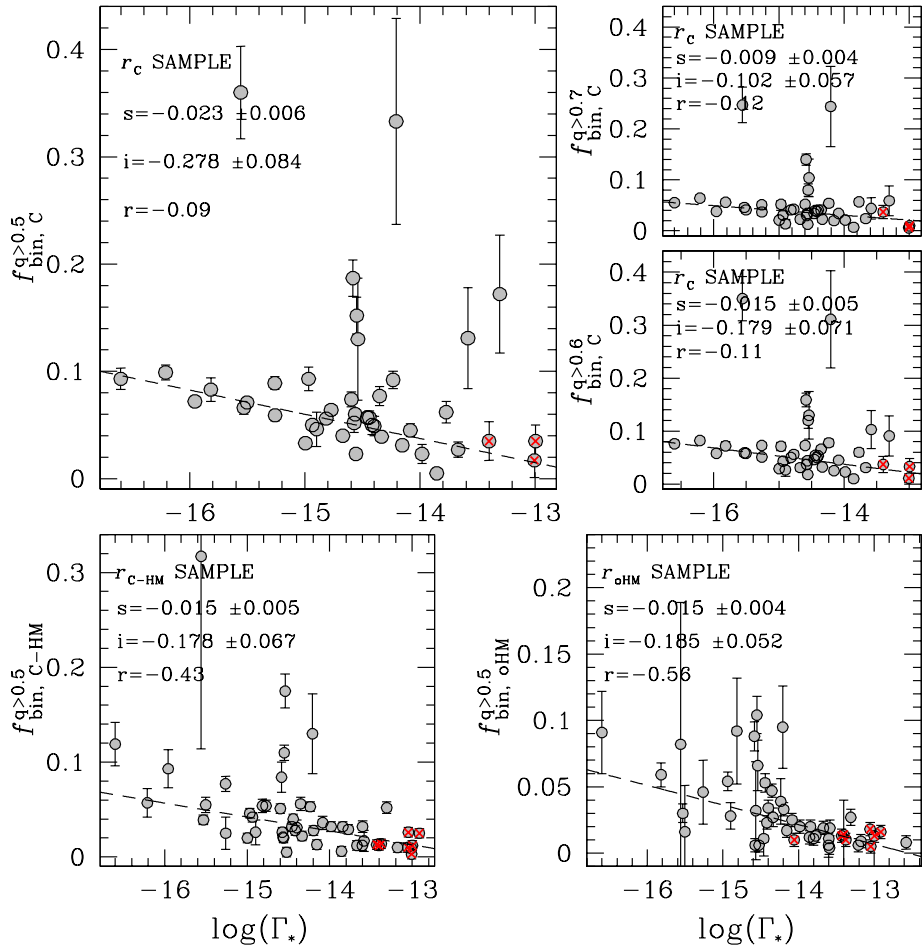}
      \caption{Fraction of binaries with $q> 0.5$,$q>0.6$, and $q>0.7$ in the
        $r_{\rm C}$ region (\textit{upper panels})  and fraction of binaries
        with $q>0.5$ in the {$r_{\rm C-HM}$} and {$r_{\rm oHM}$}
        regions (\textit{bottom panels}) as a function
	of the collisional parameter ($\Gamma_{*}$). The adopted symbols are
        already defined in Fig.~\ref{MV}.}
         \label{gammacol}
   \end{figure*}

\begin{figure*}[ht!]
   \centering
   \includegraphics[width=12.0cm]{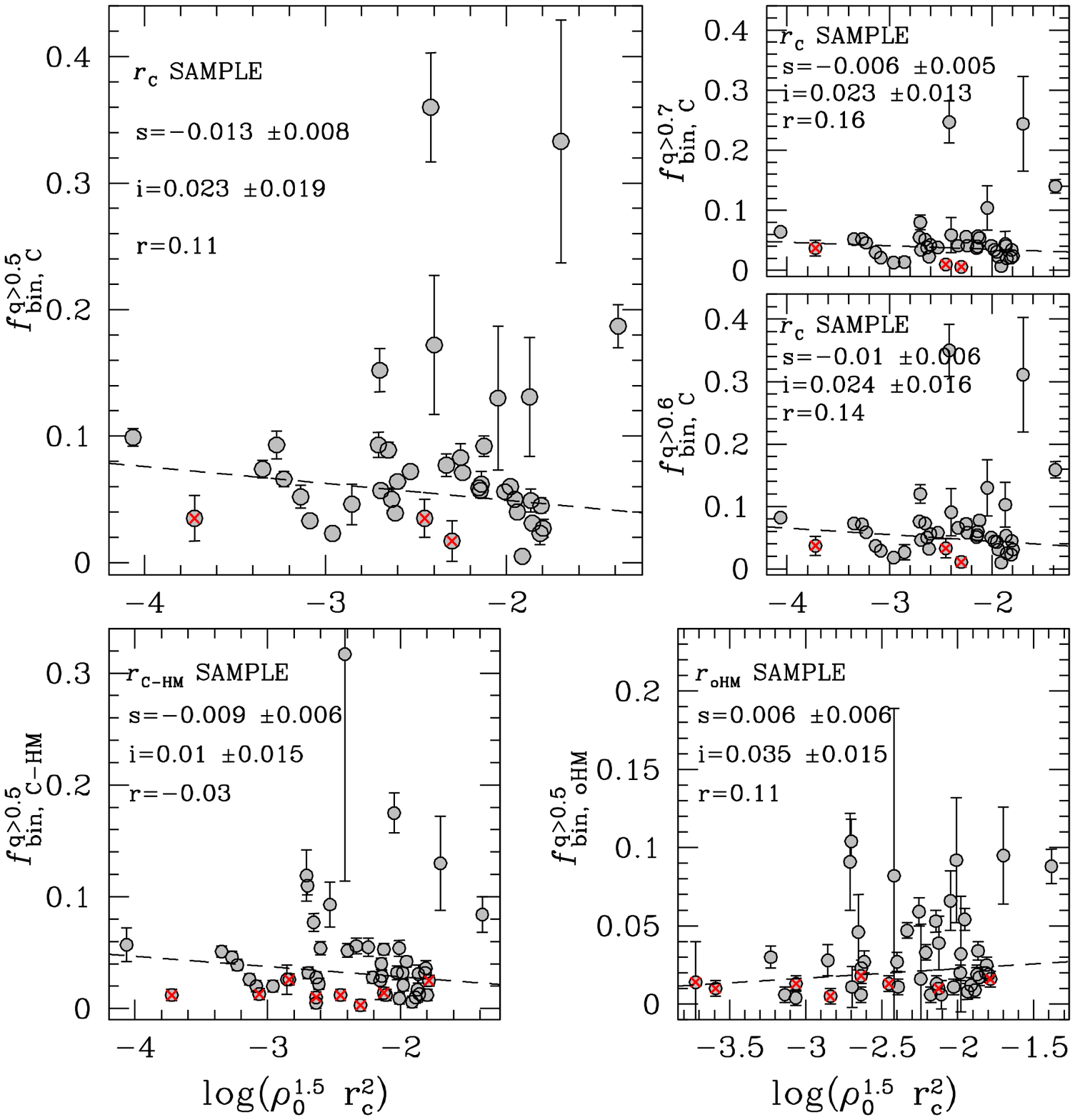}
      \caption{As in Fig.~\ref{gammacol}. In this case we used the encounter
frequency adopted by Pooley \& Hut (2006) in the 
 approximation used for virialized systems.}
         \label{gammacolN}
   \end{figure*}

Noteworthy, a similar anticorrelation  between the frequency of BSSs and the
absolute luminosity of  the parent cluster has been found by Piotto
et  al.\ (2004), Leigh, Sills, \& Knigge (2007), and Moretti et al.\ (2008). 
Interestingly enough, Fig.~\ref{BINvsBSS} shows that the fraction of
binaries is indeed correlated with the fraction of BSSs.
Sollima et al.\ (2008)  
  observed a similar correlation between the BSS specific frequency
  and the fraction of binaries in the core of 13 low-density Galactic GCs. These
  authors suggested that the evolution of primordial binaries could be
the dominant BSS formation process (see also Knigge et al.\ 2009 and
Leigh, Sills \& Knigge 2011).
However, Davies et al.\ (2004) 
provided a simple model showing that the correlation between the  BSS
frequency and the cluster mass
may be the result of the evolution of  the binary   fraction  due  to
encounters. 
 Here, we can only note that, 
figures.~\ref{gammacol}  and {gammacolN} seem to suggest a mild correlation between
binary fraction and the collisional parameter, while
there is no significant correlation between the BSS frequency and the collisional
parameter (e.\ g.\ Piotto et al.\ 2004, Davies et al.\ 2004, Leigh Sills, \& Knigge 2007, Moretti et al.\ 2008).
It is clear that the connection between binaries and BSSs is far from trivial. 
The interpretation of the correlation of binary fraction with cluster
parameters, and with BSS fraction is beyond the purposes of the present paper.
%
%

%

%
\subsubsection{${\it f}_{\rm bin}$ versus age}
\begin{figure*}[ht!]
   \centering
   \includegraphics[width=12.0cm]{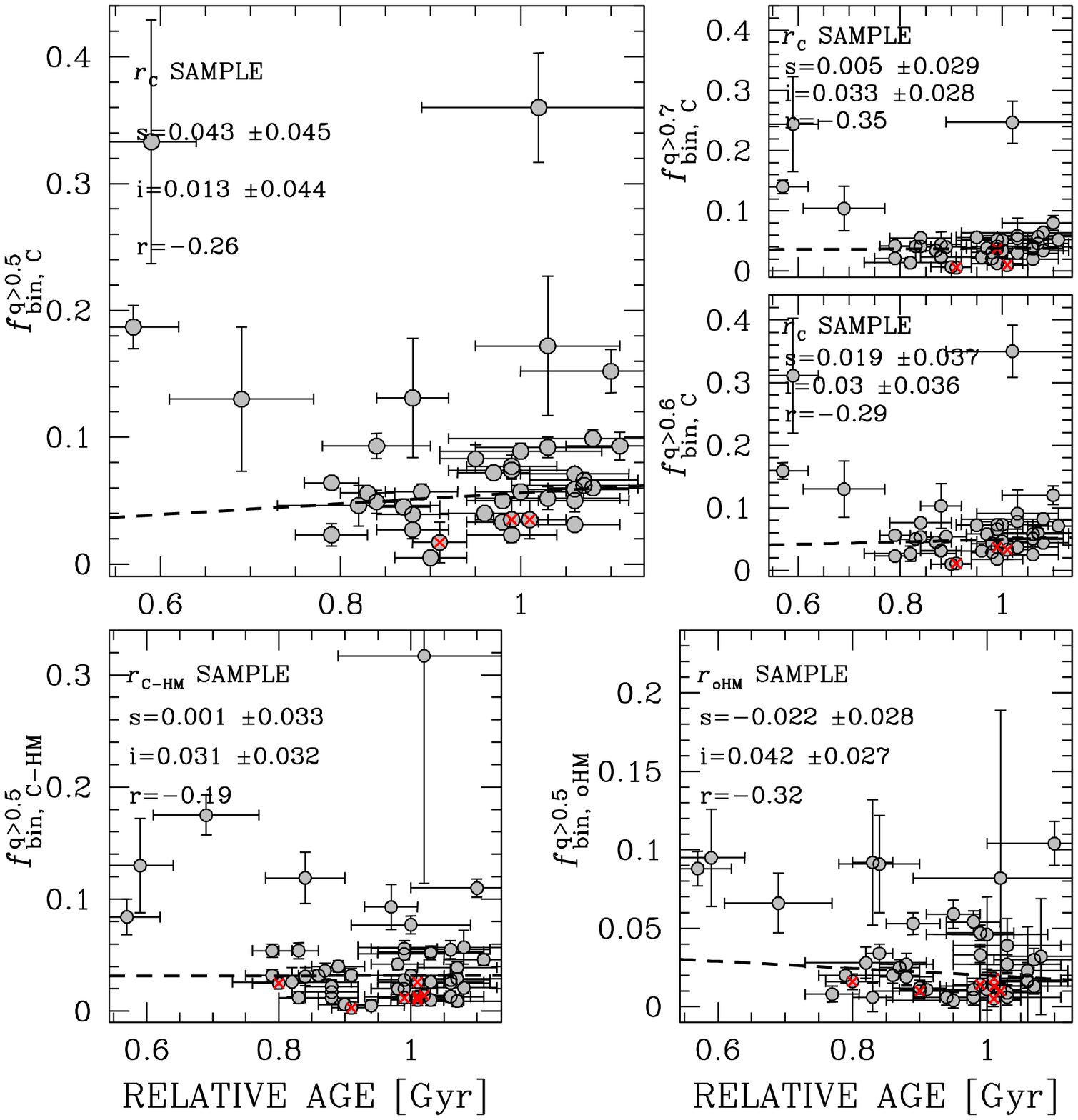}
      \caption{
Fraction of binaries with {\it q}$>$ 0.5, {\it q}$>$0.6, and {\it
  q}$>$0.7 in the 
        $r_{\rm C}$ region (\textit{upper panels})  and fraction of binaries
        with {\it q}$>$0.5 in the {$r_{\rm C-HM}$} and {$r_{\rm oHM}$}
        regions (\textit{bottom panels}) as a function
	of the relative age  measured by Mar\'\i n-Franch  et al.\ (2009). The adopted symbols are
        already defined in Fig.~\ref{MV}.}
         \label{AGE}
   \end{figure*}

\begin{figure*}[ht!]
   \centering
   \includegraphics[width=12.0cm]{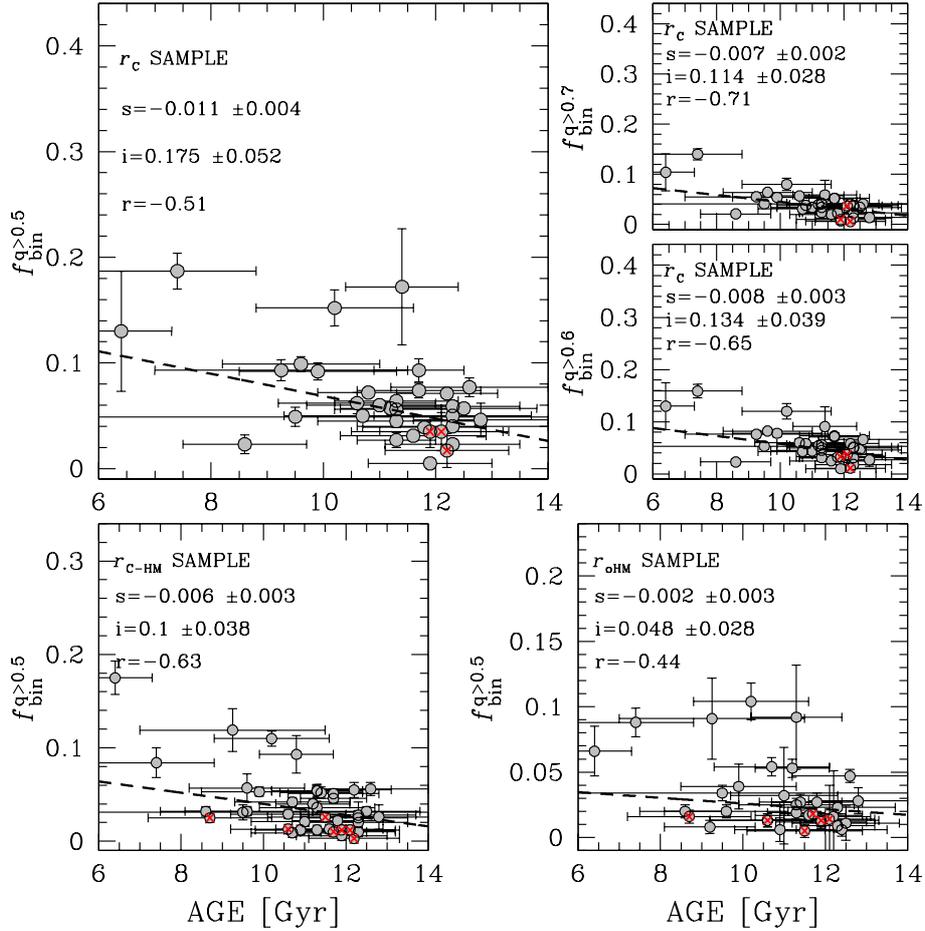}
      \caption{
         As in Fig.~\ref{AGE} but in this case we used the age measures
        from Salaris \& Weiss (2002) and De Angeli et al.\ (2005).}
         \label{AGEDA}
   \end{figure*}

Figures~\ref{AGE}  and ~\ref{AGEDA} plot
 the   fraction of binaries  with ${\it q}>0.5$
 in the core as  a function  of relative
ages by Mar\'\i n-Franch  et al.\ (2009) and the absolute ages from 
  and from Salaris \& Weiss \ (2002) and De Angeli et al.\ (2005)
  respectively. 
There is no evident trend between ages and the binary fraction.

\begin{figure*}[ht!]
   \centering
   \includegraphics[width=6.0cm]{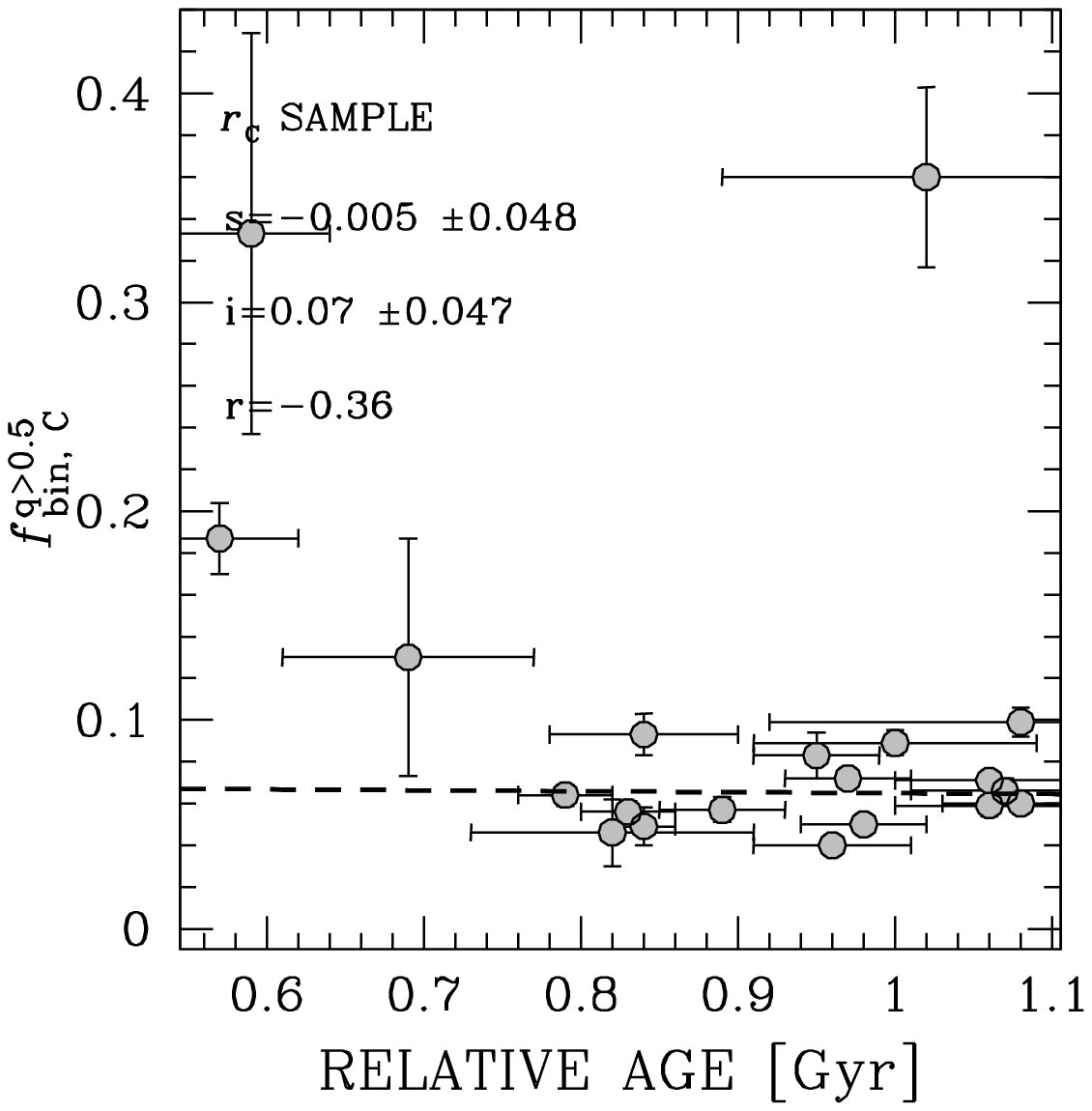}
   \includegraphics[width=6.0cm]{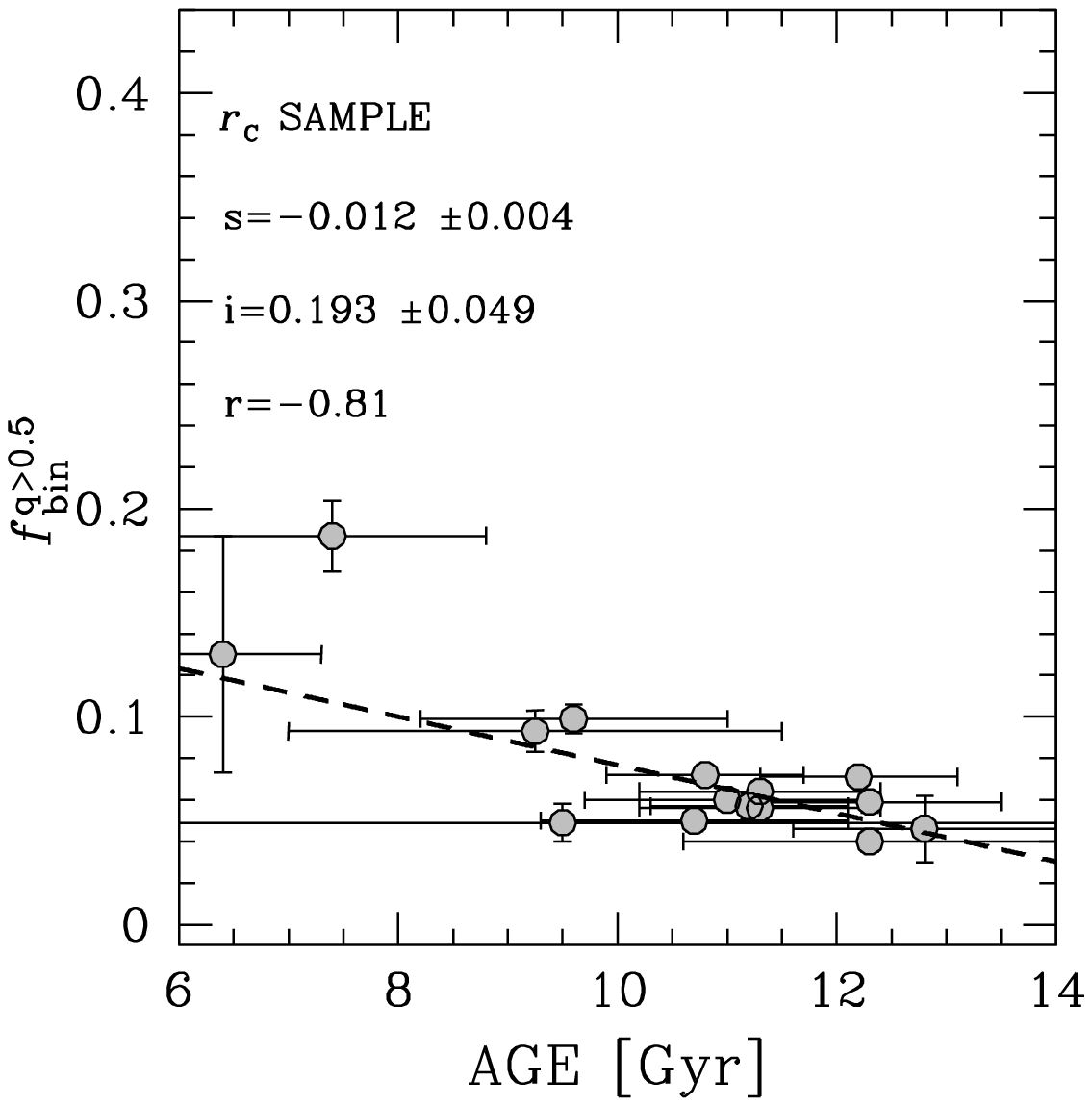}
      \caption{Fraction of binaries with {\it q}$>$ 0.5 in the $r_{\rm
          C}$ sample for low density clusters (log$(\rho_0)<2.75$) as
        a function  of the relative age from Mar\'\i
        n-Franch  et al.\ (2009) (\textit{left panel}) an absolute age
        from Salaris \& Weiss (2002) and De Angeli et al.\ (2005)
        (\textit{right panel}).} 
         \label{AGELD}
   \end{figure*}

Sollima et al.\ (2007) compared the fraction of binaries measured in
the core of thirteen clusters, with the cluster ages from Salaris \&
Weiss (2002) and De Angeli et al.\ (2005) 
and found an   anticorrelation  between age and binary
fraction  suggesting  age as   the  dominant  parameter that
determines the fraction of binaries in a GC.
Our sample of 59 GCs does not confirm such correlation.
 Sollima et al.\ (2007) sample is limited to low density clusters.
In order to verify whether the binary fraction dependence
on age is limited to low density clusters, in Fig.~\ref{AGELD} we 
plot the binary fraction for the $r_{\rm C}$ sample as a function of
the age from De Angeli et al.\ (2005) and the relative age from Marin Franch et
al.\ (2009) for  clusters with central density log$(\rho_0)<2.75$ (same central
density limit of Sollima et al.\ 2007 sample). 
 We also note that the youngest low density clusters in our sample have a
 larger binary fraction, but the fact that at least one old GCs (E3)
 hosts a large binary fraction suggests that more data are needed to
 confirm any systematic trend. 
%
%
%
\subsubsection{HB morphology}
\label{HB}
Binaries have been considered as a possible second parameter of the HB morphology by several authors.
In particular, the existence of a link between field B-type subdwarf (sdB) -- which are the counterpart in
the field of the extremely hot  horizontal branch (EHB) stars in GCs -- and binary systems
is well-established, both on observational and theoretical
grounds. A large population of binaries has been
found among field sdBs (e.\ g.\ Napiwotzki et al.\ 2004 and references therein).
However, the formation scenario of EHB stars in GCs may be different. In fact, several radial-velocity surveys for the measurement of the binary fraction
among EHB stars  have revealed a significant lack of binary systems
(Moni Bidin et al.\ 2006, 2009). 

In order to investigate possible relations between the fraction of binaries and the HB shape we
 used three different parameters: \\
1)  the median color difference between the HB and the RGB [$\Delta(V - I)$], measured by Dotter et al.\ (2010) for 60 GCs using the same CMDs of this paper; \\
2)  The HB morphology index from Mackey \& van den Bergh\ (2005);\\
3) the effective temperature of the hottest HB stars ($T_{\rm eff, HB}$), measured by Recio-Blanco et al.\ (2006).
 $\Delta(V - I)$, HB index, and $T_{\rm eff, HB}$ measures are
  available for 56, 55, and 28 GCs studied in this paper.

\begin{figure*}[ht!]
   \centering
   \includegraphics[width=11.0cm]{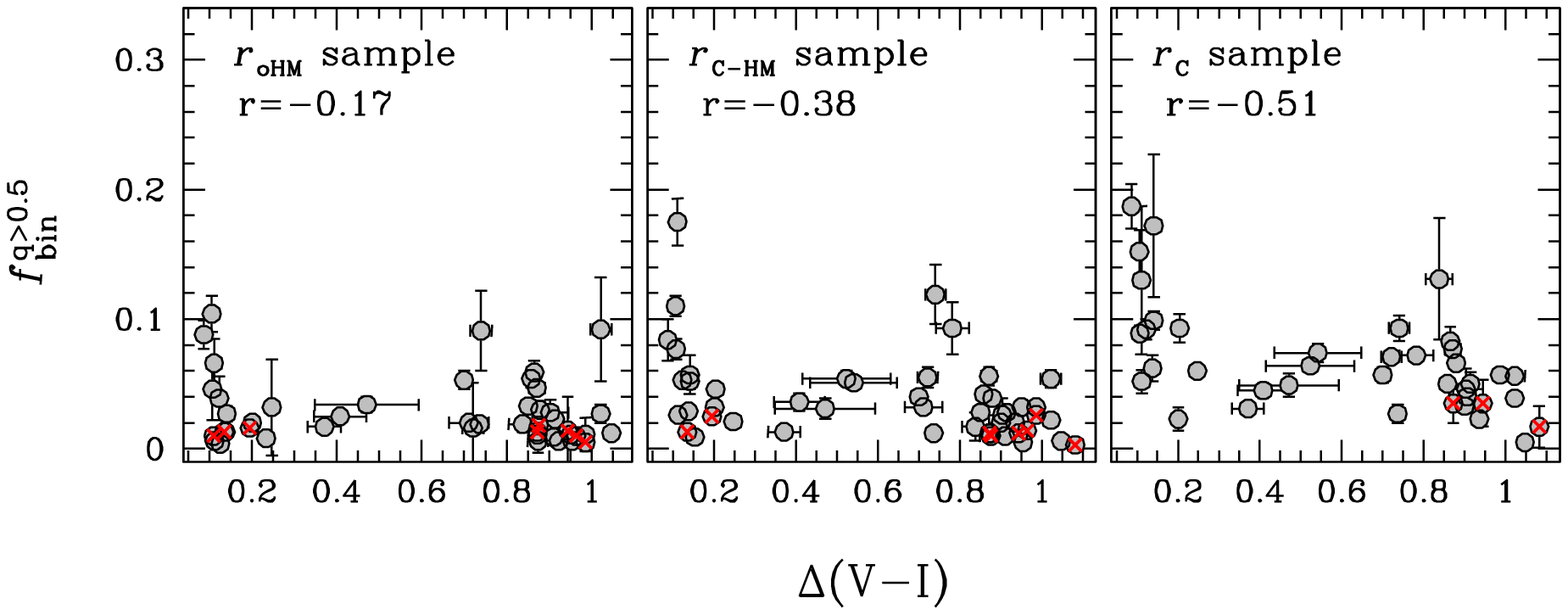}
   \includegraphics[width=11.0cm]{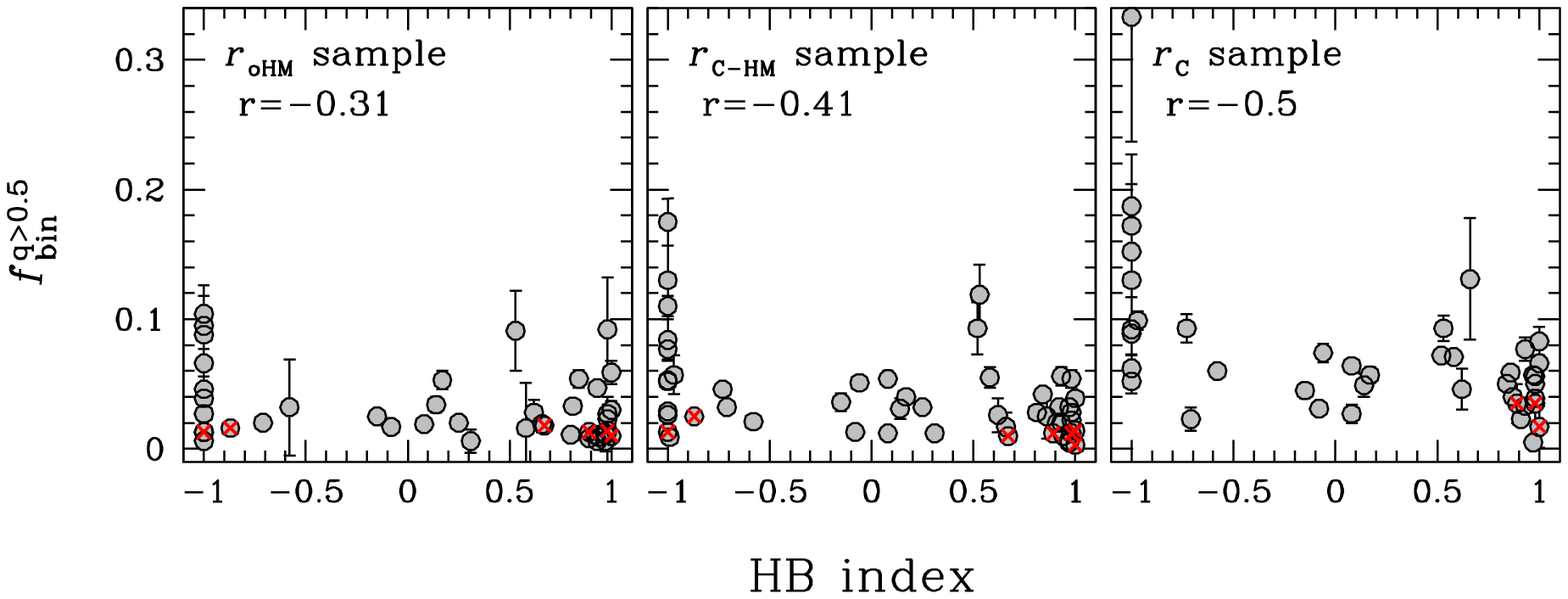}
   \includegraphics[width=11.0cm]{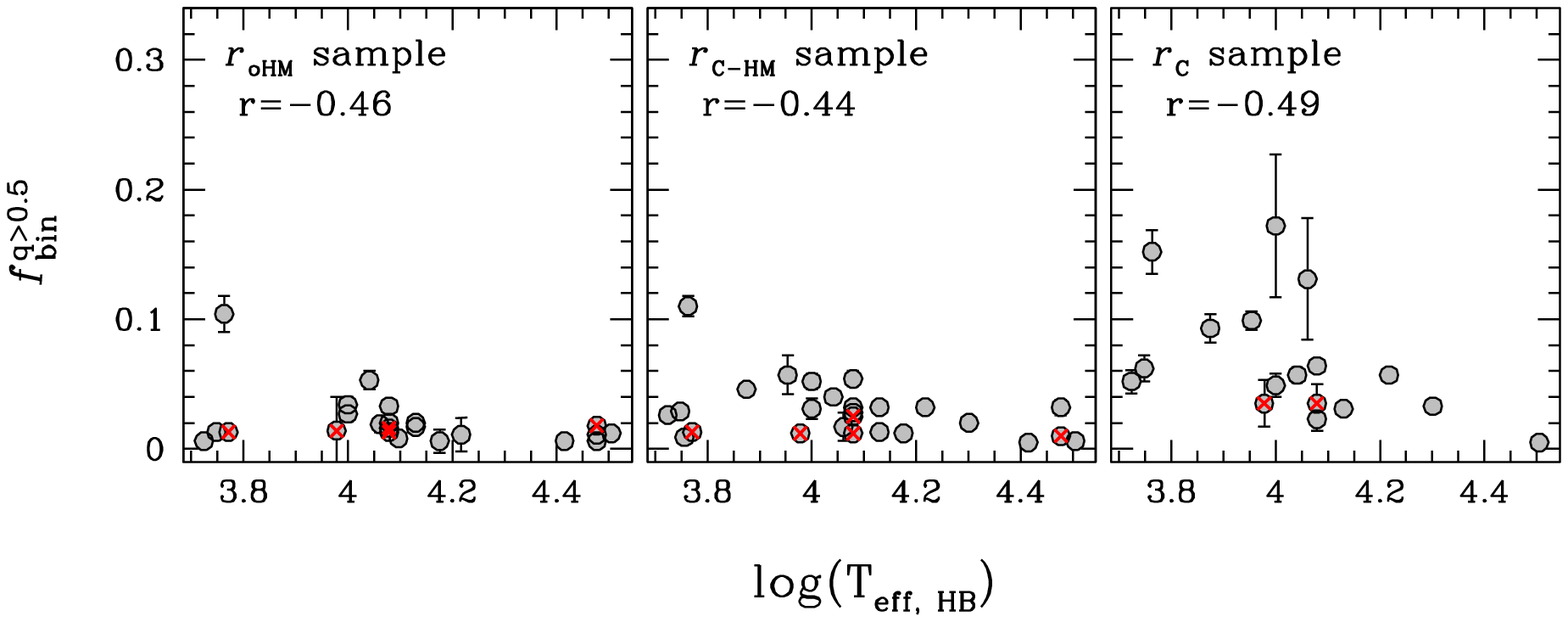}
      \caption{Fraction of binaries with {\it q}$>$ 0.5 as a function
	of the temperature of the hottest HB stars (\textit{bottom}), the HB morphology index (\textit{middle}), and the  median color difference between the HB and the RGB (\textit{top}).}
         \label{PARHB}
   \end{figure*}

Monovariate relation between the fraction of binaries with $q>0.5$ and these parameters are shown in Fig.~\ref{PARHB}.
We find no significant relations between the fraction of binaries and the HB
parameters, confirming the small or null impact of the binary
population on the HB morphology.
Similar results are obtained for binaries with $q>0.6$ and $q>0.7$.

%

\begin{center}
\begin{table*} 
\label{tabCORE}
\scriptsize{
\begin{tabular}{lllll}
\hline\hline
ID & ${\it f}_{\rm bin}$ & note & region & reference    \\
\hline
\hline
    E 3                   & 0.29$\pm$0.09          & lower limit & within 2 core radius   & Veronesi et al.\ (1996)            \\
\hline
    ARP 2                 & $>$0.08                & lower limit & within core   & Sollima et al.\ (2007)            \\
                          & 0.329-0.521            & all         & within core   & Sollima et al.\ (2007)            \\
\hline
    NGC 104 (47 Tucanae)  & 0.14$\pm$0.04          & all         & within half-mass radius    & Albrow et al.\ (2001)  \\
                          & $>$0.05                & lower limit & outside half-mass radius   & De Marchi \& Paresce \ (1995)  \\
                          & $\sim$0.02             & all         & outside    half-mass radius    & Anderson \ (1997)      \\
\hline
    NGC 288               & $>$0.10                & lower limit & 1-6 core radius                & Bolte \ (1992)         \\
                          & 0.10-0.20              & all         & within half-mass radius        & Bellazzini et al.\ (2002)         \\
                          & 0.01$^{+0.1}_{-0.0}$   & all         & outside    half-mass radiu     & Bellazzini et al.\ (2002)         \\
                          & $>$0.06                & lower limit & within core                    & Sollima et al.\ (2007)           \\
                          & 0.116-0.145              & all       & within core                    & Sollima et al.\ (2007)            \\
\hline
    NGC 362               & 0.21$\pm$0.06          & all         & within half-mass radius        & Fischer et al.\ (1993)    \\
\hline
    NGC 2808              & 0.20$\pm$0.04          & & outside half-mass radius  & Alcaino et al.\ (1998)           \\
                          & 0.04$\pm$0.01          & all & outside half-mass radius  & Milone et al.\ (2010)           \\
\hline
    NGC 3201              & $<$0.10                & upper limit & outside half-mass radius & Cote et al.\ (2007)   \\
\hline
    NGC 4590              & $>$0.09                & lower limit & within core   & Sollima et al.\ (2007)           \\
                          & 0.142-0.186              & all        & within core   & Sollima et al.\ (2007)            \\
\hline
    NGC 5053              & $>$0.08                & lower limit & within core   & Sollima et al.\ (2007)           \\
                          & 0.110-0.125              & all        & within core   & Sollima et al.\ (2007)            \\
\hline
    NGC 5139 ($\omega$ Centauri) & $<$0.05           & upper limit  &  outside half-mass radius  & Elson et al.\ (1995) \\
\hline
    NGC 5272 (M3)         & `low'                  & all          &  outside half-mass radius  & Gunn \& Griffin \ (1979) \\
                          & $\sim$0.04             & all          &  outside half-mass radius  & Pryor et al.\ (1988) \\
                          & 0.14$\pm$0.08          & all          &   inside half-mass radius  & Zhao \& Bailyn\ (2005) \\
                          & 0.02$^{+0.16}_{-0.02}$ & all          &  outside half-mass radius  & Zhao \& Bailyn\ (2005) \\
\hline
    NGC 5466              & $>$0.08                & lower limit  & within core   & Sollima et al.\ (2007)           \\
                          &  0.095-0.117              & all       & within core   & Sollima et al.\ (2007)            \\
\hline
    NGC 5897              & $>$0.07                & lower limit & within core   & Sollima et al.\ (2007)           \\
                          & 0.132-0.171              & all        & within core   & Sollima et al.\ (2007)            \\
\hline
    NGC 6101              & $>$0.09                & lower limit & within core   & Sollima et al.\ (2007)           \\
                          & 0.156-0.210              & all        & within core   & Sollima et al.\ (2007)            \\
\hline
    NGC 6121 (M4)         & $0.23^{+0.34}_{-0.23}$   & all        &  inside half-mass radius  & Cote \&  Fischer (1996) \\
                          & $\sim$0.02             & all         & outside half-mass radius & Richer et al.\ (2004) \\
\hline
    NGC 6341 (M92)        & $0.00^{+0.03}_{-0.00}$ & lower limit & outside half-mass radius & Anderson \ (1997)           \\
\hline
    NGC 6362              & $>$0.06                & lower limit & within core   & Sollima et al.\ (2007)           \\
                          & 0.118-0.127              & all        & within core   & Sollima et al.\ (2007)            \\
\hline
    NGC 6397              & $<$0.07                & upper limit  & within half-mass radius& Cool \& Bolton (2002)            \\
                          & 0.051$\pm$0.010        & all        & within half-mass radius& Davis et al.\ (2008)            \\
                          & 0.012$\pm$0.004        & all        & 1.3-2.8 half-mass radii& Davis et al.\ (2008)            \\
\hline
    NGC 6656 (M22)        &$0.03^{+0.16}_{-0.03}$    & all        & outside half-mass radius & Cote et al.\ (1996) \\
\hline
    NGC 6723              & $>$0.06                & lower limit & within core   & Sollima et al.\ (2007)           \\
                          & 0.161-0.218            & all        & within core   & Sollima et al.\ (2007)            \\
\hline
    NGC 6752              & 0.27$\pm$0.12         & all                        & within core & Rubenstein \& Bailyn (1997)       \\
                          & 0.03$\pm$0.01          & lower limit, {\it q}$>$0.5 & within core & Milone et al.\ (2010)             \\
                          & 0.02$^{+0.16}_{-0.02}$ & all                        & between core and half-mass radius  & Rubenstein \& Bailyn (1997)       \\
                          & 0.01$\pm$0.01          & lower limit, {\it q}$>$0.5 & between core and half-mass radius & Milone et al.\ (2010)             \\
\hline
    NGC 6792              & `low'                  & all & outside half-mass radius  & Catelan et al.\ (2008)           \\
\hline
    NGC 6809 (M55)        & $>$0.06                & lower limit & within core   & Sollima et al.\ (2007)            \\
                          &  0.096-0.108              & all        & within core   & Sollima et al.\ (2007)            \\
\hline
    NGC 6838 (M71)        & 0.22$^{+0.26}_{-0.12}$ & all                         & within half-mass radius  & Yan \& Mateo (1994)       \\
\hline
    NGC 6981              & $>$0.10                & lower limit & within core   & Sollima et al.\ (2007)            \\
                          &  0.281-0.399           & all          & within core   & Sollima et al.\ (2007)            \\
\hline
    NGC 7078 (M15)        & $\sim$0.07             & all          &  within half-mass radius  & Gebhardt et al.\ (1994)   \\
\hline
    NGC 7099 (M30)        & $<$0.05                & upper limit  & outside half-mass radius  &  Alcaino et al.\ (1998)   \\
\hline
   PALOMAR 12             & $>$0.18                & lower limit  & within core   & Sollima et al.\ (2007)            \\
                          & 0.408-0.506              & all        & within core   & Sollima et al.\ (2007)            \\
\hline
   PALOMAR 13             & $>$0.30$\pm$0.04       & lower limit & inside $\sim$18 core radii  & Clark et al. \ (2004)            \\
\hline
    TERZAN 7              & $>$0.21                & lower limit & within core   & Sollima et al.\ (2007)            \\
                          & 0.509-0.649              & all        & within core   & Sollima et al.\ (2007)            \\
                          &   &     &         \\
\hline
\end{tabular}
}
\caption{Collection of literature binary fraction estimates. For each
  GC we listed the measured fraction of binaries ($f_{\rm bin}$),
  specified if the latter is a lower limit, an upper limit or a
  measure of the total  fraction of binaries. We also indicate the spatial
  region where this measure was done and give the reference.}
\end{table*}
\end{center}

\section{Summary}
In this paper we have analyzed the properties of the population of MS binaries
of a sample of 59 GCs.
The main dataset consists in the ACS/WFC images of
the Globular Clusters Treasury project (GO10775, PI Sarajedini) that
allowed us to obtain uniform and deep photometry for an unprecedented
number of GCs (see Sarajedini et al.\ 2007 and
Anderson et al.\ 2008 for details).
We have also used ACS/WFC, WFC3 and WFPC2 data from the archive to obtain
proper motions when images overlapping the GO10775 data are available.
The CMDs have been corrected for the effects of differential reddening and
photometric zero point variations due to small inaccuracies in the PSF model.

We have measured the fraction of binaries with mass ratio {\it
  q}$>$0.5 and estimated the total fraction of binaries for MS stars
that are between 0.75 and 3.75 magnitudes fainter than the MS turn
off. We have found that:

\begin{itemize}
\item in nearly  all the GCs the fraction of binaries is significantly smaller
than  in the field,  where the binary fraction is larger than 0.5
  (e.\ g.\ Duquennoy et al.\ 1991, Fisher \& Marcy 1992)
 with a few relevant exceptions (E3, Palomar 1) 
where the total binary fraction is greater than $\sim$0.4.

\item
We have obtained the fraction of binaries in five intervals of {\it q}
(for {\it q}$>$0.5) and found  that the mass-ratio distribution is
generally flat.

\item
There is no evidence for a significant correlation of the binary
fraction with primary mass of the binary system.

\item
We measured the fraction of binaries in the cluster core, in the
region between the core and the half-mass radius, and outside the
half-mass radius and studied their radial distribution.
Binary stars are more centrally concentrated
than single MS stars with the fraction of binaries generally
decreasing by a factor of $\sim$2 from the center to about two core radii. 

\item
We investigated monovariate relations between the fraction of binaries
(in the ${\it r}_{\rm C}$, ${\it r}_{\rm C-HM}$, and ${\it r}_{\rm oHM}$ sample)
and the main parameters of their host GCs (absolute magnitude, HB
morphology, age, ellipticity, metallicity, collisional parameter, half
mass and core relaxation time, central concentration, central velocity
dispersion, and central luminosity density).

\item
We found a significant anticorrelation between the fraction of binaries
 in a GC and its absolute luminosity (mass).

\item
We found a marginal correlation between the cluster central density and
the central velocity dispersion.

\item
We did not find any significant relation between the binary fraction and the
HB morphological parameters.

\item
We  confirm a significant correlation between the
fraction of binaries and the fraction of BSSs, indicating that the main formation mechanism of BSSs must be related to binary evolution.

\end{itemize}

\begin{acknowledgements}
We wish to warmly tank
Ivan R.\ King without whom most of the results presented in this review would not have been possible.
 We are really gratefull to the referee for the excellent and huge work she/he did.
We also thank Andrea Bellini for carefully reading this manuscript,
Antonio Sollima for usefull discussion, and Edoardo La Gioia for
helping us in the images treatment. 
A.\ P.\ M., A.\ A., and G.\ P.\ are founded by the ministry of
science and technology of 
the Kingdom of Spain (grant AYA 2010-16717). A.\ P.\ M. and A.\ A.\ are also
founded by the Istituto de Astrofisica de Canarias (grant P3-94). 
G.\ P. and A.\ P.\ M.\ acknowledge partial support by MIUR under
the program PRIN2007 (prot.\ 20075TP5K9) and by ASI under the program
ASI-INAF I/016/07/0.
\end{acknowledgements}

\begin{figure}[ht!]
   \centering
   \includegraphics[width=8.5cm]{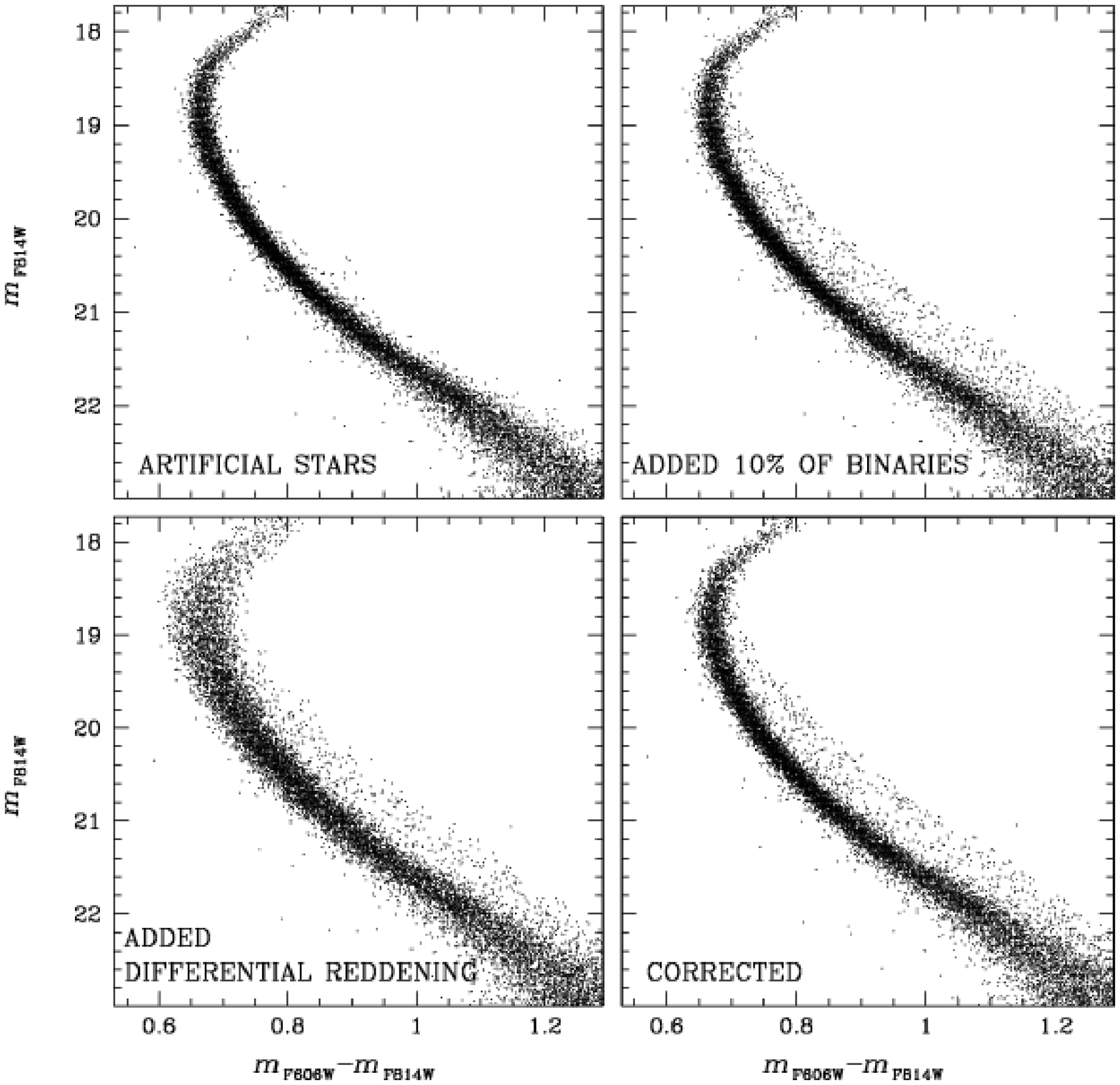}
      \caption{Artificial stars CMD for NGC 2298 (\textit{upper-left}) and
        simulated CMD with a fraction of $f_{\rm bin}^{\rm TOT}$0.10 of
        binaries added (\textit{upper-right}). Bottom panels show the
        simulated CMD after we added differential reddening ({\it left})
        and the simulated CMD after the correction for differential
        reddening (\textit{right}).}
         \label{SIMUCMD}
   \end{figure}

{\bf APPENDIX A. Reliability of the measured binary fraction.}
\label{appendice}
\\
In this appendix we investigate whether
the fraction of binaries with {\it q}$>$0.5 that we measured with
the procedure described in
Sect.~\ref{sec:highq} are reliable or are affected by any
systematic uncertainty due to the method we used.
The basic idea of this test consists of simulating a number of CMDs
with a given fraction of binaries, measuring the fraction of binaries
in each of them, and comparing the added fraction of binaries  with
the measured ones.\\
{\bf Simulation of the CMD.}\\
We started by using artificial stars to simulate a CMD made of single
stars following the procedure already described in Sect.~\ref{blends}.
To simulate binary stars to be added to the simulated CMD
we adopted the following procedure:
\begin {itemize}
\item{We selected a fraction $f_{\rm bin}^{\rm TOT}$
of  single stars  equal  to the fraction  of  binaries that we want to
add to the CMD and   derived  their  masses  by  using  the
  Dotter  et  al.\ (2007)  mass-luminosity  relation. In our
  simulations we  assumed the values of $f_{\rm bin}^{\rm TOT}=$0.05, 0.10, 0.30, and 0.50.}
\item{For each of them, we calculated the mass $\mathcal{M}_{2}=q \times \mathcal{M}_{1}$ of the
  secondary star and  obtained the corresponding  $m_{\rm F814W}$
  magnitude. Its  color was derived by the MSRL. For simplicity we
  assumed a flat mass-ratio distribution.}
\item {Finally, we summed up the F606W and F814W fluxes of the two
  components, calculated the corresponding magnitudes, added the
  corresponding photometric error, and replaced the original
  star in the CMD with this binary system.}
\end{itemize}
As an example, in the upper panels of Fig.~\ref{SIMUCMD} we show the
 artificial star CMD made of single stars only (left panel), and the
 CMD where we added a fraction $f_{\rm bin}^{\rm TOT}$=0.10 of binaries
 (right panel), for the case of NGC 2298.\\
%
{\bf Simulation of the differential reddening.}\\
To probe how well the reddening correction works, we considered a
simple model.
The simulation of the differential reddening suffered by any single
star is far from trivial as we have poor information on the structure
of the interstellar medium between us and each GC.
For simplicity, in this work we assumed that reddening variations
are related to the positions (X, Y) of each stars by the following
relations:\\

\begin{figure}[ht!]
   \centering
   \includegraphics[width=8.5cm]{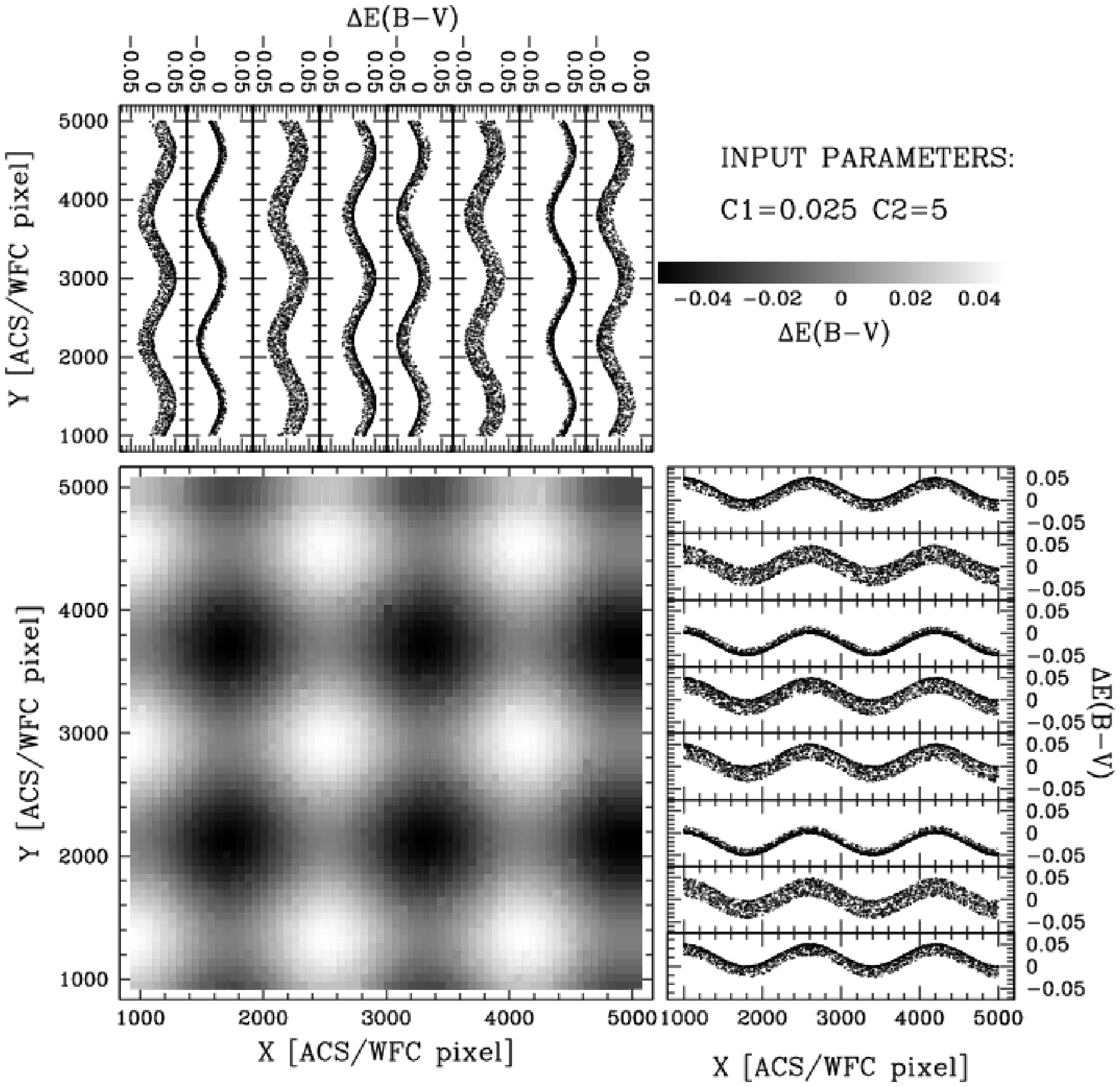}
      \caption{\textit{Bottom-left}: Map of differential reddening added to
        the simulated CMD of NGC 2298. The gray levels indicate the
        reddening  variations as indicated in the \textit{upper-right}
        panel. \textit{Upper-left} and \textit{bottom-right} panels show
        $\Delta E(B-V)$ as a function of the Y and X coordinate
        respectively for stars into 8 vertical and horizontal  slices.}
         \label{SIMURED}
   \end{figure}

$
\Delta E(B-V)=C_{1} (cos(X')+sin(Y'))
$\\
where\\
$X'=C_{2} \pi (X-X_{\rm MAX})/(X_{\rm MAX}-X_{\rm MIN})$, \\
$Y'=C_{2} \pi (Y-Y_{\rm MAX})/(Y_{\rm MAX}-Y_{\rm MIN})$.\\
Here $X_{\rm MIN, MAX}$ and $Y_{\rm MIN, MAX}$ are the minimum and the maximum
values of the coordinates X and Y, $C_{1}$ is a free parameter that
determines the maximum amplitude $E(B-V)$ variation, and $C_{2}$
governs the number of differential reddening peaks within the field of
view.
In this work, we used for each GC the value of $C_{1}$ that ranges
from 0.005 to 0.05 to account for
the observed reddening variation in all the GCs, while we arbitrarily assumed three
values of  $C_{2}$=3, 5, and 8 to reproduce three different fine-scales of
differential reddening.
As an example, in Fig.~\ref{SIMURED} we
show the map of differential reddening added to
        the simulated CMD of NGC 2298 that is obtained by assuming
        $C_{1}=0.025$ and $C_{2}=5$.

Then, we have transformed the values of $\Delta$E(B-V) corresponding to the
position of each stars into $\Delta A_{\rm F606W}$, and  $\Delta
A_{\rm F814W}$ and added these absorption variations to the F606W and
the F814W magnitudes. The CMD obtained after we added differential
reddening is shown in the bottom left panel of Fig.~\ref{SIMUCMD} for
NGC 2298. We applied to this simulated CMD the procedure to
correct for differential reddening described in Sect.~\ref{reddening} and obtain the corrected CMD  shown in the bottom right panel.
For each of these binary-enhanced simulated CMD, we also generated a CMD made of artificial stars by following the approach
described in Sect.~\ref{blends}.
In our investigation we did not account for field stars.
For each combination of the $f_{\rm  bin}^{\rm TOT}$ and $C_{2}$ we have simulated 200 CMDs with random values of the $C_{1}$.
\\

{\bf Measurements of the binary fraction.}\\
Finally, we used the procedure of
Sect.~\ref{sec:highq} to measure the fraction of binaries with mass
ratio {\it q}$>$0.5 defined as:\\
$
f_{\rm bin}^{\rm q>0.5}=\frac {N_{\rm SIMU}^{\rm B}} {N_{\rm SIMU}^{\rm A}}
 - \frac {N_{\rm ART}^{\rm B}}{N_{\rm ART}^{\rm A}}
$ \\
where  $N_{\rm SIMU}^{\rm A}$ and $N_{\rm SIMU}^{\rm B}$ are the numbers of stars in the regions {\it A} and {\it B} in the CMD,
as defined in
Fig.~\ref{reg1}
 in the binary-enhanced simulated CMD and $N_{\rm ART}^{\rm A}$ and $N_{\rm ART}^{\rm B}$ the numbers of stars in the same regions of the artificial stars CMD.
%

\begin{figure}[ht!]
   \centering
   \includegraphics[width=8.5cm]{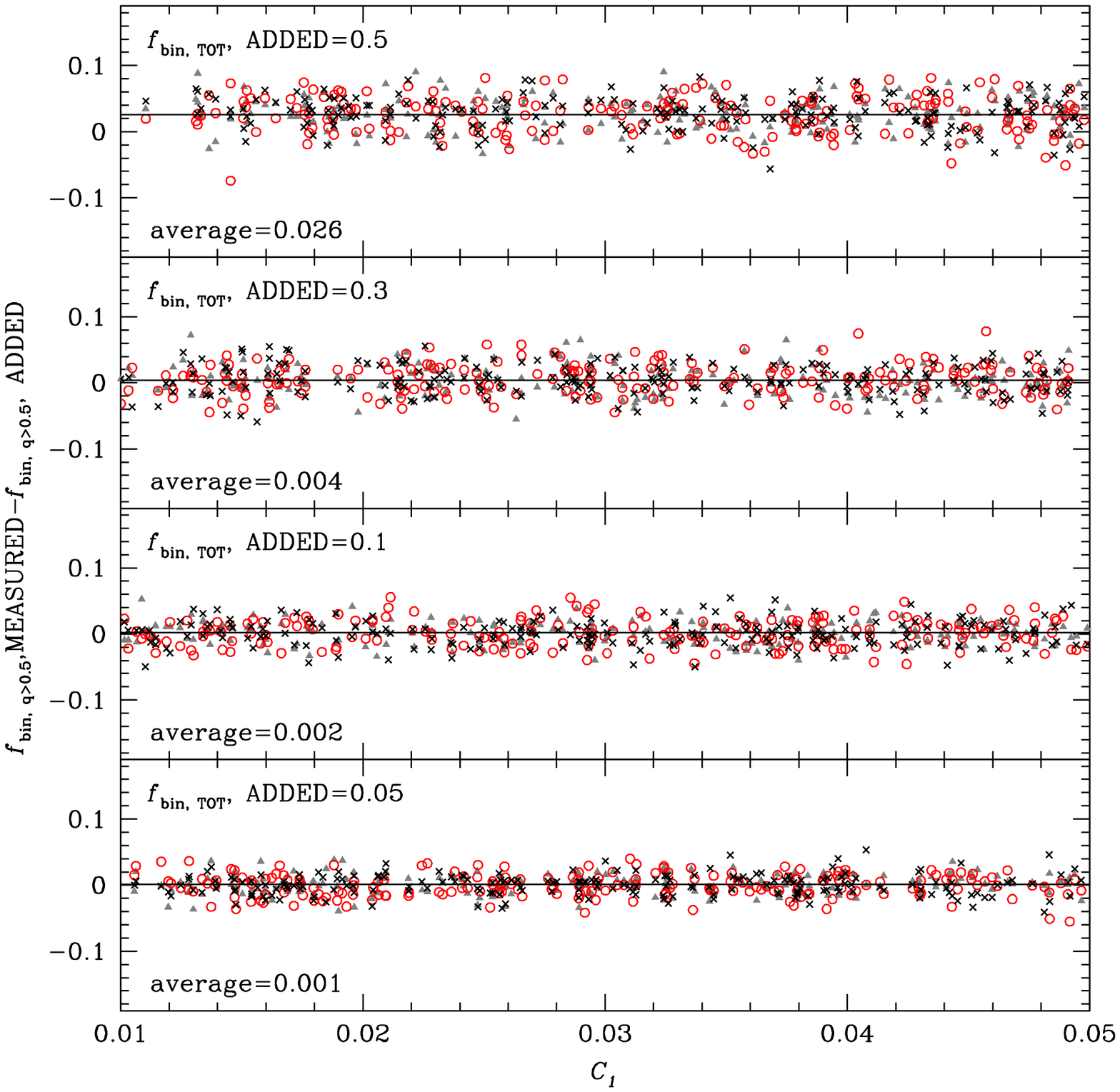}
      \caption{Difference between the measured fraction of binaries
        and the fraction of binaries in input as a function of the
        parameter $C_{1}$ for four difference values of the input
        binary fraction. Black lines indicate the average
        difference. Red circles, gray triangles and black crosses
      indicate simulations with $C_{2}=$3, 5, and 8 respectively.}
         \label{RISSIMU}
   \end{figure}

Results are shown in Fig.~\ref{RISSIMU} where we  plotted
the difference between the measured and the input fraction of binaries
        versus the parameter $C_{1}$ for four difference values of the
        input binary fraction. We found that, for input binary
        fraction of 0.05, 0.10 and 0.30, the average difference are negligible
($< 0.5\%$), as indicated by the the black lines and the numbers
        quoted in the inset.
 In the case of a large binary fraction ($f_{\rm bin}^{\rm TOT}$=0.5) the measured fraction of binaries with {\it
          q}$>$0.5 is systematically underestimated by $\sim$0.03.
        Apparently our results do not depend on the value of the
        parameter $C_{2}$.
      Simulations with $C_{2}=$3, 5, and 8 ( indicated in Fig.~\ref{RISSIMU} with red circles, gray
      triangles, and black crosses, respectively)
      give indeed the same average differences.
Our comparison between the fraction of binaries added to the simulated CMD and the measured ones demonstrate that the fraction of binaries determined in this work and listed in Table 2 are not affected by 
any significant systematic errors related to the procedure we adopted.
%

\begin{figure}[ht!]
   \centering
   \includegraphics[width=8.5cm]{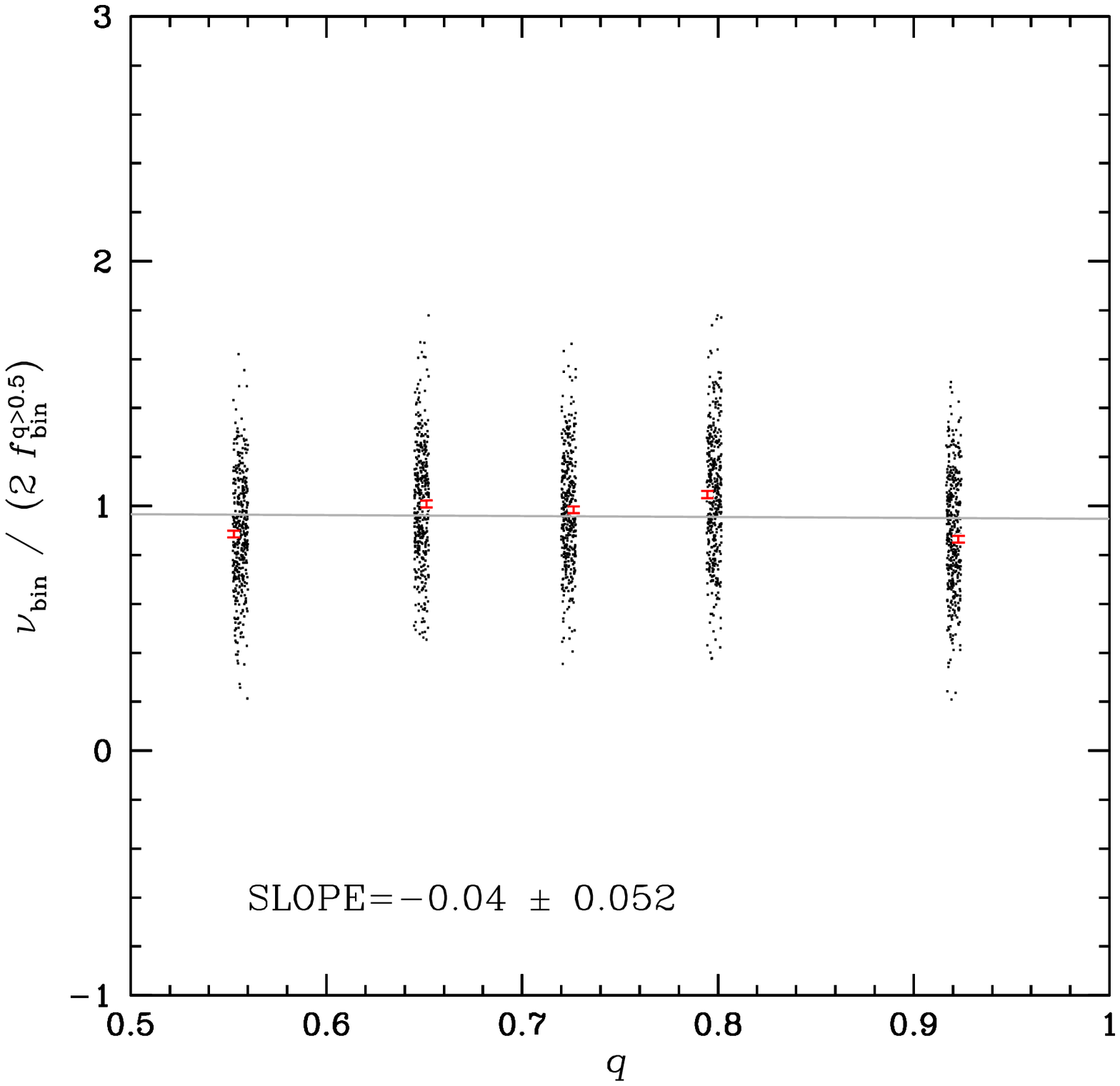}
      \caption{
	Fractions of binaries per unit {\it q}
	measured in five mass-ratio intervals as a function of {\it q}
        for all the simulated GCs. 
	To compare the {\it q} distribution in simulated clusters with
        different fraction of binaries, we divided 
        $\nu_{\rm bin}$ by two times  the fraction of binaries with {\it q}$>$0.5.
	For clarity, black points have been randomly scattered around
        the corresponding  {\it q} value. 
	The means normalized binary fractions in each mass-ratio bin are represented by red points with error bars, while the
	gray line is the best fitting line, whose slope is quoted in
        the inset. 
}
         \label{QSIMU}
   \end{figure}

We have also determined the fraction of binaries in five mass-ratio intervals by following the approach described in Sect.~\ref{sec:qdist}
 for real stars. To this aim, we have divided the region {\it B} of
 the CMD defined in Sect.~\ref{reg1} into five subregions as
 illustrated in Fig.~\ref{qsetup} for real stars. The size of each
 region is chosen in such a way that each of them covers
a portion of the CMD with almost the same area.
The resulting mass-ratio distribution is shown in Fig.~\ref{QSIMU},
where we have plotted the fraction of binaries per unit {\it q} as a
function of the mass ratio. As already done in the case of real stars,
to compare the mass-ratio distribution in simulated CMDs with
different binary fraction, we have divided $\nu_{\rm bin}$ by two
times the measured fraction of binaries with {\it q}$>$0.5. The best
fitting gray line closely reproduce the flat mass-ratio distribution
in input with $\nu_{\rm bin}$=1. 
%

\begin{figure}[ht!]
   \centering
   \includegraphics[width=8.5cm]{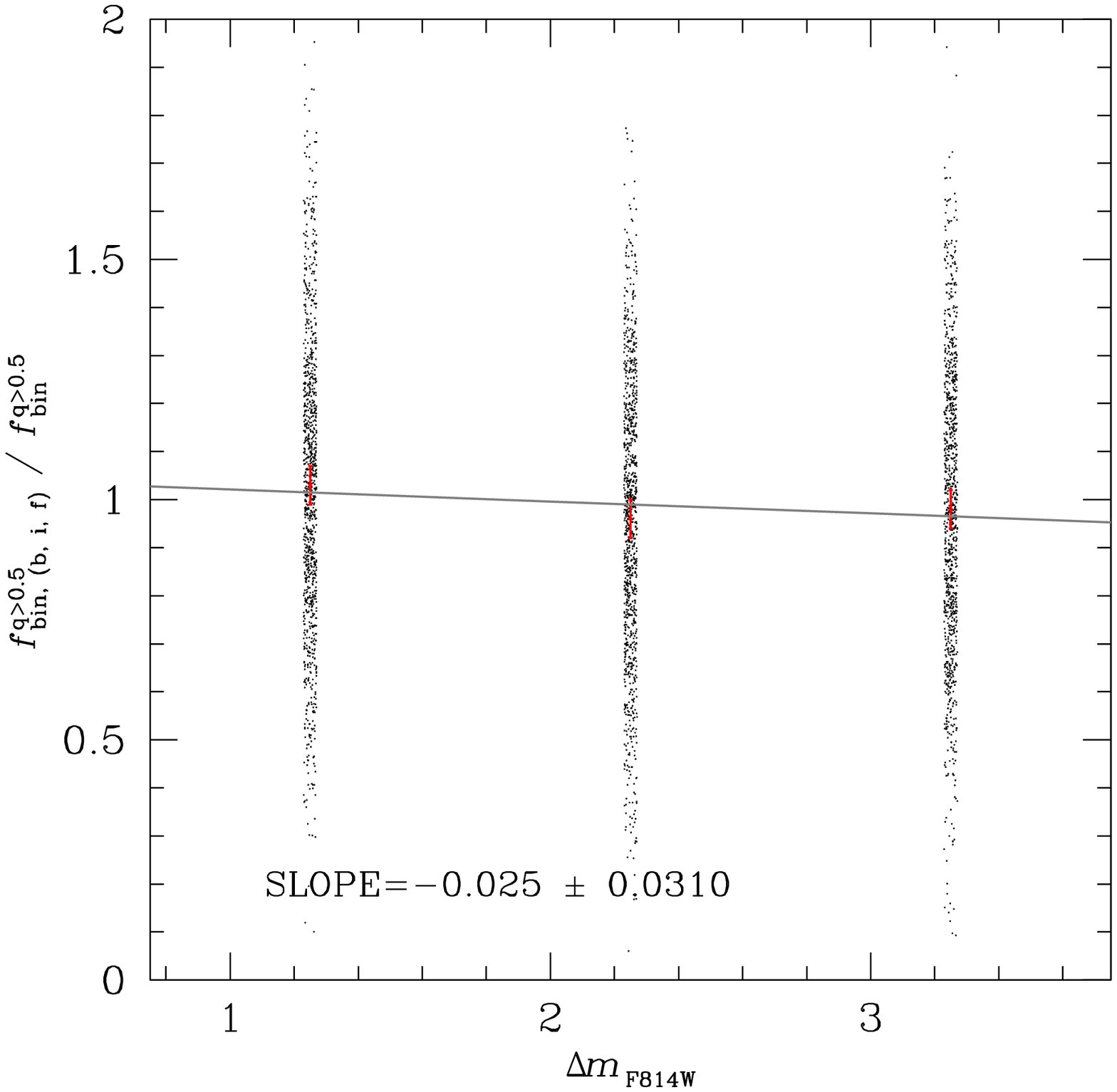}
      \caption{ Fraction of binaries with {\it q}$>$0.5 in three
        magnitude intervals as a function of $\Delta {\it m}_{\rm
          F814W}$ for all the simulated GCs. To compare the measured
        fraction of binaries  in different clusters we have divided
        the measured binary fractions in each magnitude interval by
        the value of ${\it f}^{\rm q>0.5}_{\rm bin}$ measured in the
        interval between 0.75 and 3.75 F814W magnitudes below the MS
        turn off. Red points with error bars are the means normalized binary
        fractions in each magnitude interval. The gray line with the
        quoted slope is the best-fitting least-squares line.
      }
         \label{MSIMU}
   \end{figure}

 Finally we have measured in the simulated CMDs the fraction of
binaries with ${\it q}>0.5$ in three intervals
[0.75,1.75], [1.75,2.75], and [2.75,3.75] F814W magnitudes below the
MSTO. To do this we used the procedure already described in
Sect.~\ref{magdistr} for real stars and we have normalized  the ${\it f}^{\rm
  q>0.5}_{\rm bin}$ value measured in each magnitude bin by the
fraction of binaries with  ${\it q}>0.5$ measured in the whole
interval between 0.75 and 3.75 F814W magnitudes below the MSTO.
Results are shown in Fig.~\ref{MSIMU} where we have plotted the
normalized binary fractions as a function of $\Delta {\it m}_{\rm
  F814W}$. The best-fitting gray line is nearly flat, and well
reproduces the input magnitude distribution.

%
These tests demonstrate that both the mass-ratio distribution
determined in Sect.~\ref{sec:qdist} for the 59 GCs studied in this
work and shown in Fig.~\ref{QDALL}  as well as the binary fractions
measured in different magnitude intervals in Sect.~\ref{magdistr} are  not
biased by significant systematic errors related to the procedure we
adopted. 

\bibliographystyle{aa}

\end{document}